# Robust Atom Interferometry with Double Bragg Diffraction

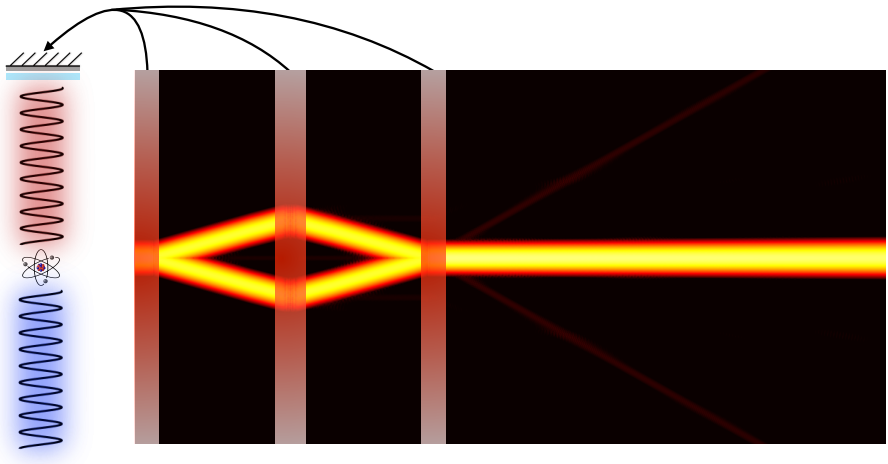

**M. Sc. Rui Li**

geboren in Jinan, China

2026

**Referent:**

Dr. Naceur Gaaloul
Institut für Quantenoptik
Leibniz Universität Hannover

**Korreferent:**

Prof. Dr. Klemens Hammerer
Institut für Theoretische Physik
Leibniz Universität Hannover

**Korreferent:**

Prof. Dr. Enno Giese
Institut für Angewandte Physik
Technische Universität Darmstadt

**Tag der Promotion:**

26.01.2026

# ABSTRACT

This thesis develops a general theoretical and numerical framework for achieving high-contrast atom interferometry based on double Bragg diffraction (DBD). While DBD offers intrinsic symmetry, reduced sensitivity to internal-state systematics, and suitability for microgravity experiments, its performance has long been limited by imperfect diffraction and contrast loss. This work overcomes these limitations by constructing an analytic Hamiltonian description of DBD—including Doppler effects and polarization imperfections—and by deriving reduced two- and five-level models via a truncated Magnus-expansion approach. These models clarify the origin of AC-Stark shifts, polarization-induced errors, and Doppler selectivity, and they provide accurate predictions for realistic input momentum distributions.

Building on this theoretical foundation, the thesis introduces a tri-frequency laser scheme with dynamically tunable detuning and evaluates different detuning-control strategies using a five-level S-matrix formalism. Linear detuning sweeps and optimal-control pulses are shown to provide near-ideal beam-splitter and mirror performance, respectively, ensuring robust contrast across a wide range of experimental imperfections. Complementary full three-dimensional simulations using the GPU-accelerated *Universal Atom Interferometer Simulator* (UATIS) incorporate interacting Bose–Einstein condensates and realistic optical potentials, revealing transverse effects and polarization-induced distortions that extend the predictions of the one-dimensional non-interacting models. Taken together, this thesis establishes a coherent theoretical and numerical framework demonstrating that, with appropriate detuning control, double-Bragg atom interferometers can reach the robustness required for precision inertial sensing and future space-based quantum tests of fundamental physics.

# ZUSAMMENFASSUNG

Diese Dissertation entwickelt einen allgemeinen theoretischen und numerischen Rahmen für hochkontrastreiche Atominterferometrie auf Basis der doppelten Bragg-Diffraktion (DBD). Obwohl DBD intrinsische Symmetrie, reduzierte Empfindlichkeit gegenüber systematischen Störeinflüssen und besondere Eignung für Mikroschwerelosigkeit bietet, wurde ihre Leistungsfähigkeit lange durch unvollständige Diffraktion und Kontrastverluste begrenzt. Die Arbeit überwindet diese Einschränkungen durch eine analytische Hamilton-Beschreibung der DBD – einschließlich Dopplereffekten und Polarisationsfehlern – sowie durch reduzierte Zwei- und Fünf-Niveau-Modelle auf Grundlage einer abgeschnittenen Magnus-Expansion. Diese Modelle klären die Ursprünge von AC-Stark-Verschiebungen, polarisationsbedingten Fehlern und Dopplerselektivität und liefern präzise Vorhersagen für realistische Impulsverteilungen.

Aufbauend auf diesem theoretischen Fundament führt die Dissertation ein dreifrequentes Laserschema mit dynamisch einstellbarer Detuning-Führung ein und bewertet verschiedene Detuning-Kontrollstrategien mithilfe eines fünfniveaubasierten S-Matrix-Formalismus. Lineare Detuning-Sweeps und optimal-kontrollierte Pulse ermöglichen nahezu ideale Strahlteiler-bzw. Spiegeloperationen und sichern robusten Kontrast unter vielfältigen experimentellen Imperfektionen. Ergänzende GPU-beschleunigte 3D-Simulationen mit dem *Universal Atom Interferometer Simulator* (UATIS) berücksichtigen wechselwirkende Bose–Einstein-Kondensate und realistische optische Potenziale und zeigen transversale Effekte sowie polarisationsinduzierte Verzerrungen, die über eindimensionale Modelle hinausgehen. Insgesamt demonstriert die Dissertation, dass doppelt-Bragg-basierte Atominterferometer mit geeigneter Detuning-Kontrolle die nötige Robustheit für präzise inertiale Messungen und zukünftige weltraumgestützte Quantentests grundlegender Physik erreichen können.

**Schlagwörter:** Doppelte Bragg-Diffraktion, Atominterferometrie, S-Matrix-Theorie, adiabatische Passage, Optimierungskontrolltheorie, AC-Stark-Verschiebungen, Doppler-Detuning, ultrakalte Atome, Bose–Einstein-Kondensat, quantumpräzise Sensorik.

**DEDICATION**

This dissertation is dedicated to my mother and father,
and to my loved one, whose constant love, patience, and support
have guided me through every step of this journey.

# ACKNOWLEDGMENTS

This thesis would not have been accomplished without the support of many friends, colleagues, family and loved ones. Among all the people, I must chiefly acknowledge Klemens Hammerer and Naceur Gaaloul for the guidance, the generous helps and numerous fruitful discussions along my PhD study, to whom I own a great amount of gratitude. I would also thank Enno Giese for kindly agreeing to review my thesis and Fei Ding for chairing my defense.

Now, let us rewind time to the moment I completed my Master's studies at TU Kaiserslautern (RPTU) in Michael Fleischhauer's group at the end of 2020. The COVID-19 pandemic had just begun, and I was searching for PhD positions in Germany when Axel Pelster forwarded me an email advertising an opening at Leibniz University Hannover. That message ultimately led to my connection with the Leibniz University, the Institute of Quantum Optics, and the Theory of Quantum Sensing (T-SQUAD) group led by Naceur. I submitted my application and was fortunate to be offered the position.

Upon my arrival, I was warmly welcomed by the group members at the time—Naceur, Florian, Sina, Annie, Matty, Jan-Niclas, Stefan, Gabriel, and Christian. I still vividly remember our first group lunch at the Mexican burrito restaurant near E-Damm. At the beginning of my PhD, I owe special thanks to Florian Fitzek, from whom I learned Fortran and became familiar with the Fortran-based Universal Atom Interferometer Simulator (UATIS). This codebase later evolved—through the dedicated work of Stefan Seckmeyer—into the main computational tool responsible for producing many of the exact numerical simulations presented in this thesis.

After attending the first plenary group meeting of AG Rasel (of which T-SQUAD was a subgroup at the time), I realized how many outstanding experimentalists were working

here—performing atom-interferometry experiments in almost every environment imaginable: on the ground (Q-Port, Q-Gyro, VLBAI), on sounding rockets (MAIUS), in the drop tower (QUANTUS), and even aboard the International Space Station (CAL/BECCAL). I was first assigned to support the QUANTUS-1 team, where I met my experimental collaborators Simon Kanthak, Mikhail Cheredinov, and Sven Abend, as well as the head of AG Rasel, Ernst Maria Rasel. Through this collaboration, I gained my first direct insight into how a double-Bragg atom interferometry experiment is performed in the laboratory. Initially, my main task was to perform 3D simulations of the optical dipole potentials using parameters supplied by the experimental team, and to compare the simulated results with the measured data.

Later, seeking a deeper physical understanding of the simulation results, and through an introduction by Naceur during the first Denmark Seminar after the pandemic, I had the opportunity to meet Klemens and attend his Theoretical Quantum Optics Lecture. After one of the lectures—in which the Magnus expansion was introduced—I showed Klemens some of the simulation results for double Bragg diffraction in the experimentally relevant parameter regime. At that time, I could not make sense of the Rabi frequency scan: the dynamics were clearly non-sinusoidal at small Rabi frequencies, and higher-order diffraction processes appeared once the Rabi frequency crossed certain thresholds. This discussion was one of the first important steps toward the theoretical developments that ultimately became the Chapter 3 of this thesis. After the completion of the full theoretical framework of double Bragg diffraction—including AC-Stark shifts, polarization imperfections, and Doppler detuning—we sought to enhance the efficiency of DBD pulses beyond what adiabatic passage techniques could achieve. This motivated the application of optimal control theory (OCT), for which Victor Martinez had the expertise and the software tool. The collaboration proceeded smoothly, and we accumulated a lot of good results in a relatively short time addressing different experimental imperfections. At the same time, we also shared the high-efficiency pulses we had developed with our colleagues in the QUANTUS-1 collabora-tion. Remarkably, they worked in the experiment exactly as predicted—a deeply rewarding moment that connected the theory directly with laboratory reality. For this, I would like to sincerely thank all experimental colleagues who made this "physics-in-your-face," hands-on experience possible, in particular Sven, Simon, Mikhail, and Ekim. Their openness, patience, and enthusiasm greatly enriched my understanding of real-world atom-interferometry experiments and strengthened the collaboration between theory and experiment. If time permits, I am confident that we will be able to publish a collaborative work in which theory and experiment go hand in hand.

After tackling the dynamics of a single double-Bragg pulse, it is natural to pursue a full theoretical description of the complete Mach-Zehnder interferometer sequence. In this context, I developed a preliminary theoretical model and applied it to the Q-Port experiment, where it helped explain the beating behavior observed in the measured interrogation-time T-scan fringes of the 3-by-3 BEC arrays. This work contributed to the publication of Knut Stolzenberg and Christian Struckman. Later, with the guidance of Klemens and Naceur, we extended this preliminary theoretical model into a general framework capable of describing double-Bragg-diffraction interferometers under external acceleration. This development ultimately formed the basis of Chapter 4 of this thesis. I am grateful to Knut and Christian for many insightful discussions regarding the Q-Port experiments, and to Dennis Schlippert for his support throughout this collaboration.

Moreover, I would like to express my gratitude to the members of the AG Hammerer group, with whom I had the pleasure of interacting during the final stage of my PhD through numerous group meetings, journal clubs, and group activities. In particular, I thank Patrick, Ivan, Maja, Marius, Kasper, Tim, Erin, Julian, Ruolin, and many others for their stimulating discussions and their collegial support. I also extend my sincere thanks to the administrative staff who assisted me throughout my doctoral studies—including Birgit Gemmeke, Birgit Ohlendorf, and Anne-Dore Göldner-Pauer—for their patience, kindness, and invaluable help.

Last but certainly not least, I wish to express my deepest gratitude to my family and my loved ones. Their unwavering support, patience, and encouragement have sustained me throughout every stage of this journey. My parents' steadfast belief in me has been a continuous source of strength, and the love and understanding of my girlfriend, Haiyue, have made even the most demanding moments manageable. Without their constant support—both near and far—this work would not have been possible.

# Author Contributions

Parts of this thesis have appeared elsewhere as peer-reviewed publications or on preprint servers. Here I summarize my personal contributions to these works. Please note that the contributions of other authors without my involvement are not explicitly listed. In the order of the corresponding chapters, these are:

**Chapter 2 (from Sec. 2.2) and Chapter 3:** **Rui Li**, V. J. Martínez-Lahuerta, S. Seckmeyer, Klemens Hammerer, and Naceur Gaaloul, *Robust double Bragg diffraction via detuning control*, Phys. Rev. Research **6**, 043236 (2024).

**Author contribution: R.L.** and **K.H.** conceived the idea. **R.L.** performed the analytical modeling and the exact numerical calculations with UATIS Python package developed by **S.S.**. **V.M.-L.** performed the optimizations with OCT using Q-CTRL's Boulder Opal package. **R.L.** and **V.M.-L.** drafted the manuscript with the participation of all co-authors.

**Chapter 4:** **Rui Li**, Víctor José Martínez-Lahuerta, Naceur Gaaloul, and Klemens Hammerer, *High-contrast double Bragg interferometry via detuning control*, AVS Quantum Sci. **8**, 014402 (2026) (**Editor's Pick**).

**Author contribution: R.L.**, **K.H.** and **N.G.** conceived the idea. **R.L.** performed the analytical modeling and the numerical validation. **V.M.-L.** performed the optimizations with OCT. **R.L.** drafted the manuscript with the participation of all co-authors.

**Additional publications (partially contributing):**

- K. Stolzenberg, C. Struckmann, S. Bode, **R. Li**, A. Herbst, V. Vollenkemper, D. Thomas, A. Rajagopalan, E. M. Rasel, N. Gaaloul, and D. Schlippert, *Multi-Axis Inertial Sensing with 2D Matter-Wave Arrays*, Phys. Rev. Lett. **134**, 143601 (2025).

# Abbreviations and Acronyms

**AI**           Atom interferometry

**BEC**          Bose–Einstein condensate

**SBD**          Single Bragg diffraction

**DBD**          Double Bragg diffraction

**H.c.**         Hermitian conjugate

**COM**          Center-of-mass

**QPN**          Quantum projection noise

**QFT**          Quantum field theory

**SQL**          Standard quantum limit

**UFF**          Universality of free fall

**UCR**          Universality of clock rates

**LMT**          Large momentum transfer

**DKC**          Delta-kick collimation/cooling

**LMT**          Large momentum transfer

**MZI**          Mach–Zehnder interferometer

**MZAI**         Mach–Zehnder atom interferometer

**RWA**          Rotating-wave approximation

**TOF**          Time of flight

| | |
|---|---|
| **C-DBD** | Conventional double Bragg diffraction |
| **CD-DBD** | Constant-detuning-mitigated double Bragg diffraction |
| **DS-DBD** | Detuning-sweep-mitigated double Bragg diffraction |
| **OCT** | Optimal control theory |
| **OCT-DBD** | Optimal-control-theory-assisted double Bragg diffraction |
| **BS** | Beam splitter |
| **M** | Mirror |
| **OCT-BS** | Optimal-control-theory-designed beam splitter |
| **OCT-M** | Optimal-control-theory-designed mirror |
| **UATIS** | Universal atom interferometer simulator |
| **VLBAI** | Very-long-baseline atom interferometer |
| **MAIUS** | Materiewelleninterferometer unter Schwerelosigkeit (*German*) |
| **QUANTUS** | Quantengase unter Schwerelosigkeit (*German*) |
| **CAL** | Cold Atom Laboratory |
| **ISS** | International Space Station |
| **CSSAI** | China Space Station atom interferometer |
| **CSS** | China Space Station |

# Table of Contents

# List of Figures

# List of Tables

# Chapter One

# Introduction

## 1.1 A Century of Quantum Mechanics and the Rise of Atom Interferometry (1925–2025)

THE year 2025 marks one hundred years since Werner Heisenberg's seminal *Umdeutung* (reinterpretation) paper "*Über quantentheoretische Umdeutung kinematischer und mechanischer Beziehungen* (On the quantum-theoretical reinterpretation of kinematical and mechanical relationships)" [1], which laid the foundations of matrix mechanics and signaled the birth of modern quantum mechanics. Over the past century, quantum mechanics has evolved from a theoretical model trying to explain the atomic spectra into the governing theory of microscopic world, stemming from which the technology of manipulating matter with coherent light. Among these, atom interferometry has emerged as one of the most striking demonstrations of quantum coherence on macroscopic scales.

### 1.1.1 Foundations of quantum mechanics (1900–1930)

The conceptual roots of quantum theory trace back to the turn of the twentieth century:

- **1900 – Max Planck** introduced the quantum hypothesis to explain blackbody radiation, marking the beginning of quantized energy levels [2–5].

- **1905** – **Albert Einstein** established the photon concept by explaining the photoelectric effect [6].

- **1913** – **Niels Bohr** formulated the quantized atomic model in a series of three landmark papers—later collectively referred to as "The Trilogy"—which successfully explained the discrete spectral lines of hydrogen [7–9].

- **1924** – **Louis de Broglie** postulated wave–particle duality, associating a wavelength $\lambda = h/p$ to any massive particle, thereby predicting matter-wave interference [10, 11].

- **1925** – **Werner Heisenberg** developed matrix mechanics [1], which marked the birth of modern quantum theory.

- **1926** – **Erwin Schrödinger** introduced wave mechanics [12], soon formulated into what became known as the "Schrödinger equation". Together with the probabilistic interpretation of the wavefunction proposed by **Max Born** [13], the Schrödinger equation provided a complete dynamical description of quantum systems.

- **1927** – **Heisenberg** established the uncertainty principle [14], establishing the fundamental limit to the simultaneous precision with which position and momentum can be known.

De Broglie's matter-wave hypothesis [10, 11] and the ensuing development of wave mechanics laid the conceptual foundation for modern atom interferometry, establishing the wave–particle duality that underpins interference phenomena with atoms and molecules.

## 1.1.2 Matter-wave experiments and quantum field theory developments (1930–1980)

The subsequent decades witnessed the experimental confirmation and theoretical expansion of quantum mechanics across multiple platforms:

- **1927** − **Davisson and Germer**, and independently **Thomson**, demonstrated electron diffraction, confirming de Broglie's prediction of matter waves [15–18].

- **1927–1970s** – **Dirac, Tomonaga, Schwinger, Feynman,** and **Dyson** unified quantum mechanics with special relativity through the development of quantum electrodynamics (QED) [19–33]. Subsequently, this achievement was extended to the full framework of quantum field theory (QFT) by **Yang, Goldstone, Glashow, Salam, Weinberg, Higgs** and their collaborators, leading to the description of the electroweak and strong interactions and ultimately completing the Standard Model [34–40]. Together, these developments established the modern understanding of elementary particles and their interactions, gauge symmetry and spontaneous symmetry breaking, and the physical significance of both Abelian and non-Abelian phases.

- **1974** − **Rauch**, **Zeilinger**, **Werner, Greenberger** and collaborators realized neutron interferometry, extending wave–particle duality to massive, charge-neutral particles [41–46].

### 1.1.3 The quantum optics era and birth of atom interferometry (1980–2000)

The advent of laser cooling and trapping in the 1980s brought unprecedented control over atomic motion and coherence, paving the way for atom interferometry:

- **1985** – **Chu, Cohen-Tannoudji**, and **Phillips** developed laser cooling techniques [47–52], later enabling the first production of Bose-Einstein condensate (BEC) in the laboratory by **Cornell, Wieman**, and **Ketterle** [53, 54].

- **1991** − **Kasevich** and **Chu** realized the first light-pulse atom interferometer, demonstrating coherent beam splitting using stimulated Raman transitions [55].

- **1990s** – The development of atomic Bragg diffraction [56, 57] and Bloch oscillations [58, 59] in optical lattices enabled future realization of large-momentum-transfer (LMT) atom interferometry. These advances significantly enhanced sensitivity while reducing decoherence due to internal state coupling, paving the way for next-generation precision interferometers.

These advances established atom interferometry as a precision probe of gravitation, rotation, and fundamental constants, directly manifesting the quantum superposition principle at macroscopic scales.

## 1.1.4 Quantum technologies and the second quantum revolution (2000–2025)

The 21st century has seen atom interferometry evolve from a laboratory curiosity into a central tool in quantum metrology and tests of fundamental physics:

- **2000s** – Atom interferometers achieved unprecedented sensitivities, enabling measurements of the local gravitational acceleration $g$ at the $10^{-9}$ level [60–64], determinations of the fine-structure constant $\alpha$ at the $10^{-10}$ level [65–67], and quantum tests of the universality of free fall (UFF) by measuring the Eötvös ratio down to the $10^{-12}$ level [68–73].

- **2010s** – Space-borne quantum sensors (e.g., MAIUS, QUANTUS, CAL on the ISS, and CSSAI on the CSS etc.) demonstrated Bragg and Raman interferometers in microgravity [74–82].

- **2020s** – Large-scale LMT interferometers are being developed for gravitational-wave detection [83–89], and searches for new physics beyond the Standard Model such as ultralight dark matter [90–94].

Atom interferometry now stands as a vivid realization of de Broglie's century-old vision of matter waves. By coherently splitting and recombining atomic wave packets, it enables the measurement of inertial and gravitational effects through relative phase shifts with unprecedented precision—bridging the superposition principle of quantum mechanics with general relativistic phenomena and offering new avenues to probe physics beyond the Standard Model.

### 1.1.5 Outlook

A century after Heisenberg's pioneering work [1], the understanding of quantum mechanics continues to evolve and develop. Atom interferometry, positioned at the interface of quantum mechanics and gravity, embodies the culmination of a century of progress—from Planck's quantized energy to macroscopic quantum superposition. Its ongoing development, including double Bragg diffraction and optimal control strategies explored in this thesis, will play a vital role in the application of quantum science, enabling precision tests of fundamental physics and the realization of advanced quantum technologies for civilian use, such as quantum gyroscope and quantum inertial navigation system, both on Earth and in space.

## 1.2 The Mystery of Quantum Mechanics: Wave-Particle Duality

> We choose to examine a phenomenon which is impossible, *absolutely* impossible, to explain in any classical way and which has in it the heart of *Quantum Mechanics.* In reality, it contains the only mystery.
>
> *R. P. Feynman et al. [95]*
>
> *The Feynman Lectures on Physics,*
>
> *Vol. III, Chap. 1, 1965*

The wave-particle duality of matter distinguishes the world of quantum mechanics from the realm of classical mechanics. To some extent, the puzzling and counterintuitive nature of quantum mechanics can be traced back to this duality— or "complementarity," as Niels Bohr termed it [96]. In classical world, particles and waves have distinct properties. Usually, one describes a single particle as a point mass associated with their center of mass (COM) position $\vec{x}_{COM}$ and momentum $\vec{p}_{COM}$, while a wave (such as water waves or sound waves) is characterized by a wave vector $\vec{k}$ perpendicular to its wavefront and a frequency $\omega$. In quantum world, the fundamental entities, such as photons or electrons, which are building blocks of the classical world, cannot be fully described using the position (particle-like) or wave vector (wave-like) attributes alone. They manifest as particles or waves according to the experimental circumstances.

As an example, let us consider "a phenomenon which is impossible to explain in any classical way and which lies at the heart of quantum mechanics"—the well-known double-slit *gedanken* (thought) experiment from Richard Feynman's celebrated *Lectures on Physics* [95], illustrated in Fig. 1.1(a). In this experiment, electrons are emitted from an electron gun toward a double slit, and their subsequent detection reveals the quintessential dual nature of matter waves. Behind the slits, a light source and two photo-detectors are placed to determine which slit the electron passes through. A detection screen further downstream collects the electrons and records their arrival positions. Depending on whether the light source is turned on or off, the observed pattern on the screen changes dramatically. When the light is on, the which-path (*German: welcher Weg*) information is recorded through detectors $D_1$ and $D_2$, and the resulting distribution consists of two distinct intensity peaks given by adding up the probability-amplitude-squared, $|\phi_1|^2$ and $|\phi_2|^2$, of electrons passing through each individual hole, i.e.,

$$P = |\phi_1|^2 + |\phi_2|^2 \equiv P_1 + P_2, \tag{1.1}$$

as shown in Fig. 1.1(b). In contrast, when the light source is off and no which-path information is obtained, the electrons produce an *interference* pattern, given by squaring the sum of the two probability amplitudes, i.e,

$$P_{12} = |\phi_1 + \phi_2|^2 = |\phi_1|^2 + |\phi_2|^2 + 2\Re(\phi_1^*\phi_1) \equiv P_1 + P_2 + 2\sqrt{P_1 P_2}\cos\delta, \qquad (1.2)$$

as shown in Fig. 1.1(c). The last term in Eq. (1.2) is the "interference term", where $\delta = \arg(\phi_2) - \arg(\phi_1)$ is the phase difference between the two complex amplitudes $\phi_1$ and $\phi_2$. In the first case, the electrons behave like classical particles, producing a pattern on the screen similar to the random spray from a poorly aimed machine gun. In the second case, the electrons behave like waves whose pattern on the screen resembles that of water waves passing through two narrow slits. This striking difference underscores one of the central mysteries of quantum mechanics: the act of measurement fundamentally alters the outcome of an experiment. In atom interferometry, no measurement is performed during the interferometer sequence; instead, the detection occurs only at the very end, after the final light pulse. It therefore belongs to the second scenario, where the wave nature of matter is revealed through interference. In the following section, we will illustrate how an atom interferometer operates and what physical quantities its interference pattern encodes.

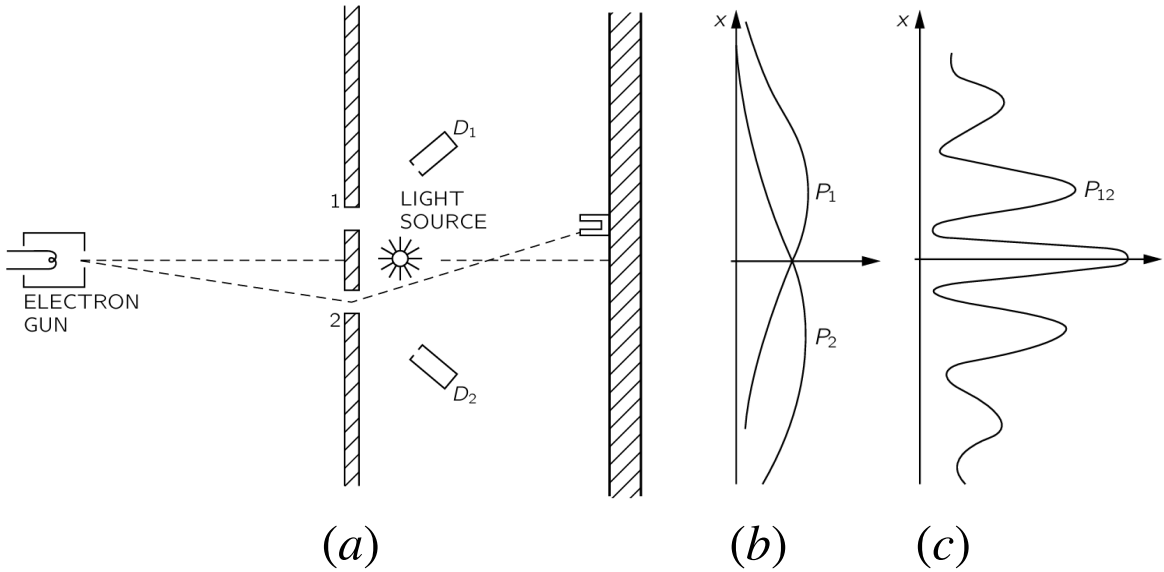

**Figure 1.1** Feynman's two-slit experiment on electrons with a light source after the slit (figures adapted from Ref. [95]). (a) A electron gun shoot electrons with the same energy towards a narrow double-slit with hole 1 and hole 2, behind the slit, a light source and two photo-detectors, $D_1$ and $D_2$ behind hole 1 and 2, are placed to detect which path the electron has taken via collecting the photons scattered by the electron passing a particular hole. In the far field down the stream, there is a screen with a movable electron detector that record the position of the arrival electron passing the double slit. (b) The pattern of arrival electrons on the final screen with the light source on: no interference. (c) The pattern of arrival electrons on the final screen with the light source off: showing interference.

## 1.3 Interfering Matter with Light: Light-Pulse Atom Interferometry

In this section, we introduce the original matter-wave Mach–Zehnder interferometer using stimulated Raman transitions, first demonstrated by Kasevich and Chu [55], and provide qualitative insights into its operating principles and measured signals, without delving into detailed derivations. Before doing so, however, it is instructive to briefly recall the optical Mach–Zehnder interferometer, illustrated in Fig. 1.2, which was independently developed in 1891-1892 by the Austrian physicist Ludwig Mach and the Swiss physicist Ludwig Zehn-der [97, 98], and serves as a pedagogical analog to its matter-wave counterpart.

### 1.3.1 An optical Mach–Zehnder interferometer

The optical Mach–Zehnder interferometer, illustrated in Fig. 1.2, is a device designed to measure the relative phase shift between two collimated beams derived from a single coherent light source. Such a phase shift may arise from the insertion of an optically dense medium or a change in the optical path length along one arm of the interferometer. The basic operating principle is as follows: a light beam from a single coherent source (ideally a monochromatic laser with wave number $k_L$, shown at the bottom left of Fig. 1.2) is split by a 50:50 beam splitter into two beams of equal intensity and fixed relative phase. These beams then propagate along two distinct paths—denoted as the "upper" and "lower" arms—before being reflected by mirrors (located in the upper-left and lower-right corners of Fig. 1.2) and redirected toward a second 50:50 beam splitter, where they are recombined and superposed before entering the two detectors at the final output ports, labeled port-1 and port-2 in the upper-right corner of Fig. 1.2. The detectors can record the intensity of the light which is proportional to the number of photons arrived, which we denote as $I_1$ and $I_2$.

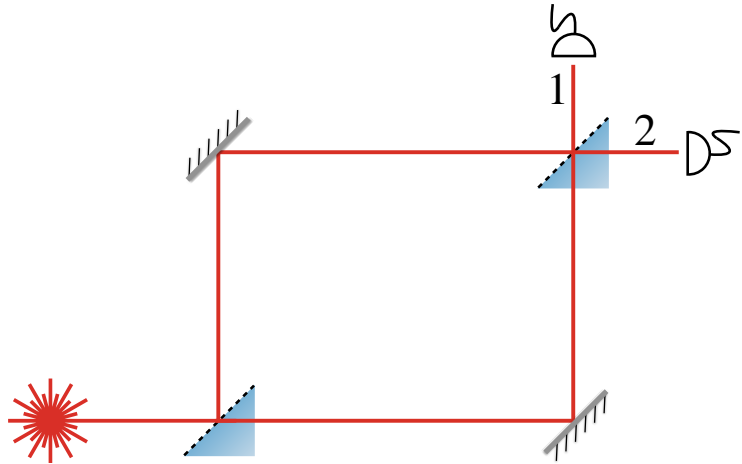

**Figure 1.2** Schematic of an optical Mach–Zehnder interferometer using two 50:50 beam splitters (light-blue triangles) and two totally reflective mirrors (gray planes) to split, redirect, and recombine coherent light beams.

First, let us analyze the simplest case. If no sample is inserted and both arms have identical optical path lengths, then the intensity at port-1 is maximum (constructive interference), or $I_1 = I_{max}$, and the intensity at port-2 is zero (destructive interference), or $I_2 = 0$. This result follows directly from the reflection and transmission processes at the two beam splitters. In order for the light to enter output port-1, the beam on the upper path have to undergo three reflections, each causing a phase jump of $\pi$, while the beam on the lower path only go through one reflection, leading to a total phase difference of $2\pi$ and the two paths at output port-1 interfere constructively. In contrast, when light exits through output port 2, the upper and lower beams acquire a total relative phase difference of $\pi$, leading to destructive interference. This phase difference arises because, although both beams are reflected twice, their second reflections occur at different optical boundaries (see the upper-right beam splitter in Fig. 1.2): the upper path reflects from the boundary to the optically denser medium, resulting in a $\pi$ phase jump, whereas the lower path reflects from the boundary to the optically rarer medium, where no phase shift occurs.

For the general case, when the optical path lengths of the two arms differ—due, for instance, to a movement of the mirrors relative to the beam splitter or to a change in the refractive index ($n > 1$) along one of the paths—the intensities at the two output ports satisfy

$$\frac{I_1 - I_2}{I_1 + I_2} = \cos(\Delta\phi),\tag{1.3}$$

where

$$\Delta\phi = k_L\left(\int_{\text{upper}} n(s)ds - \int_{\text{lower}} n(s)ds\right) \equiv k_L \oint n(s)ds\tag{1.4}$$

is the accumulated phase difference between the two arms, with $s$ denoting the geometric coordinate along the two paths.

### 1.3.2   A matter-wave counterpart: Kasevich-Chu atom interferometer

Now, suppose that instead of using a single light source to probe the optical phase difference between the two arms of the interferometer, we employ a freely falling cloud of atoms as a probe of the matter-wave phase, forming what is known as an *atom interferometer*. This concept was first realized experimentally by Kasevich and Chu to measure the local gravitational acceleration [55, 99]. In a Kasevich–Chu atom interferometer, the optical phase difference $\Delta\phi$ measured in the previous Mach–Zehnder interferometer is replaced by the relative matter-wave phase accumulated along distinct atomic trajectories. These trajectories are coherently manipulated using sequences of counter-propagating laser pulses that drive two-photon Raman transitions between internal hyperfine states. Each pulse acts as either a beam splitter or a mirror for the atomic wave packets by coherently transferring photon recoil momentum to the atoms. As a result, the interferometric phase arises from the accumulated action along each path, including contributions from kinetic and potential energies as well as the laser phases imprinted by the light–matter interactions. Similar to the optical case, this accumulated phase difference determines the atomic populations in the output ports and constitutes the measurable interferometric signal.

In the following, we briefly describe the inner workings of the Kasevich-Chu atom interferometer illustrated in Fig. 1.3, and show what is interfering and what physical quantities its "interference pattern" encode. In its simplest implementation, the Kasevich–Chu atom interferometer employs a sequence of three laser pulses forming a $\pi/2$–$\pi$–$\pi/2$ configuration, analogous to the beam splitter–mirror–beam splitter sequence in the optical Mach–Zehnder interferometer. At $t = 0$, the first $\pi/2$ pulse coherently splits the atomic wave packet into two wave packets with a well-defined momentum difference of $\hbar\boldsymbol{k}_{\text{eff}}$, initiating their free evolution along spatially separated trajectories. After a time interval $T$, a $\pi$ pulse exchanges the momenta of the two wave packets, acting as a mirror and redirecting them toward one another. Following another free evolution period of duration $T$, the final $\pi/2$ pulse recombines the two trajectories, allowing the relative phase accumulated between them to be mapped onto measurable internal-state (or momentum-state) populations, $P_1 \equiv P(|g_1, \boldsymbol{p}\rangle)$ and $P_2 \equiv P(|g_2, \boldsymbol{p} + \hbar\boldsymbol{k}_{\text{eff}}\rangle)$, which are given by

$$P_1 = \frac{1}{2} + \frac{1}{2}\cos(\Delta\Phi), \tag{1.5}$$

$$P_2 = \frac{1}{2} - \frac{1}{2}\cos(\Delta\Phi). \tag{1.6}$$

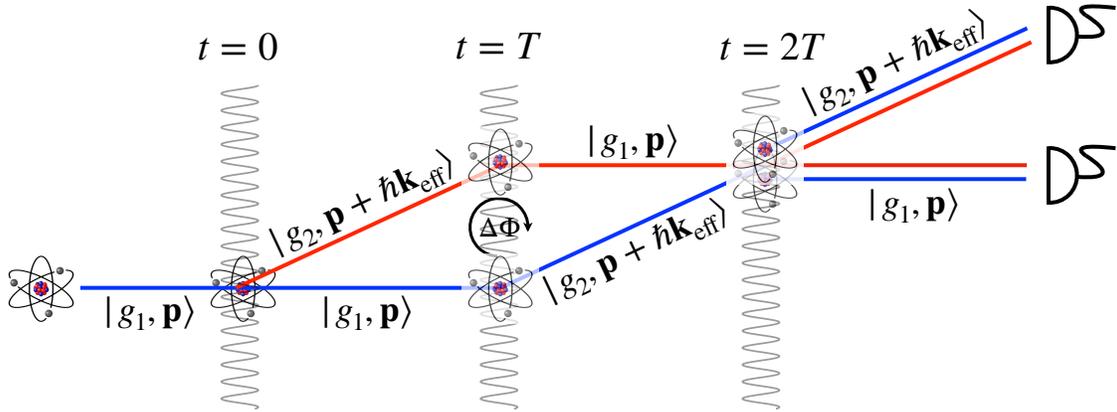

**Figure 1.3** Space-time diagram of Kasevich-Chu atom interferometer with light-matter interactions (wiggled gray lines) realizing coherent beam splitters and mirrors for the atoms. Atomic trajectories from the upper arm are colored in red, while that from the lower arm are colored in blue. $\Delta\Phi$ stands for the accumulated phase shift between the two arms.

The resulting population difference thus encodes the phase shift $\Delta\Phi$ arising from inertial effects such as local acceleration, which can be derived from semi-classical calculations [100–103] to be

$$\Delta\Phi = \frac{1}{\hbar} \int (\mathcal{L}^{(+)} - \mathcal{L}^{(-)}) dt = \boldsymbol{k}_{\text{eff}} \cdot \boldsymbol{a} T^2, \tag{1.7}$$

where $\mathcal{L}^{(+)}$ ($\mathcal{L}^{(-)}$) stands for the effective Lagrangian along the upper (lower) arm of the interferometer (see Chap. 2.1), $\boldsymbol{k}_{\text{eff}}$ denotes the effective two-photon wave vector and $\boldsymbol{a}$ the acceleration of the atoms relative to the laboratory frame. This simple relation forms the cornerstone of modern atom interferometry, linking quantum interference of atomic state populations directly to inertial and gravitational phenomena.

### 1.3.3 Sensitivity of the Kasevich-Chu atom interferometer

The sensitivity of a Kasevich-Chu atom interferometer to inertial effects, such as an external acceleration $\boldsymbol{a}$, is determined by the phase shift relation $\Delta\Phi = \boldsymbol{k}_{\text{eff}} \cdot \boldsymbol{a}T^2$ (Eq. (1.7)), where $\boldsymbol{k}_{\text{eff}}$ is the effective wave vector associated with the light-induced momentum transfer, and $T$ is the interrogation time between successive light pulses. This expression shows that the interferometric phase—and thus the measurable signal—is proportional to both the projection of the effective wave vector $\boldsymbol{k}_{\text{eff}}$ along the linear acceleration $\boldsymbol{a}$ and the square of the interrogation time. Hence, the sensitivity improves linearly with the effective momentum transfer and quadratically with interrogation time $T$, motivating the development of large-momentum-transfer (LMT) atom interferometers operating at long interrogation times.

Moreover, in the quantum regime, the phase resolution of an atom interferometer is fundamentally limited by the so-called *quantum projection noise* (QPN) [104]. This noise arises from the discrete nature of atomic detection statistics: when measuring the output populations of $N$ uncorrelated atoms, the finite number of atoms introduces statistical fluctuations in the measured population signals, leading to an intrinsic phase uncertainty. For a two-port interferometer with probabilities $P_1 = \cos^2(\Delta\Phi/2)$ and $P_2 = \sin^2(\Delta\Phi/2)$, the corresponding binomial detection statistics yield a mid-fringe (where $P_1 = \cos^2(\pi/4) = 1/2 = \sin^2(\pi/4) = P_2$) phase uncertainty [104]

$$\delta(\Delta\Phi_{\text{QPN}}) = 2\frac{\delta N_1}{N} = 2\frac{\sqrt{NP_1(1-P_1)}}{N} = \frac{1}{\sqrt{N}}, \tag{1.8}$$

where $\delta N_1$ stands for the standard deviation of the detected number of atoms in state $|g_1, \boldsymbol{p}\rangle$. Assuming $\boldsymbol{k}_{\text{eff}}$ is aligned with the acceleration, the QPN-limited acceleration sensitivity can be expressed as

$$\delta a_{\text{QPN}} = \frac{\delta(\Delta\Phi_{\text{QPN}})}{|\boldsymbol{k}_{\text{eff}}|T^2} = \frac{1}{\sqrt{N}k_{\text{eff}}T^2}, \tag{1.9}$$

where $N$ is the total number of (uncorrelated) detected atoms, $T$ is the interrogation time, and $k_{\text{eff}} = |\boldsymbol{k}_{\text{eff}}|$ [100–103, 105]. This relation defines the *standard quantum limit* (SQL) for

the sensitivity of a Kasevich–Chu atom interferometer.[1]

To surpass the SQL, quantum-enhanced measurement strategies based on entangled or spin-squeezed atomic ensembles have been proposed and experimentally demonstrated [106–109]. These approaches reduce the phase uncertainty below the standard $1/\sqrt{N}$ scaling and can, in principle, approach the *Heisenberg limit*, where the collective phase uncertainty scales as $\delta(\Delta\Phi) \propto 1/N$, yielding an acceleration sensitivity

$$\delta a_{\mathrm{H}} = \frac{1}{N k_{\mathrm{eff}} T^2}. \tag{1.10}$$

However, realizing such performance requires exquisite control of atomic correlations and suppression of decoherence, which remain active and challenging frontiers in quantum metrology and atom interferometry [110].

In practice, the overall sensitivity of an atom interferometer is determined by the interplay between this fundamental quantum noise limit and a variety of technical constraints, including imperfect fringe contrast, laser phase noise, vibration-induced phase jitter, optical wavefront aberrations, and finite atomic coherence times, among others.

The other route to further improving sensitivity is to move beyond the standard momentum transfer of $k_{\mathrm{eff}} = 2\hbar k_L$ and to target significantly larger momentum separations between the interferometer arms while preserving high contrast. Among the available techniques for realizing LMT beam splitters, double Bragg diffraction is particularly appealing due to its intrinsic symmetry, which provides robustness against laser phase noise and enables seamless integration with subsequent Bloch oscillations. This combination can boost the momentum transfer into the regime of hundreds to thousands of $\hbar k_L$ without a fundamental upper limit [111]. These features make double Bragg diffraction especially suitable for applications in microgravity and space-based missions, where long interrogation times are achievable.

---

[1]In practice, imperfect fringe contrast $\mathcal{C} < 1$ and non-unit detection efficiency $\eta < 1$ degrade this limit approximately as $\delta a \approx \frac{1}{\mathcal{C}\sqrt{\eta N}\, k_{\mathrm{eff}}\, T^2}$.

## 1.4   Thesis Outline

This thesis is structured as follows. **Chapter 2** reviews the phase of de Broglie matter waves and derives the classical path-integral expression for the phase shift of an atom interferometer in a weak gravitational potential. Applying this formalism to a Mach–Zehnder interferometer based on double Bragg diffraction (DBD), we obtain an analytic phase model that identifies the distinct physical contributions during each interferometer stage. The discussion includes the role of light–atom interactions, the interpretation of the measured interferometric phase, and subtleties related to gravitational redshift. The chapter concludes with a derivation of the DBD Hamiltonian under imperfect polarizations using optical-dipole approximations.

**Chapter 3** develops a general theoretical framework for single DBD pulses, incorporating time-dependent Rabi frequencies, AC-Stark shifts, polarization errors, and Doppler detuning. Using a reduced effective Hamiltonian from a truncated Magnus expansion, several detuning-control strategies are introduced—constant detuning (CD-DBD), linear detuning sweeps (DS-DBD), and optimal-control pulses (OCT). Their robustness is benchmarked against exact numerical simulations using the Universal Atom Interferometer Simulator (UATIS).

**Chapter 4** extends the analysis to the full DBD-based Mach–Zehnder interferometer. A tri-frequency retro-reflective scheme is proposed to compensate gravitational Doppler shifts, and a five-level S-matrix description is derived for the full pulse sequence. The contrast performance of the four detuning-control strategies is evaluated through $T$-scan fringe simulations under realistic imperfections such as momentum width, center-of-mass offsets, polarization errors, and lattice-depth fluctuations.

**Chapter 5** presents numerical methods for simulating three-dimensional interferometers using interacting Bose–Einstein condensates. The realistic optical dipole potential of the guiding beam is modeled, the impact of polarization errors on guided expansion is quantified, and the applicability of the one-dimensional DBD theory is assessed. A full 3D simulation of a DBD-based Mach–Zehnder interferometer illustrates the capabilities of UATIS for designing next-generation precision sensors.

# Chapter Two

# Fundamentals of Double Bragg

# Diffraction Atom Interferometers

WHILE Raman-based interferometers have become the standard for precision inertial sensing, their use of two internal states makes them intrinsically susceptible to differential AC-Stark shifts [61, 112–115], Zeeman shifts [61, 105, 116], and decoherence from spontaneous emission [117, 118]. Bragg diffraction was introduced to overcome these limitations by manipulating atomic momentum states within a single internal level, thereby suppressing internal-state–related systematics. Building on this concept—and inspired by the double-diffraction technique developed for Raman interferometers [119]—double Bragg diffraction (DBD) symmetrically couples atoms to the $\pm 2\hbar k_L$ momentum states using two counter-propagating optical lattices with orthogonal polarizations. This realizes an inversion-symmetric, large-momentum-transfer (LMT) beam splitter that surpasses single-Bragg or single-Raman schemes. The resulting symmetry provides *intrinsic robustness* against laser phase noise and makes DBD exceptionally well suited for high-precision and space-borne atom interferometry, as will be discussed in the following sections.

## 2.1 Ideal Double Bragg Atom Interferometers under Linear Accelerations

### 2.1.1 The phase of a de Broglie matter wave

The working principle of an atom interferometer is that via coupling the atom represented by a de Broglie matter wave $\Psi = \exp(-i\beta(\vec{x}, t))$ to external interaction fields, the interaction will encode path- or state-dependent phases to the matter wave otherwise absent in the interaction-free case. In case of gravitational interaction, the phase of the matter wave with a rest mass $m$ due to gravity is proportional to its proper time

$$\beta(\vec{x}, t) \equiv \beta_g(\vec{x}, t) = mc^2 \tau(\vec{x}, t)/\hbar. \tag{2.1}$$

In general relativity, the proper time $\tau$ is determined by the space-time metric in Cartesian coordinates $x^\mu = (ct, x, y, z)$ via the line element:

$$c^2 d\tau^2 = g_{\mu\nu} dx^\mu dx^\nu. \tag{2.2}$$

In the limit of a weak gravitational field, such as around the earth, and a velocity $|\vec{v}| \ll c$, the metric coefficients reduces to

$$g_{00} \simeq 1 + \frac{2}{c^2}\phi, \ g_{jj} \simeq -1, \ g_{ij} = 0 \tag{2.3}$$

where $i, j = 1, 2, 3$ and $i \neq j$ and $\phi = \phi(\vec{r})$ is the Newtonian potential with $\phi(\vec{r})/c^2 \ll 1$. Hence, in the non-relativistic limit, the proper time element follows from the relation

$$d\tau = \sqrt{\frac{1}{c^2} g_{\mu\nu} \frac{dx^\mu}{dt} \frac{dx^\nu}{dt}} dt \approx \sqrt{1 - \frac{2}{c^2}\left(\frac{1}{2}\vec{v}^2 - \phi\right)} dt \approx dt - \frac{1}{mc^2}\left(\frac{1}{2}m\vec{v}^2 - m\phi\right) dt \tag{2.4}$$

$$\equiv dt - \frac{1}{mc^2}\mathcal{L}_g dt \tag{2.5}$$

where $\mathcal{L}_g = \frac{1}{2}m\vec{v}^2 - m\phi$ denotes the classical Lagrangian of a particle in the Newtonian gravitational potential. As a result, the de Broglie wave $\Psi$ takes the form

$$\Psi \equiv \exp(-i\beta_g(\vec{x}, t)) = \exp\left(-i\frac{mc^2}{\hbar}\int d\tau\right) = \exp\left(-i\frac{mc^2}{\hbar}t\right)\exp\left(\frac{i}{\hbar}\int \mathcal{L}_g dt\right). \tag{2.6}$$

The appearance of the classical action and, in particular, of the kinetic and the potential energies in the phase $\beta_g(\vec{r}, t)$ of the de Broglie wave is a consequence of the proper time, i.e., non-relativistic residues from general relativity. For other types of interaction, we assume that the accumulated phase of the matter wave is given by the time integral of the total Lagrangian $\mathcal{L}$. It should be noted that the rest mass contribution $\Delta\phi_0 = -(mc^2t/\hbar)$ in Eq. (2.6), in an interferometer with two paths of identical coordinate time $t$, is not of interest due to the resulting phase shifts from this term are identical. Therefore, the phase shift $\alpha$ of an atom interferometer with two arms are given by

$$\alpha = \frac{1}{\hbar} \int (\mathcal{L}^{(+)} - \mathcal{L}^{(-)})dt \equiv \frac{1}{\hbar} \oint \mathcal{L} dt \tag{2.7}$$

where $\mathcal{L}^{(+)}$ and $\mathcal{L}^{(-)}$ stand for the Lagrangian along the upper and lower arms, respectively.

### 2.1.2 Symmetric double Bragg diffraction interferometer

In this subsection, we outline the inner workings of the symmetric DBD-based interferometer. For the sake of clarity, we consider the one-dimensional (1D) motion of the atom along $z$-axis. The space-time diagram of an ideal DBD-based interferometer is sketched in Fig. 2.1. At time $t = 0$, a short laser pulse of two counter-propagating optical lattices (see Fig. 2.2) driving symmetric double Bragg diffraction excites the atom in ground state $|0\rangle \equiv |g, p = 0\rangle$ via two-photon transitions to an equal superposition of momentum states $|g, \pm 2\hbar k_L\rangle$. The parameters of the pulse (shape, duration, intensity and carrier-envelop phase etc.) is chosen such that it creates a full population inversion from the initial state $|0\rangle$ to the target symmetric state $|1\rangle \equiv (|g, +2\hbar k_L\rangle + |g, -2\hbar k_L\rangle)/\sqrt{2}$. It should be noted that, although this first pulse acting as a beam splitter in position space, it is a rather a $\pi$-pulse from the view of the Rabi oscillation between the initial state $|0\rangle$ and final state $|1\rangle$.

At time $t = T$, the second laser pulse interacts with the atom as a mirror operation. It drives a full population inversion between the states $|g, +2\hbar k_L\rangle$ and $|g, -2\hbar k_L\rangle$. After the pulse, the atom in the upper arm moving in state $|g, +2\hbar k_L\rangle$ goes into state $|g, -2\hbar k_L\rangle$, and a net momentum loss of $-4\hbar k_L$ is transferred via a four-photon process, similarly, the atom in the lower arm moving in state $|g, -2\hbar k_L\rangle$ makes a transition into state $|g, +2\hbar k_L\rangle$, and a net momentum gain of $+4\hbar k_L$ is transferred via the reversed four-photon process. The parameters of pulse are adjusted such that the atom makes the transitions with absolute certainty.

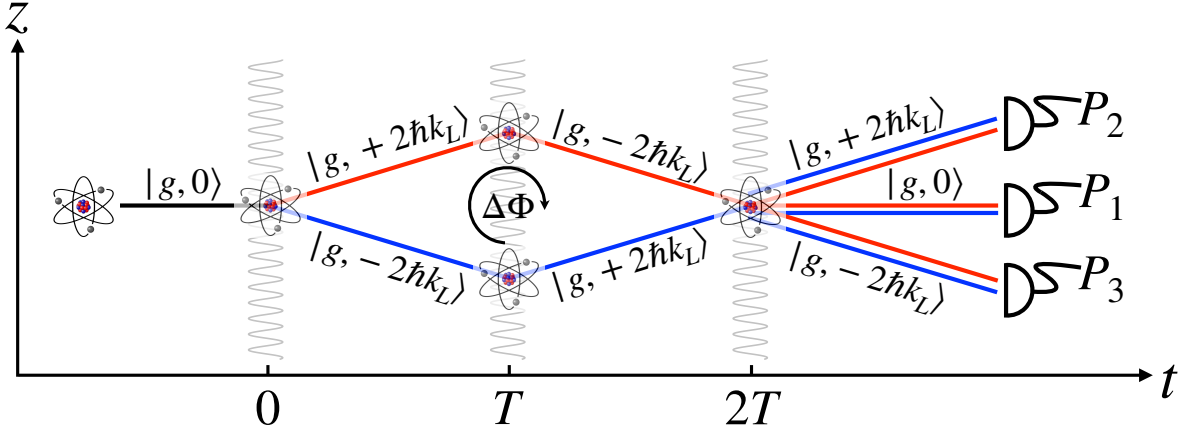

**Figure 2.1** Space-time diagram of an ideal DBD-based Mach-Zehnder interferometer. The three laser pulses of the Mach-Zehnder sequence are illustrated by grey wiggle lines at time $t = 0$, $T$, $2T$, respectively. The space-time trajectories of the atom from upper and lower arms of the interferometer are colored in red and blue, respectively. After $t = 2T$, with a clear spatial separation of the three momentum ports, the atoms are imaged on the camera from which the relative populations $P_1$, $P_2$, and $P_3$ in different momentum ports can be extracted. $\Delta\Phi$ represents the interferometer phase shift between the two arms.

At time $t = 2T$, a third laser pulse shines on the atom and the two interferometer arms recombine. For the purpose of this pulse is to recreate the ideal beam-splitter, the parameters of this pulse is chosen as identical to the first pulse at $t = 0$. As a result, depending on the relative phase shift $\Delta\Phi$ accumulated between the two arms of the interferometer from $t = 0$ to $t = 2T$, the atom will be split into three momentum ports unless $\Delta\Phi = 0$ (modulo $2\pi$) and the atom remains in the symmetric state $|1\rangle$ right before the application of the third pulse, in which case the atom will all end up in a single momentum port as the initial state $|g, 0\rangle$.

### 2.1.3 Classical trajectories of the DBD-based interferometer

In what follows, we derive the phase shift $\Delta\Phi$ for a DBD-based interferometer subject to a constant linear acceleration $\vec{a} = a\hat{z}$. For simplicity, we adopt a semi-classical approach analogous to Ref. [120], where the phase shift is obtained by solving the classical Euler–Lagrange equations for the interferometer arms and evaluating the corresponding classical trajectories. These trajectories are then substituted into the Lagrangian of Eq. (2.7) to compute the accumulated phase. The classical Lagrangian of a DBD-based interferometer in the presence of a linear acceleration is given by

$$\mathcal{L}_{DBD} = \frac{1}{2}mv^2 - V_a(z) - V_{lp}(z,t) \tag{2.8}$$

where $V_a(z) = -maz$ is the linear potential corresponding to a constant acceleration $a = -\frac{1}{m}\frac{\partial V_a(z)}{\partial z}$ and $V_{lp}(z,t)$ is an effective potential taking into account the light-atom interaction during the three DBD light pulses at times $t = 0$, $T$, $2T$ whose explicit form is given by

$$V_{lp}(z,t) = \sum_{i=0}^{2} \delta(t - iT) J_i(z) \tag{2.9}$$

where the short light pulse centered at $T_i$ is approximated by a Dirac delta function $\delta(t - T_i)$ with factors [120]

$$J_i(z) \equiv \begin{cases} -\hbar[2k_L z + \phi(iT)], & |g, 0\rangle \to |g, +2\hbar k_L\rangle \text{ or } |g, -2\hbar k_L\rangle \to |g, 0\rangle \\ +\hbar[2k_L z + \phi(iT)], & |g, 0\rangle \to |g, -2\hbar k_L\rangle \text{ or } |g, +2\hbar k_L\rangle \to |g, 0\rangle \\ -\hbar[4k_L z + 2\phi(iT)], & |g, -2\hbar k_L\rangle \to |g, 0\rangle \to |g, +2\hbar k_L\rangle \\ +\hbar[4k_L z + 2\phi(iT)], & |g, +2\hbar k_L\rangle \to |g, 0\rangle \to |g, -2\hbar k_L\rangle \end{cases} \tag{2.10}$$

corresponding to momentum transfer $\pm 2\hbar k_L$ (for the first and last beam-splitter pulses) and $\pm 4\hbar k_L$ (for the mirror pulse) indicated by the plus and minus sign at time $t = iT$. Substituting the Lagrangian (2.8) into Euler-Lagrangian equation, one gets the classical equation of motion

$$\dot{p}(t) = ma + \sum_{i=0}^{2} \delta(t - iT) p_i \tag{2.11}$$

with momentum transfer in the $i$th laser pulse

$$p_i \equiv \begin{cases} +2\hbar k_L, & |g,0\rangle \to |g,+2\hbar k_L\rangle \text{ or } |g,-2\hbar k_L\rangle \to |g,0\rangle \\ -2\hbar k_L, & |g,0\rangle \to |g,-2\hbar k_L\rangle \text{ or } |g,+2\hbar k_L\rangle \to |g,0\rangle \\ +4\hbar k_L, & |g,-2\hbar k_L\rangle \to |g,+2\hbar k_L\rangle \\ -4\hbar k_L, & |g,+2\hbar k_L\rangle \to |g,-2\hbar k_L\rangle \end{cases} \tag{2.12}$$

resulting from the transitions $|g,0\rangle \to |g,\pm 2\hbar k_L\rangle$ (or $|g,\mp 2\hbar k_L\rangle \to |g,0\rangle$) and transitions $|g,\mp 2\hbar k_L\rangle \to |g,\pm 2\hbar k_L\rangle$ indicated by the effective momentum transfer, respectively. Integration of the Euler-Lagrangian equation yields

$$p(t) = p(0) + mat + \sum_{i=0}^{2} \Theta(t - iT) p_i \equiv p_a(t) + \sum_{i=0}^{2} \Theta(t - iT) p_i \tag{2.13}$$

or the velocity trajectory

$$v(t) \equiv \dot{z}(t) = v(0) + at + \sum_{i=0}^{2} \Theta(t - iT) \frac{p_i}{m} \equiv v_a(t) + \sum_{i=0}^{2} \Theta(t - iT) \frac{p_i}{m}, \tag{2.14}$$

where $v_a(t) \equiv v(0) + at$ represents velocity of the atom in the presence of constant acceleration $\vec{a} = a\hat{z}$ but the absence of the light pulses. Integrating the velocity, one obtains the classical spatial trajectories

$$z(t) = z_a(t) + \sum_{i=0}^{2} \int_{-\infty}^{t} dt' \, \Theta(t' - iT) \frac{p_i}{m} \tag{2.15}$$

with $z_a(t)$ describing the motion of the atom solely in the linear potential or the so-called *mid-point trajectory*:

$$z_a(t) \equiv z(0) + v(0)t + \frac{1}{2}at^2 \tag{2.16}$$

associated with the Lagrangian $\mathcal{L}_a = \frac{1}{2}mv^2 - V_a(z)$ in the presence of linear acceleration but the absence of the light pulses.

### 2.1.4 Semi-classical derivation of the interferometric phase shift

Now, we calculate the relative phase shift $\Delta\Phi$ of the upper arm $(+)$ and lower arm $(-)$ of an ideal DBD-based interferometer from $t = 0$ to $t = 2T$ according to Eq. (2.7) with $\mathcal{L}$ replaced by the classical DBD Lagrangian Eq. (2.8).

$$\Delta\Phi = \frac{1}{\hbar} \int_0^{2T} \left( \mathcal{L}_{DBD}^{(+)}(t) - \mathcal{L}_{DBD}^{(-)}(t) \right) dt \tag{2.17}$$

$$= \frac{1}{\hbar} \int_0^{2T} \left( \frac{1}{2}m \left[ v^{(+)}(t)^2 - v^{(-)}(t)^2 \right] - \left[ V_a(z^{(+)}(t)) - V_a(z^{(-)}(t)) \right] \right.$$

$$\left. - \left[ V_{lp}(z^{(+)}(t), t) - V_{lp}(z^{(-)}(t), t) \right] \right) dt \tag{2.18}$$

where $v^{(\pm)}(t)$ and $z^{(\pm)}(t)$ are given by Eq. (2.14) and Eq. (2.15) with the corresponding momentum transfer $p_i$ (see Eq. (2.12)) for the upper $(+)$ and lower $(-)$ trajectories at the $i$th laser pulse, respectively. For the convenience of later discussion, we first name the three parts shown in Eq. (2.18) as "phase shift due to kinetic energy ($\Delta\Phi_{kin}$)", "phase shift due to linear potential with constant acceleration $a$ ($\Delta\Phi_a$)" and "phase shift due to light potential ($\Delta\Phi_L$)", respectively, and proceed to calculate them one by one for a clear physical picture.

*Phase shift due to kinetic energy.* The phase shift due to kinetic energy can be calculated as following

$$\Delta\Phi_{kin} = \frac{m}{2\hbar} \int_0^{2T} \left[ v^{(+)}(t) + v^{(-)}(t) \right] \left[ v^{(+)}(t) - v^{(-)}(t) \right] dt \tag{2.19}$$

$$= \frac{m}{2\hbar} \frac{8\hbar k_L}{m} \left( \int_0^T v_a(t)dt - \int_T^{2T} v_a(t)dt \right) \tag{2.20}$$

$$= 4k_L \left( v(0)t|_0^T + \frac{1}{2}at^2|_0^T - v(0)t|_T^{2T} - \frac{1}{2}at^2|_T^{2T} \right) \tag{2.21}$$

$$= -4k_L aT^2 \tag{2.22}$$

In Table 2.1, we summarize the classical velocity trajectories and their combinations as well as kinetic phase shift in different time intervals for a clear view. From the table, one can read that the accumulated phase shift due to kinetic energy between the upper and lower paths is $4k_L v(0)T + 2k_L aT^2$ in the first half of the interferometer (between 0 and $T$) and $-4k_L v(0)T - 6k_L aT^2$ in the second half (between $T$ and $2T$). Thus, the overall phase shift due to kinetic energy is $\Delta\Phi_{kin} = -4k_L aT^2 \equiv -k_{eff} aT^2$ with $k_{eff} = 4k_L$ being the effective momentum transfer (in units of $\hbar$) between the two arms after the first beam splitter pulse.

*Phase shift due to linear potential with constant acceleration.* The phase shift due to

|  | Between 0 and $T$ | Between $T$ and $2T$ |
|---|---|---|
| $v^{(+)}(t)$ | $v_a(t) + \frac{2\hbar k_L}{m}$ | $v_a(t) - \frac{2\hbar k_L}{m}$ |
| $v^{(-)}(t)$ | $v_a(t) - \frac{2\hbar k_L}{m}$ | $v_a(t) + \frac{2\hbar k_L}{m}$ |
| $v^{(+)}(t) + v^{(-)}(t)$ | $2v_a(t)$ | $2v_a(t)$ |
| $v^{(+)}(t) - v^{(-)}(t)$ | $\frac{4\hbar k_L}{m}$ | $-\frac{4\hbar k_L}{m}$ |
| $(v^{(+)} + v^{(-)})(v^{(+)} - v^{(-)})$ | $\frac{8\hbar k_L}{m}v_a(t)$ | $-\frac{8\hbar k_L}{m}v_a(t)$ |
| $\Delta\Phi_{kin}$ | $4k_L v(0)T + 2k_L aT^2$ | $-4k_L v(0)T - 6k_L aT^2$ |
| Total $\Delta\Phi_{kin}$ | $-4k_L aT^2$ | |

**Table 2.1** A table summary of the classical velocity trajectories and phase shift due to kinetic energy between different times for a DBD interferometer.

constant linear acceleration $\vec{a} = a\hat{z}$ can be derived as

$$\Delta\Phi_a = -\frac{1}{\hbar}\int_0^{2T}\left[V_a(z^{(+)}(t)) - V_a(z^{(-)}(t))\right]dt \tag{2.23}$$

$$= \frac{ma}{\hbar}\int_0^{2T}\left[z^{(+)}(t) - z^{(-)}(t)\right]dt \tag{2.24}$$

$$= \frac{ma}{\hbar}\left\{\int_0^T\frac{4\hbar k_L}{m}t\,dt + \int_T^{2T}\left[\frac{4\hbar k_L}{m}t - \frac{8\hbar k_L}{m}(t-T)\right]dt\right\} \tag{2.25}$$

$$= \frac{ma}{\hbar}\left\{\frac{2\hbar k_L}{m}T^2 + \frac{2\hbar k_L}{m}T^2\right\} \tag{2.26}$$

$$= 4k_L aT^2 \tag{2.27}$$

with the help of the analytical expressions of the upper and lower spatial trajectories summarized in Table 2.2. We observe that, unlike the kinetic-energy contribution, the phase shift due to a potential linear in the spatial coordinate, $V(z) = -maz$, is identical in both halves of the DBD interferometer. For the first half ($0 < t < T$), the relative coordinate between the arms evolves as $\Delta z(t) = 4\hbar k_L t/m$, and the phase shift yields

$$\Delta\Phi_a^{(1)} = -\frac{1}{\hbar}\int_0^T V\big(\Delta z(t)\big)\,dt = -\frac{1}{\hbar}\int_0^T\big(-ma\,\Delta z(t)\big)\,dt = 2k_L aT^2. \tag{2.28}$$

For the second half ($T < t < 2T$), the arms exchange roles, but the relative separation

| | Between 0 and $T$ | Between $T$ and $2T$ |
|---|---|---|
| $z^{(+)}(t)$ | $z_a(t) + \frac{2\hbar k_L}{m}t$ | $z_a(t) + \frac{2\hbar k_L}{m}t - \frac{4\hbar k_L}{m}(t-T)$ |
| $z^{(-)}(t)$ | $z_a(t) - \frac{2\hbar k_L}{m}t$ | $z_a(t) - \frac{2\hbar k_L}{m}t + \frac{4\hbar k_L}{m}(t-T)$ |
| $z^{(+)}(t) - z^{(-)}(t)$ | $\frac{4\hbar k_L}{m}t$ | $\frac{4\hbar k_L}{m}t - \frac{8\hbar k_L}{m}(t-T)$ |
| $\Delta\Phi_a$ | $2k_L aT^2$ | $2k_L aT^2$ |
| Total $\Delta\Phi_a$ | $4k_L aT^2$ | |

**Table 2.2** A table summary of the upper and lower spatial trajectories and phase shift due to linear potential with constant acceleration $\vec{a} = a\hat{z}$ between different times for a DBD interferometer.

evolves symmetrically, giving the same result,

$$\Delta\Phi_a^{(2)} = -\frac{1}{\hbar}\int_T^{2T} V\big(\Delta z(t)\big)\,dt = 2k_L aT^2. \tag{2.29}$$

Thus, the total phase shift from a potential linear in the spatial coordinate is added up to be $4k_L aT^2$ which is the same in magnitude but opposite in sign compared to the total phase shift due to kinetic energy given by $-4k_L aT^2$ (see Eq. (2.22)).

*Phase shift due to light potential.* The phase shift due to light potential, according to Eq. (2.18), is

$$\Delta\Phi_L = -\frac{1}{\hbar}\int_0^{2T} \left[ V_{lp}(z^{(+)}(t),t) - V_{lp}(z^{(-)}(t),t) \right] dt \tag{2.30}$$

with $V_{lp}(z,t)$ given by Eq. (2.9). Since there are Dirac Delta functions involved in the definition of $V_{lp}(z,t)$, one needs to adjust the integral's lower and upper bounds to avoid those unanalytical points on time-axis, i.e., $\int_0^{2T} \to \lim_{\epsilon\to 0}\int_{0-\epsilon}^{2T+\epsilon}$, with which we obtain the final phase shift due to light potential:

$$\Delta\Phi_L = -\frac{1}{\hbar}\lim_{\epsilon\to 0}\int_{0-\epsilon}^{2T+\epsilon}\left[V_{lp}(z^{(+)}(t),t) - V_{lp}(z^{(-)}(t),t)\right]dt \tag{2.31}$$

$$= -\frac{1}{\hbar}\lim_{\epsilon\to 0}\int_{0-\epsilon}^{2T+\epsilon}\sum_{i=0}^{2}\delta(t-iT)\left[J_i(z^{(+)}(t)) - J_i(z^{(-)}(t))\right]dt \tag{2.32}$$

$$= -\frac{1}{\hbar}\sum_{i=0}^{2}\left[J_i(z^{(+)}(iT)) - J_i(z^{(-)}(iT))\right] \tag{2.33}$$

$$= 4k_L\left[z_a(0) - z^{(+)}(T) - z^{(-)}(T) + z_a(2T)\right] + 2\left[\phi(0) - 2\phi(t) + \phi(2T)\right] \tag{2.34}$$

$$= 4k_L\left\{z_a(0) - 2z_a(T) + z_a(2T)\right\} + 2\left\{\phi(0) - 2\phi(T) + \phi(2T)\right\} \tag{2.35}$$

$$= 4k_L a T^2 + 2\ddot{\phi}T^2. \tag{2.36}$$

In deriving the last equation, we assume that the relative phase evolution of the two lasers is at most quadratic in time, e.g., up to a linear frequency chirp,

$$\phi(t) = \Delta\phi_0 - \Delta\omega\,t - \tfrac{1}{2}a_L t^2. \tag{2.37}$$

For lasers with constant frequencies, the second-order derivative of laser phase vanishes ($\ddot{\phi} = 0$), and only the phase from the midpoint trajectory ("midpoint phase") remains. Thus, the DBD interferometer phase shift reduces to

$$\Delta\Phi = \Delta\Phi_{kin} + \Delta\Phi_a + \Delta\Phi_L = 4k_L a T^2 \equiv k_{\text{eff}} a T^2, \tag{2.38}$$

where $k_{\text{eff}}$ stands for the effective wave number associated with the net momentum transfer imparted by the first beam-splitter pulse.

## 2.1.5 Physical interpretation of the phase shift

There are ongoing debates about the physical interpretation of the interferometric phase shift and its physical origin [120–129]. Here, we briefly outline the different viewpoints without attempting to settle the controversy. The starting point of the above derivation is the assumption that the phase evolution of a de Broglie matter wave is proportional to the proper time elapsed along the atomic trajectory (see Eq.(2.6)). Under the weak-field Newtonian approximation, this reduces to the classical action along the path taken by the atom in the gravitational potential. From this perspective, the matter-wave phase shift $\beta_g$ can be regarded as a manifestation of proper time evolution: measuring the phase is equivalent to measuring the proper time accumulated along the trajectory. This interpretation underlies tests of the Universality of Clock Rates (UCR) [130] or gravitational red shift, in which the proper time of different atomic species or distinct trajectories of the same atom can be compared via their interferometric phases.

However, it is equally important to emphasize that the construction of atom interferometers relies critically on light–matter interactions, which serve as beam splitters and mirrors to split and recombine the atomic wave packets in order to create interference between distinct paths. The contribution of these "quantum mechanical" interactions is typically incorporated into the "general relativistic" action integral by assumption, raising questions about how to consistently separate "free evolution" (including phases due to potential and kinetic energy) from "laser-induced" phase terms in the overall interpretation. From Eq.(2.38), one can identify that the total phase shift of a DBD interferometer comprises three distinct contributions: the kinetic phase $\Delta\Phi_{kin}$ from Eq.(2.22), the potential phase $\Delta\Phi_a$ from Eq.(2.27), and the light–matter interaction phase $\Delta\Phi_L$ from Eq.(2.36). For acceleration $a > 0$, both $\Delta\Phi_a$ and $\Delta\Phi_L$ contribute positively, whereas $\Delta\Phi_{kin}$ is negative and equal in magnitude to $\Delta\Phi_a$, or equivalently, to the midpoint phase contained in $\Delta\Phi_L$. Therefore, the way these three contributions are grouped determines how partial phase cancellations occur, leading to different possible physical interpretations of the measured phase shift. On one hand, one can consider the grouping order

$$\Delta\Phi = [\Delta\Phi_{kin} + \Delta\Phi_a] + \Delta\Phi_L = [-4k_L a T^2 + 4k_L a T^2] + 4k_L a T^2 + 2\ddot{\phi}T^2 \tag{2.39}$$

$$= 4k_L a T^2 + 2\ddot{\phi}T^2 = \Delta\Phi_L. \tag{2.40}$$

Since the two terms inside the bracket of Eq. (2.39), $\Delta\Phi_{kin}$ and $\Delta\Phi_a$, cancel each other,

one arrives at the physical interpretation that the measured interferometric phase shift $\Delta\Phi$ originates solely from the light-matter interaction phase $\Delta\Phi_L$ [131]. On the other hand, one may alternatively group the terms as [121]

$$\Delta\Phi = [\Delta\Phi_{kin} + \Delta\Phi_L] + \Delta\Phi_a = [-4k_L a T^2 + 4k_L a T^2 + 2\dddot{\phi}T^2] + 4k_L a T^2 \qquad (2.41)$$

$$= 2\dddot{\phi}T^2 + \Delta\Phi_a, \qquad (2.42)$$

which leads to a different physical interpretation of the same measured interferometric phase shift $\Delta\Phi$ as partly due to the laser phase acceleration $2\dddot{\phi}T^2$ from the light–matter interaction phase, and partly from the potential phase $\Delta\Phi_a$. If the laser phase acceleration vanishes, the measured interferometric phase shift reduces entirely to the potential-dependent term, $\Delta\Phi = \Delta\Phi_a$, implying that the phase shift directly reflects the gravitational redshift of the matter-wave clock traveling along the two paths, which is given by

$$\Delta\omega/\omega_0 = \Delta\phi/c^2, \qquad (2.43)$$

where $\Delta\omega$ is the observed clock frequency shift (red shift) and $\omega_0$ is the reference frequency of the clock in the absence of the scalar potential $\Delta\phi$, e.g., for Newtonian gravitational potential $\Delta\phi = gz$.

The two viewpoints—one emphasizing the phase imprinted through light–matter interaction, the other attributing the phase shift to the gravitational potential difference—are mathematically equivalent yet conceptually distinct. They reflect an ongoing discussion in the interpretation of phase origin in matter-wave interferometry and its connection to tests of the UCR (or gravitational red shift), and more generally, Einstein's equivalence principle [130, 132–136].

## 2.2 Double Bragg Diffraction Hamiltonian with Ideal Polarizations

We now move beyond the semi-classical treatment and develop a microscopic description of double Bragg diffraction. Specifically, we derive the single-particle Hamiltonian under generic time-dependent effective two-photon Rabi frequencies and detunings, while explicitly accounting for realistic experimental imperfections such as polarization errors. This framework establishes the theoretical foundation of Chap. 3. Physically, double Bragg diffraction can be viewed as a large-momentum-transfer technique in atom interferometry that arises from simultaneously driving two single Bragg diffraction (SBD) processes in opposite directions using counter-propagating optical lattices L1 and L2, as illustrated in Fig. 2.2.

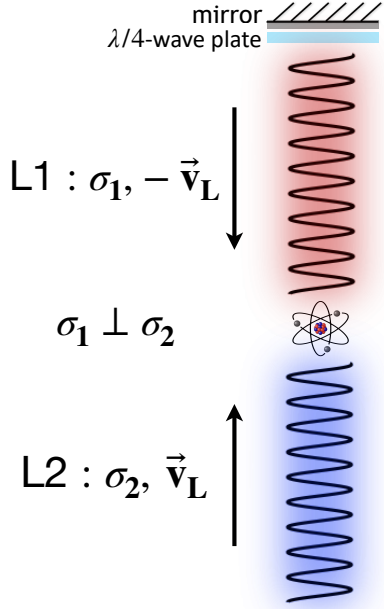

**Figure 2.2** Schematic of double Bragg diffraction with retro-reflective twin-lattice setup.

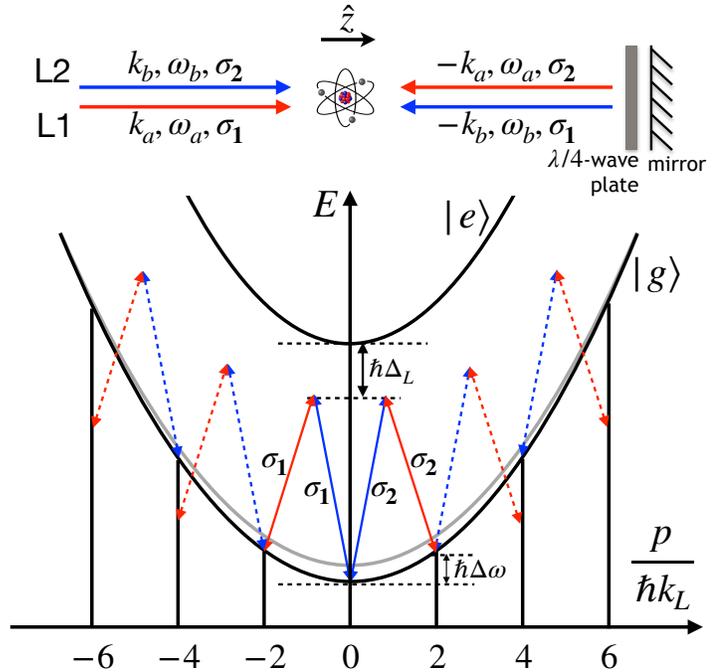

**Figure 2.3** Energy levels and resonances in double Bragg diffraction with (or without) AC Stark shift in solid black (or gray) curves.

To suppress unwanted cross couplings, the two beam pairs L1 and L2 with polarizations $\boldsymbol{\sigma}_1$ and $\boldsymbol{\sigma}_2$ must ideally be orthogonal to each other, satisfying $\boldsymbol{\sigma}_i^\dagger \boldsymbol{\sigma}_j = \delta_{i,j}$. This condition ensures that only the intended double Bragg processes contribute. To illustrate this, consider Fig. 2.3. Suppose an atom absorbs a red photon from L1, thereby gaining momentum $\hbar k_a$ in the positive $z$ direction. If the atom then undergoes stimulated emission into a blue photon from L2, it receives an additional momentum $-\hbar k_b$, corresponding to a net transfer of $\hbar(k_a - k_b)$. This process represents an unwanted cross couplings, since conventional double Bragg diffraction is designed to drive transitions with momentum transfer $\pm\hbar(k_a + k_b)$. By enforcing strictly orthogonal polarizations between L1 and L2, these spurious diffraction channels are suppressed. The more general case with imperfect polarization overlap ($\boldsymbol{\sigma}_1^\dagger \boldsymbol{\sigma}_2 \neq 0$), which leads to additional standing-lattice diffractions, will be treated in the next section.

Here, we derive the effective dipole potential experienced by the atoms for DBD from the two pairs of input and retro-reflected light fields with frequencies $\omega_a$ and $\omega_b$ and orthonormal polarizations $\boldsymbol{\sigma}_1$ and $\boldsymbol{\sigma}_2$ shown in Fig. 2.2. We assume their wave numbers are approximately the same and are given by the lattice wave number $k_L$, i.e. $|\vec{k}_a| \approx |\vec{k}_b| \equiv k_L = 2\pi/\lambda_L$. With this assumption, the input and retro-reflected light fields propagating in $z$-direction can be written in phasor notation as

$$E_{in} = E_0\left\{e^{i(k_L z - \omega_a t + \phi_a)}\boldsymbol{\sigma}_1 + e^{i(k_L z - \omega_b t + \phi_b)}\boldsymbol{\sigma}_2\right\} \tag{2.44}$$

$$E_{retro} = e^{i\pi}E_0\left\{e^{i(-k_L z - \omega_a t + \phi_a)}\boldsymbol{\sigma}_2 + e^{i(-k_L z - \omega_b t + \phi_b)}\boldsymbol{\sigma}_1\right\} \tag{2.45}$$

where $E_0 = E_0(t)$ is the real electric field amplitude of the two input lasers which can take a temporal pulse shape, $\phi_a$ and $\phi_b$ are the initial laser phases that are manually controllable by the experimental settings and the extra phase $e^{i\pi}$ in Eq.(2.45) is due to the reflection at the surface of the mirror (sitting at $z = 0$ without loss of generality) in Fig. 2.2. The total electric field strength is therefore given by the sum

$$
\begin{aligned}
E_{total} &= E_{in} + E_{retro} \\
&= E_0\Big\{\big(e^{i(k_L z - \omega_a t + \phi_a)} - e^{i(-k_L z - \omega_b t + \phi_b)}\big)\boldsymbol{\sigma}_1 + \big(e^{i(k_L z - \omega_b t + \phi_b)} - e^{i(-k_L z - \omega_a t + \phi_a)}\big)\boldsymbol{\sigma}_2\Big\}.
\end{aligned}
\tag{2.46}
$$

The electric dipole potential experienced by an atom with an induced electric dipole moment $\boldsymbol{\mu}_{ind} = \alpha \boldsymbol{E}$ ($\alpha$ is the polarizability of the atom and we assume $\alpha > 0$) is given by

$$U_{dip} = -\boldsymbol{\mu}_{ind}^* \cdot \boldsymbol{E} = -\alpha \boldsymbol{E}^* \cdot \boldsymbol{E} = -\alpha \boldsymbol{E}^\dagger \boldsymbol{E}. \tag{2.47}$$

Plug in the total electric field $E_{total}$ for $\boldsymbol{E}$, we have

$$
\begin{aligned}
U_{dip} &= -\alpha |E_{total}|^2 \\
&= -\alpha E_0^2 \Big\{ \big[ e^{-i(k_L z - \omega_a t + \phi_a)} - e^{i(k_L z + \omega_b t - \phi_b)} \big] \big[ e^{i(k_L z - \omega_a t + \phi_a)} - e^{-i(k_L z + \omega_b t - \phi_b)} \big] (\boldsymbol{\sigma}_1^\dagger \boldsymbol{\sigma}_1) + \\
&\quad \big[ e^{-i(k_L z - \omega_b t + \phi_b)} - e^{i(k_L z + \omega_a t - \phi_a)} \big] \big[ e^{i(k_L z - \omega_b t + \phi_b)} - e^{-i(k_L z - \omega_a t + \phi_a)} \big] (\boldsymbol{\sigma}_2^\dagger \boldsymbol{\sigma}_2) \Big\}
\end{aligned}
\tag{2.48}
$$

where terms proportional to $(\boldsymbol{\sigma}_1^\dagger \boldsymbol{\sigma}_2)$ and $(\boldsymbol{\sigma}_2^\dagger \boldsymbol{\sigma}_1)$ are neglected due to their orthogonality. Since polarization vectors are normalized, the above equation can be further simplified as

$$U_{dip} = -\alpha E_0^2 \Big\{ 4 - \big[ e^{-i[2k_L z + (\omega_b - \omega_a)t - (\phi_b - \phi_a)]} + c.c. \big] - \big[ e^{-i[2k_L z - (\omega_b - \omega_a)t + (\phi_b - \phi_a)]} + c.c. \big] \Big\} \tag{2.49}$$

$$= -\alpha E_0^2 \Big\{ 4 - 2\cos(2k_L z + \Delta\omega t - \Delta\phi) - 2\cos(2k_L z - \Delta\omega t + \Delta\phi) \Big\} \tag{2.50}$$

$$= -4\alpha E_0^2 \Big\{ \sin^2(k_L z - \frac{\Delta\omega}{2}t + \frac{\Delta\phi}{2}) + \sin^2(k_L z + \frac{\Delta\omega}{2}t - \frac{\Delta\phi}{2}) \Big\} \tag{2.51}$$

$$= 2\hbar\Omega(t) \Big\{ \cos^2(k_L z - \frac{\Delta\omega}{2}t + \frac{\Delta\phi}{2}) + \cos^2(k_L z + \frac{\Delta\omega}{2}t - \frac{\Delta\phi}{2}) - 2 \Big\} \tag{2.52}$$

$$= 2\hbar\Omega(t) \Big\{ \cos^2(k_L z - \frac{\Delta\omega}{2}t + \frac{\Delta\phi}{2}) + \cos^2(k_L z + \frac{\Delta\omega}{2}t - \frac{\Delta\phi}{2}) \Big\} + \text{Const}_z \mathbf{1} \tag{2.53}$$

where we have defined the laser frequency difference $\Delta\omega \equiv \omega_b - \omega_a$, the relative phase of the lasers $\Delta\phi \equiv \phi_b - \phi_a$ from line (2.49) to line (2.50) and introduced the time-dependent Rabi frequency as $2\hbar\Omega(t) \equiv 4\alpha E_0(t)^2$ from line (2.51) to line (2.52). After neglecting the constant part (with respect to $z$), we obtain the effective dipole potential of the double Bragg diffraction:

$$V_{\text{DBD}} = 2\hbar\Omega(t) \Big\{ \cos^2(k_L z - \frac{\Delta\omega}{2}t + \frac{\Delta\phi}{2}) + \cos^2(k_L z + \frac{\Delta\omega}{2}t - \frac{\Delta\phi}{2}) \Big\} \tag{2.54}$$

and the corresponding DBD Hamiltonian is given by

$$H_{\text{DBD}} = \frac{\hat{p}^2}{2m} + 2\hbar\Omega(t) \Big\{ \cos^2(k_L \hat{z} - \frac{\Delta\omega}{2}t + \frac{\Delta\phi}{2}) + \cos^2(k_L \hat{z} + \frac{\Delta\omega}{2}t - \frac{\Delta\phi}{2}) \Big\}. \tag{2.55}$$

It should be noted that above Hamiltonian describes the superposition of two counter-propagating lattice potentials with lattice velocity $v_L = \pm \Delta\omega/(2k_L)$, hence the double Bragg diffraction can be understood as the "superposition" of two single Bragg diffractions to the left and to the right. Alternatively, it can be regarded as a standing lattice with a time-modulated amplitude:

$$V_{\text{DBD}} = 2\hbar\Omega(t)\left\{ \frac{1}{2}\cos\left(2k_L\hat{z} - (\Delta\omega t - \Delta\phi)\right) + \frac{1}{2}\cos\left(2k_L\hat{z} + (\Delta\omega t - \Delta\phi)\right) \right\} + \text{Const}_z \mathbf{1} \tag{2.56}$$

$$= 2\hbar\Omega(t)\cos(\Delta\omega t - \Delta\phi)\cos(2k_L\hat{z}) + \text{Const}_z \mathbf{1}, \tag{2.57}$$

where we have used trigonometric identities $\cos^2(x) = (1 + \cos[2x])/2$ from Eq. (2.54) to Eq. (2.56) and $\cos(\alpha + \beta) + \cos(\alpha - \beta) = 2\cos(\alpha)\cos(\beta)$ from Eq. (2.56) to Eq. (2.57). Thus, the controllable relative phase difference $\Delta\phi$ merely acts as a carrier–envelope phase and cannot imprint interferometric phases onto different arms—underpinning the *intrinsic robustness* of DBD against laser phase noise—unless the condition $1/\Delta\omega \geq \tau_p$ is satisfied, where $\tau_p$ denotes the characteristic pulse duration (e.g., the temporal width of a Gaussian pulse). Consequently, the only viable way to imprint interferometric phases is to introduce a third frequency, thereby generating two distinct values of $\Delta\omega$ in the optical potential. These frequencies can then be employed to compensate the opposite Doppler shifts associated with the upward and downward diffraction processes in non-degenerate geometries. This case will be addressed in Chapter 4 in the context of double Bragg accelerometers or gravimeters.

## 2.3 Double Bragg Diffraction Hamiltonian with Imperfect Polarizations

### 2.3.1 Polarization errors in the retro-reflective DBD scheme

Before deriving the total Hamiltonian with the polarization imperfections, we first give a generic characterization of the polarization errors due to non-orthogonality of the beam polarizations in the retro-reflection scheme. We assume the two input polarizations $\sigma_1$ and $\sigma_2$ are chosen such that they interchange with each other after the retro-reflection (see Fig 2.3). Therefore, the $\lambda/4$-wave plate's fast axis, denoted as $\hat{f}$, must align perfectly along the angle bisector of the angle formed by the two unit polarization vectors as shown in Fig. 2.4. The

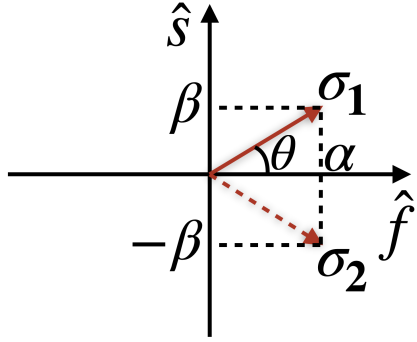

**Figure 2.4** Polarization Jones vectors before ($\sigma_1$) and after ($\sigma_2$) a double pass of the quarter-wave plate. $\hat{f}$ ($\hat{s}$) stands for the fast (slow) axis of the wave plate.

normalized Jones vector of the two polarizations are given by

$$\boldsymbol{\sigma_1} = \alpha \hat{f} + \beta \hat{s} \tag{2.58}$$

$$\boldsymbol{\sigma_2} = \alpha \hat{f} - \beta \hat{s} \tag{2.59}$$

where $\hat{f}$ ($\hat{s}$) denote the fast (slow) axis of the $\lambda/4$-wave plate and $|\alpha|^2 + |\beta|^2 = 1$. The polarization error $\varepsilon_{pol}$ is defined by the Hermitian product of the two polarization vectors

$$\varepsilon_{pol} \equiv \boldsymbol{\sigma_1^\dagger}\boldsymbol{\sigma_2} = \alpha^*\alpha - \beta^*\beta = |\alpha|^2 - |\beta|^2 = \boldsymbol{\sigma_2^\dagger}\boldsymbol{\sigma_1}, \tag{2.60}$$

which is a real scalar within the interval $[-1, 1]$. Due to the $U(1)$ gauge freedom of the Jones vectors—specifically, the equivalence $\boldsymbol{\sigma_1} \sim e^{i\pi}\boldsymbol{\sigma_1}$—the value of $\varepsilon_{pol}$ can always be restricted to the interval $[0, 1]$ by a suitable gauge transformation. Hence, one may equivalently define

$$\varepsilon_{pol} = |\boldsymbol{\sigma_2^\dagger}\boldsymbol{\sigma_1}|. \tag{2.61}$$

Here, $\varepsilon_{pol} = 0$ corresponds to perfectly orthogonal polarizations (the ideal case), while $\varepsilon_{pol} = 1$ represents fully parallel polarizations (the worst case).

Equivalently, the polarization error can be expressed in terms of the angular misalignment between the Jones vector $\boldsymbol{\sigma_1}$ and the fast axis $\hat{f}$ of the quarter-wave plate. Let $\theta$ denote the angle between $\boldsymbol{\sigma_1}$ and $\hat{f}$ (see Fig. 2.4), then the polarization error takes the form

$$\varepsilon_{\mathrm{pol}} = |\cos^2(\theta) - \sin^2(\theta)| = |\cos(2\theta)| \equiv \left| \cos\left[ 2\left( \frac{\pi}{4} + \theta_{\mathrm{pol}} \right) \right] \right|, \tag{2.62}$$

where $\theta_{\mathrm{pol}}$ represents the deviation of the wave plate's fast axis from its ideal orientation at $\pi/4$ relative to the incident linear polarization.

In the experiment, $\theta_{\mathrm{pol}}$ is adjusted by finely rotating the $\lambda/4$ wave plate to maximize the power in the desired circular polarization component while minimizing the orthogonal one, typically monitored via a polarimeter. Even small angular misalignments of a few degrees can yield polarization errors at the percent level ($\theta_{\mathrm{pol}} \sim 0.01\pi$ corresponds to $\varepsilon_{\mathrm{pol}} \sim 0.06$), significantly affecting the double diffraction efficiency and, consequently, the interferometric contrast, which will be addressed separately in Chap. 3 and Chap. 4.

## 2.3.2 DBD Hamiltonian under polarization errors

For the case of imperfect input polarizations, i.e., $\boldsymbol{\sigma_1^\dagger \sigma_2} \neq 0$, an additional standing-lattice contribution arises in the total DBD Hamiltonian. In this situation, the terms proportional to $\boldsymbol{\sigma_1^\dagger \sigma_2}$ that were previously neglected in Eq. (2.48) must be reinstated, yielding

$$
\begin{aligned}
U_{pol} = & -\alpha E_0^2 (\boldsymbol{\sigma_1^\dagger \sigma_2}) \Big\{ \left( e^{-i(k_L z - \omega_a t + \phi_a)} - e^{i(k_L z + \omega_b t - \phi_b)} \right) \left( e^{i(k_L z - \omega_b t + \phi_b)} - e^{i(-k_L z - \omega_a t + \phi_a)} \right) \\
& + \left( e^{-i(k_L z - \omega_b t + \phi_b)} - e^{i(k_L z + \omega_a t - \phi_a)} \right) \left( e^{i(k_L z - \omega_a t + \phi_a)} - e^{i(-k_L z - \omega_b t + \phi_b)} \right) \Big\} \\
= & -\alpha E_0^2 (\boldsymbol{\sigma_1^\dagger \sigma_2}) \Big\{ 2e^{-i[(\omega_b - \omega_a)t - (\phi_b - \phi_a)]} + 2e^{i[(\omega_b - \omega_a)t - (\phi_b - \phi_a)]} - 2e^{-i2k_L z} - 2e^{i2k_L z} \Big\} \\
= & \, 4\alpha E_0^2 (\boldsymbol{\sigma_1^\dagger \sigma_2}) \big[ \cos(2k_L z) - \cos(\Delta\omega t - \Delta\phi) \big],
\end{aligned}
\tag{2.63}
$$

where we have used the same abbreviations $\Delta\omega \equiv \omega_b - \omega_a$ for the frequency difference and $\Delta\phi \equiv \phi_b - \phi_a$ for the relative phase of the input laser beams. Recalling the definition of the polarization error from the previous subsection $\varepsilon_{pol} = |\boldsymbol{\sigma_1^\dagger \sigma_2}|$, one can group the above expression into position-dependent and position-independent parts. By promoting

the position variable $z$ to the operator $\hat{z}$, and reinstating the temporal dependence of the electric field amplitude $E_0 = E_0(t)$, we arrive at a standing-lattice potential induced by polarization error:

$$V_{pol} = 4\alpha E_0(t)^2 \varepsilon_{pol} \cos\left(2k_L\hat{z}\right) - 4\alpha E_0(t)^2 \varepsilon_{pol} \cos\left(\Delta\omega t - \Delta\phi\right) \tag{2.64}$$

$$\equiv 2\hbar\Omega(t)\varepsilon_{pol}\cos\left(2k_L\hat{z}\right) + \text{Const}_z \mathbf{1}, \tag{2.65}$$

where we have reintroduced the time-dependent Rabi frequency $2\hbar\Omega(t) = 4\alpha E_0(t)^2$. Combining Eq. (2.65) with the polarization-error-free Hamiltonian (2.55) and neglecting the position-independent part, one finally obtains the general DBD Hamiltonian with imperfect polarizations:

$$H_{\text{DBD}}^{pol} = \frac{\hat{p}^2}{2m} + 2\hbar\Omega(t)\left\{\cos(\Delta\omega\, t - \Delta\phi) + \varepsilon_{pol}\right\}\cos(2k_L\hat{z}), \tag{2.66}$$

with laser frequency difference $\Delta\omega = \omega_b - \omega_a$[1], initial relative phase (carrier-envelop phase) $\Delta\phi = \phi_b - \phi_a$, and a polarization error $\varepsilon_{pol} = |\boldsymbol{\sigma}_1^\dagger \boldsymbol{\sigma}_2| = |\boldsymbol{\sigma}_1^* \cdot \boldsymbol{\sigma}_2|$. A finite $\varepsilon_{pol}$ leads to an additional standing-lattice contribution in the double Bragg diffraction Hamiltonian, which can degrade both the beam-splitter and mirror pulse efficiencies, and consequently the overall interferometric contrast if not properly mitigated.

---

[1]It should be noted that, for generic time-dependent frequencies $\omega_{a,b} = \omega_{a,b}(t)$, $\Delta\omega(t)t \equiv \phi(t_i) + \int_{t_i}^{t}[\omega_b(t) - \omega_a(t)]\,dt$ denotes the physically accumulated phase difference with an adjusted initial relative phase $\phi(t_i) = \int_0^{t_i}[\omega_b(t) - \omega_a(t)]\,dt$.

# Chapter Three

# Robust Double Bragg Diffraction via Detuning Control

## 3.1  Motivation

$\mathrm{A}$TOM interferometry (AI) has become one of the most powerful techniques for precision measurements of gravity, acceleration, and fundamental constants. Its performance largely depends on achieving large momentum transfer between atomic wave packets while maintaining high interferometric contrast. Among the various techniques for realizing LMT in atom interferometry, double Bragg diffraction stands out for its intrinsic symmetry, robustness and natural compatibility to subsequent Bloch oscillations. By coupling atoms to symmetric momentum states through two counter-propagating optical lattices, DBD doubles the interferometric scale factor compared to single Bragg diffraction while preserving coherence within a single internal state. This makes it particularly attractive for applications in microgravity environments and space missions, where the degeneracy between opposite diffraction processes is naturally maintained and long interrogation times become accessible.

Despite these advantages, realistic imperfections such as polarization errors, AC-Stark shifts, and Doppler detuning can significantly degrade DBD pulse efficiency and fringe contrast, thereby limiting interferometric sensitivity. To address these challenges, this chapter develops a comprehensive theoretical framework for DBD based on the Magnus expansion, enabling accurate modeling in the quasi-Bragg regime while retaining physical insights.

Within this framework, we explore several detuning-control strategies—including constant detuning, linear detuning, and optimal control theory (OCT) approaches—to mitigate various experimental imperfections. These methods pave the way toward high-fidelity and high-contrast DBD atom interferometers, advancing their use in next-generation quantum sensors and precision measurements.

## 3.2 Background and Original Contributions

The concept of double Bragg diffraction was first introduced by Rasel's group within the QUANTUS consortium [137] to achieve symmetric large-momentum-transfer (LMT) beam splitters without involving multiple internal atomic states. In contrast to single Bragg diffraction (SBD) and Raman transitions, double Bragg diffraction couples atoms to opposite momentum states $\pm 2\hbar k_L$ within a single internal ground-state manifold. This is achieved using two counter-propagating optical lattices with orthogonal polarizations, as discussed in detail in Chap. 2.2. This approach effectively doubles the interferometric scale factor $\mathbf{k}_{\text{eff}}$ and suppresses many systematic phase errors that arise from AC-Stark shifts and Zeeman shifts in Raman-based schemes.

The theoretical foundation of DBD was first established by Giese and collaborators [138, 139], who used a perturbative method of averaging to adiabatically eliminate higher-order momentum states. Their formalism has successfully explained the first experimental demonstrations of DBD interferometers [137], and it remains the prevailing description in the literature. However, this approach relies on the adiabatic and constant-Rabi-frequency assumptions, which limit its applicability to experimentally interesting regimes where pulse shapes, AC-Stark shifts, Doppler effects, and polarization imperfections play significant roles. Moreover, previous theoretical descriptions of DBD have been developed in a dressed-state picture, in which the direct connection between the predicted populations of the dressed states and the experimentally measured quantities defined in the bare momentum states is obscured.

In this work, we develop a new theoretical framework for DBD based on the Magnus expansion [140, 141], which retains physical transparency while naturally incorporating time-dependent Rabi frequencies, polarization errors, and Doppler detuning. This approach captures the essential dynamics of DBD in the quasi-Bragg regime, where most practical atom interferometers operate, and allows the derivation of compact effective Hamiltonians that describe both ideal and imperfect experimental conditions.

Building upon this framework, we introduce and benchmark detuning-control strategies to mitigate imperfections and enhance the robustness of DBD beam splitters. In particular, we identify three optimization approaches—constant detuning, linear detuning sweeps, and artificial-intelligence-assisted (AI-assisted) optimal control pulse (OCT-pulse)—and show that they dramatically improve double diffraction efficiency under realistic conditions.

The original contributions of this chapter are twofold:

- A physically intuitive and analytically tractable Magnus-expansion model for DBD in the quasi-Bragg regime.

- The proposal and numerical validation of robust detuning-control strategies, including an AI-assisted OCT protocol, that achieve near-unity transfer efficiencies and resilience against Doppler and polarization-induced imperfections.

Together, these developments extend the theoretical understanding of DBD beyond the adiabatic pulse limit and perfect polarization assumption, and provide practical design principles for high-efficiency LMT beam splitters.

## 3.3 Theoretical Framework

### 3.3.1 Symmetry of the double Bragg Hamiltonian

Double Bragg diffraction (DBD) describes the coherent interaction between atoms and two pairs of counter-propagating laser beams with distinct frequencies and polarizations. Each laser pair, characterized by frequencies $\omega_a$ and $\omega_b$, drives an independent moving optical lattice that couples atomic momentum states separated by multiples of $2\hbar k_L$. Ideally, the two lattices are orthogonally polarized, such that they interact with the atoms without mutual interference. Fig. 4.2 illustrates both an experimental realization via a retro-reflection setup and the corresponding energy diagram relevant to DBD.

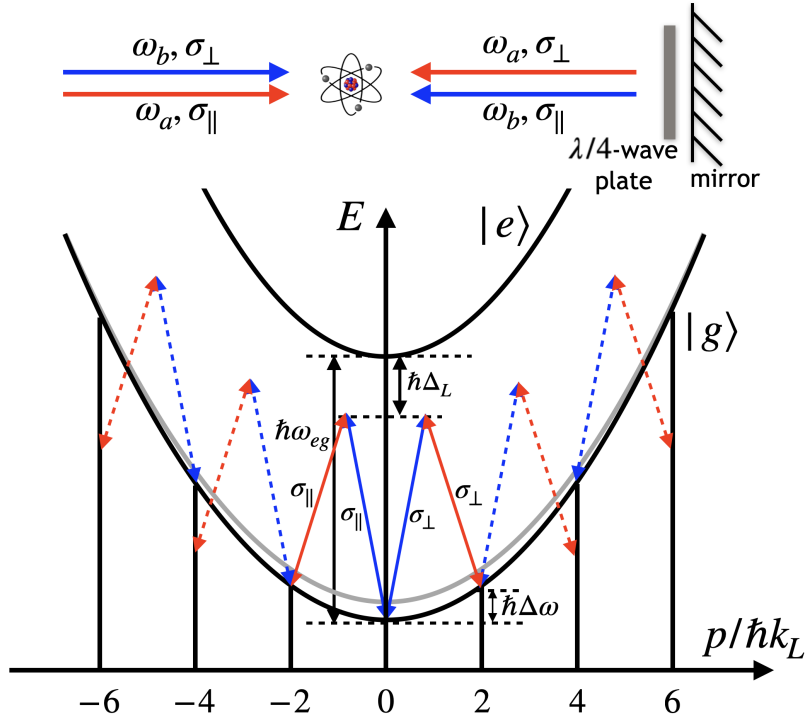

**Figure 3.1** Realization of double Bragg diffraction via a retro-reflection setup (upper panel) and the corresponding energy diagram (lower panel) showing the first-order resonance condition. The frequency difference between the two laser pairs is $\Delta\omega = \omega_b - \omega_a = 4\,\omega_{\rm rec} + \Delta$, where $\Delta$ is the detuning from resonance. The polarizations $\sigma_\perp$ and $\sigma_\parallel$ are ideally orthogonal. $\Delta_L$ denotes the single-photon detuning from the excited state $|e\rangle$. The solid (gray) dispersion curves illustrate the ground-state energy with (without) AC-Stark shift corrections.

The one-dimensional single-particle Hamiltonian describing DBD under polarization errors has been derived in Eq. (2.66) of Chapter 2. Neglecting the carrier-envelop phase, the DBD Hamiltonian reads

$$H_{\mathrm{DBD}}^{pol} = \frac{\hat{p}^2}{2m} + 2\hbar\Omega(t)\Big\{ \cos(\Delta\omega(t)\,t) + \varepsilon_{pol} \Big\} \cos(2k_L\hat{z}),$$ (3.1)

where $\varepsilon_{pol} = |\boldsymbol{\sigma}_\perp^\dagger \boldsymbol{\sigma}_\parallel|$ quantifies the polarization error due to imperfect orthogonality between the two beam polarizations, and $\Omega(t) = \Omega_a(t)\Omega_b(t)/(2\Delta_L)$ is the effective two-photon Rabi frequency with $\Omega_a(t)$ and $\Omega_b(t)$ denoting the single-photon Rabi frequencies. The detuning $\Delta_L$ is the single-photon detuning from the atomic transition. The time-dependent frequency difference between $\omega_a$ and $\omega_b$ is

$$\Delta\omega(t) = \omega_b - \omega_a = 4\,\omega_{\mathrm{rec}} + \Delta(t),$$ (3.2)

where $\omega_{\mathrm{rec}} = \hbar k_L^2/(2m)$ is the single-photon recoil frequency, and $\Delta(t)$ represents a small tunable detuning offset [1].

As shown in Fig. 4.2, the DBD setup can be realized experimentally using retro-reflection of two laser beams with linear polarizations $\boldsymbol{\sigma}_\perp$ and $\boldsymbol{\sigma}_\parallel$ [137, 142]. A $\lambda/4$ wave plate placed in the retro path converts the incoming linear polarization into circular and back, effectively swapping the two polarizations upon reflection. By adjusting the fast axis of the $\lambda/4$-wave plate, one can control the overlap and orthogonality between the two counter-propagating lattices. Ideally, the two optical lattices are orthogonal and independently drive single Bragg diffraction processes in opposite directions, thereby forming a symmetric DBD configuration.

---

[1]For generic time-dependent laser frequencies, $\Delta\omega(t)t = [4\omega_{rec} + \Delta(t)]t \equiv \phi(t_i) + \int_{t_i}^{t}[4\omega_{rec} + \Delta_{lab}(t)]\,dt$ denotes the physically accumulated relative laser phase difference. Hence, $\Delta(t)$ discussed in the thesis stands for a fictitious detuning parameter. However, it has a one-to-one correspondence to the laboratory detuning $\Delta_{lab}(t) = \frac{d\Delta(t)}{dt}t + \Delta(t)$ with a properly adjusted initial relative laser phase $\phi(t_i) = \int_0^{t_i}[4\omega_{rec} + \Delta_{lab}(t)]\,dt$. Every time-dependent detuning protocol $\Delta(t)$ optimized in the thesis corresponds to setting a physical detuning $\Delta_{lab}(t)$ in the laboratory.

### 3.3.2 Parity symmetry and reduction of the Hilbert space

The Hamiltonian (3.1) exhibits a fundamental *parity symmetry* under spatial inversion, defined by $\hat{\mathbf{P}} : \hat{z} \to -\hat{z}$. It follows that $[\hat{\mathbf{P}}, H(t)] = 0$ at all times, implying conservation of parity. Consequently, the full Hilbert space can be decomposed into even and odd subspaces,

$$\mathcal{H} = \mathcal{H}_{\text{even}} \oplus \mathcal{H}_{\text{odd}}. \tag{3.3}$$

For an atom initially at rest relative to the mirror, i.e., starting with quasi-momentum near $p = 0$ with an infinitesimally small momentum width, only the even-parity states participate in the dynamics. These are spanned by the basis

$$|n\rangle = \begin{cases} |0\hbar k_L\rangle, & n = 0, \\ \dfrac{1}{\sqrt{2}}\big(|2n\hbar k_L\rangle + |-2n\hbar k_L\rangle\big), & n > 0, \end{cases} \tag{3.4}$$

so that the wave function evolves entirely within $\mathcal{H}_{\text{even}}$.

Expressed in this basis, the DBD Hamiltonian takes the compact form

$$\begin{aligned} H(t) = \sum_{n=0}^{\infty} 4n^2 \hbar \omega_{\text{rec}} |n\rangle\langle n| + \hbar \Omega(t) C(t, \varepsilon_{\text{pol}}) \Big\{ & \sqrt{2}\big(|0\rangle\langle 1| + |1\rangle\langle 0|\big) \\ & + \sum_{n=1}^{\infty} \big(|n\rangle\langle n+1| + |n+1\rangle\langle n|\big) \Big\}, \end{aligned} \tag{3.5}$$

where $C(t, \varepsilon_{\text{pol}}) = \cos[\Delta\omega(t)t] + \varepsilon_{\text{pol}}$. We have separated $H(t)$ into a diagonal kinetic part and an off-diagonal coupling part,

$$H(t) = H_0 + H_1(t), \tag{3.6}$$

with

$$H_0 = \sum_{n=0}^{\infty} 4n^2 \hbar \omega_{\text{rec}} |n\rangle\langle n|, \tag{3.7}$$

and

$$H_1(t) = \hbar \Omega(t) C(t, \varepsilon_{\text{pol}}) \Big\{ \sqrt{2}\big(|0\rangle\langle 1| + |1\rangle\langle 0|\big) + \sum_{n=1}^{\infty} \big(|n\rangle\langle n+1| + |n+1\rangle\langle n|\big) \Big\}. \tag{3.8}$$

### 3.3.3  Transformation to the interaction picture

To separate the physically interesting dynamics driven by $H_1(t)$ from the fast dynamics driven by $H_0$, we move to the interaction picture with respect to $H_0$. The transformed Hamiltonian becomes

$$\bar{H}(t) = e^{iH_0 t/\hbar} H_1(t) e^{-iH_0 t/\hbar}$$

$$= \hbar\Omega(t) C(t, \varepsilon_{\text{pol}}) \Bigg\{ \sqrt{2}\big(e^{-i4\omega_{\text{rec}}t}|0\rangle\langle 1| + e^{i4\omega_{\text{rec}}t}|1\rangle\langle 0|\big)$$

$$+ \sum_{n=1}^{\infty} \big(e^{i4(2n+1)\omega_{\text{rec}}t}|n+1\rangle\langle n| + e^{-i4(2n+1)\omega_{\text{rec}}t}|n\rangle\langle n+1|\big) \Bigg\}, \qquad (3.9)$$

where the basis states $|n\rangle$ are given by Eq. (3.4).

The form of Eq. (3.9) explicitly shows that DBD couples momentum states separated by $2\hbar k_L$ through multi-photon transitions mediated by the two optical lattices. The oscillatory exponential factors encode the detunings between successive momentum orders, while the term $\varepsilon_{\text{pol}}$ accounts for imperfect lattice orthogonality that introduces additional coupling terms resembling those of a residual standing-wave potential. This Hamiltonian serves as the starting point for the quasi-Bragg regime analysis and for the application of the Magnus expansion developed in the following sections.

### 3.3.4  Magnus expansion and effective two-level Hamiltonian

The interaction Hamiltonian $\bar{H}(t)$ given in Eq. (3.9) provides the starting point for deriving an effective description of double Bragg diffraction in the absence of Doppler detuning. To proceed, we employ the *Magnus expansion*, a systematic technique for approximating the unitary evolution of systems governed by a time-dependent Hamiltonian [140, 141].

For a general time-dependent Hamiltonian $\bar{H}(t)$, the time evolution operator $U(t, 0)$ satisfies the Schrödinger equation

$$i\hbar\frac{d}{dt}U(t, 0) = \bar{H}(t)\,U(t, 0), \qquad (3.10)$$

with formal solution expressed as

$$U(t, 0) = \exp\Big[\sum_{i=1}^{+\infty} G_i(t)\Big]. \tag{3.11}$$

Here, $G_i(t)$ denotes the $i^{\text{th}}$ term in the Magnus series. The first two contributions, which capture the leading-order dynamics, read

$$G_1(t) = -\frac{i}{\hbar}\int_0^t \bar{H}(t_1)\,dt_1, \tag{3.12}$$

$$G_2(t) = \Big(-\frac{i}{\hbar}\Big)^2\frac{1}{2!}\int_0^t dt_1\int_0^{t_1} dt_2\big[\bar{H}(t_1),\bar{H}(t_2)\big]. \tag{3.13}$$

In practice, we truncate the Magnus expansion at second order. Since the Magnus series is non-perturbative and its convergence is not guaranteed, we choose to truncate at the second-order which gives rise to non-zero AC-Stark-shift contributions up to four-photon processes while neglecting higher-order (six-photon transitions and above) terms. We then define an effective Hamiltonian

$$H_{\text{eff}}(t) \equiv i\hbar\frac{d}{dt}\big[G_1(t) + G_2(t)\big], \tag{3.14}$$

which provides direct physical insight into the system dynamics. Once $H_{\text{eff}}$ is obtained, the full time evolution can be reconstructed through a Dyson series [33]:

$$U(t, 0) = \mathcal{T}\exp\Big[-\frac{i}{\hbar}\int_0^t H_{\text{eff}}(t)\,dt\Big]. \tag{3.15}$$

We further simplify the dynamics by working in the so-called quasi-Bragg regime, where the Rabi frequency and detuning satisfy $\Omega(t), |\Delta| \ll 8\,\omega_{rec}$. In this limit, higher-order transitions and fast oscillating terms can be adiabatically eliminated, leaving an effective description in terms of two coupled even-momentum states. This reduction yields an effective two-level-system (TLS) Hamiltonian:

$$H_{\text{eff}} = \hbar\begin{pmatrix} \frac{\Omega^2}{\omega_{rec}}\big(\frac{1}{4}\epsilon_{pol} - \frac{1}{2}\epsilon_{pol}^2\big) & \frac{\sqrt{2}}{2}\Omega\big\{e^{i\Delta t} + e^{-i(\Delta+8\omega_{rec})t} + 2\,\epsilon_{pol}e^{-i4\omega_{rec}t}\big\} \\ \frac{\sqrt{2}}{2}\Omega\big\{e^{-i\Delta t} + e^{i(\Delta+8\omega_{rec})t} + 2\,\epsilon_{pol}e^{i4\omega_{rec}t}\big\} & \frac{\Omega^2}{\omega_{rec}}\big(-\frac{3}{64} - \frac{1}{4}\epsilon_{pol} + \frac{5}{12}\epsilon_{pol}^2\big) \end{pmatrix},$$
$$\tag{3.16}$$

expressed in the truncated even-momentum basis $\{|0\rangle, |1\rangle\}$ introduced in Eq. (3.4). Here $\Delta = \Delta\omega(t) - 4\,\omega_{rec}$ is the detuning from the conventional DBD resonance, and $\Omega = \Omega(t)$ is the time-dependent Rabi frequency.

The diagonal elements of Eq. (3.16), originating from the second-order Magnus term, represent light-induced energy shifts of the momentum states. These shifts, which depend quadratically on the Rabi frequency and on polarization error $\epsilon_{pol}$, correspond to the time-dependent *AC-Stark shift* ($\Delta_{AC}$) in first-order DBD. The off-diagonal couplings, by contrast, arise at first order in the Magnus expansion and capture the effective transitions between the momentum states.

It is worth noting that the above procedure involves a certain degree of double-counting in the treatment of the AC-Stark shift. We have also explored alternative derivations, which lead to slightly different numerical prefactors in front of the polarization-error-dependent terms, yet the overall physical picture and qualitative conclusions remain unaffected. The purpose of the effective two-level Hamiltonian is therefore not to replace the full description given in Eq. (3.1), but rather to provide an analytically transparent framework for capturing the essential physics of resonance shifts and develop physical intuitions for guiding the design of detuning-control strategies in DBD.

### 3.3.5 Five-level description with Doppler detuning

To understand the dynamics of DBD for atomic wave packets with a finite momentum spread, we assume an initial Gaussian wave function in momentum space,

$$\psi(p) = \left(2\pi\sigma_p^2\right)^{-1/4} \exp\left(-\frac{(p-p_0)^2}{4\sigma_p^2}\right), \qquad (3.17)$$

where $p_0$ is the center-of-mass (COM) momentum relative to the retro-reflection mirror in Fig. 4.2, and $\sigma_p$ is the momentum width. The latter can be associated with an effective one-dimensional temperature $mk_BT \equiv \sigma_p^2$ along the DBD axis. The corresponding initial state in the plane-wave basis is

$$|\psi(t=0)\rangle = \int \psi(p) \, |p\rangle \, dp, \tag{3.18}$$

with $\langle p'|p\rangle = \delta(p - p')$. Since this state is a superposition of momentum eigenstates, it suffices to determine the evolution of each $|p\rangle$ under the DBD Hamiltonian (3.1), and then superpose the results:

$$|\psi(t)\rangle = \hat{U}(t,0) \, |\psi(0)\rangle = \int \psi(p) \, \hat{U}(t,0)|p\rangle \, dp. \tag{3.19}$$

The Hamiltonian (3.1) is invariant under discrete translations $\hat{\mathbf{T}}_n : \hat{z} \to \hat{z} + n\pi/k_L$, which allows momentum space to be partitioned into Brillouin zones, $\mathcal{P} = \cup_{n=-\infty}^{+\infty}[-\hbar k_L + 2n\hbar k_L, \ \hbar k_L + 2n\hbar k_L)$. We consider atoms with initial momentum $p \in [-\hbar k_L, \ \hbar k_L)$, i.e. in the first Brillouin zone. For DBD up to $\pm 4\hbar k_L$ momentum transfer, the relevant basis contains five states:

$$|1\rangle = |p\rangle,$$
$$|2 \text{ or } 3\rangle = \tfrac{1}{\sqrt{2}}(|p + 2\hbar k_L\rangle \pm |p - 2\hbar k_L\rangle),$$
$$|4 \text{ or } 5\rangle = \tfrac{1}{\sqrt{2}}(|p + 4\hbar k_L\rangle \pm |p - 4\hbar k_L\rangle). \tag{3.20}$$

For $p \neq 0$, these states do not belong to the even or odd parity subspaces introduced earlier. We therefore refer to them as symmetric $(+)$ and anti-symmetric $(-)$ states. In this truncated basis, the Hamiltonian including polarization errors $\epsilon_{pol}$ and Doppler detuning reads

$$H_{Doppler}^{pol}/\hbar = \begin{pmatrix} \frac{p^2}{2m\hbar} & \sqrt{2}\Omega(t)C(t,\epsilon_{pol}) & 0 & 0 & 0 \\ \sqrt{2}\Omega(t)C(t,\epsilon_{pol}) & \frac{p^2}{2m\hbar} + 4\omega_{rec} & 4\frac{p}{\hbar k_L}\omega_{rec} & \Omega(t)C(t,\epsilon_{pol}) & 0 \\ 0 & 4\frac{p}{\hbar k_L}\omega_{rec} & \frac{p^2}{2m\hbar} + 4\omega_{rec} & 0 & \Omega(t)C(t,\epsilon_{pol}) \\ 0 & \Omega(t)C(t,\epsilon_{pol}) & 0 & \frac{p^2}{2m\hbar} + 16\omega_{rec} & 8\frac{p}{\hbar k_L}\omega_{rec} \\ 0 & 0 & \Omega(t)C(t,\epsilon_{pol}) & 8\frac{p}{\hbar k_L}\omega_{rec} & \frac{p^2}{2m\hbar} + 16\omega_{rec} \end{pmatrix}, \tag{3.21}$$

where $C(t,\epsilon_{pol}) = \cos\left[(4\,\omega_{rec} + \Delta(t))t\right] + \epsilon_{pol}$.

The off-diagonal elements $(2, 3)$ and $(3, 2)$ encode a Doppler-induced coupling $4p\,\omega_{rec}/(\hbar k_L) = 2k_L v$, which directly couples the symmetric and anti-symmetric states $|2\rangle$ and $|3\rangle$. In the bare momentum basis, $|p \pm 2\hbar k_L\rangle = (|2\rangle \pm |3\rangle)/\sqrt{2}$, this manifests as opposite phase evolutions of the two components. Therefore, for $p \neq 0$, we choose to characterize the efficiency of double Bragg diffraction in terms of the populations in the bare momentum states. The AC-Stark shifts of state $|2\rangle$ and $|3\rangle$ due to the virtual population of $|4\rangle$ and $|5\rangle$ is automatically taken care of by solving the 5-level Hamiltonian (3.21) in real-time.

**Extension beyond five levels.** The above five-level truncation can be generalized systematically. For momentum transfers $\pm 2n\hbar k_L$ ($n \geq 1$), we define symmetric and anti-symmetric states

$$
\begin{aligned}
|n, +\rangle &= \frac{|p + 2n\hbar k_L\rangle + |p - 2n\hbar k_L\rangle}{\sqrt{2}}, \\
|n, -\rangle &= \frac{|p + 2n\hbar k_L\rangle - |p - 2n\hbar k_L\rangle}{\sqrt{2}}.
\end{aligned}
\tag{3.22}
$$

Ordering the basis as $\{|p\rangle, |1, +\rangle, |1, -\rangle, |2, +\rangle, |2, -\rangle, \dots\}$, the non-zero Hamiltonian matrix elements are

$$
\begin{aligned}
\langle p|H(t)|p\rangle &= \frac{p^2}{2m}, \quad \langle p|H(t)|1, +\rangle = \sqrt{2}C(t, \epsilon_{pol})\hbar\Omega(t), \\
\langle n, +|H(t)|n, +\rangle &= \langle n, -|H(t)|n, -\rangle = \frac{p^2}{2m} + 4n^2\hbar\omega_{rec}, \\
\langle n, +|H(t)|n, -\rangle &= 4n\frac{p}{\hbar k_L}\hbar\omega_{rec}, \\
\langle n, +|H(t)|(n+1), +\rangle &= \langle n, -|H(t)|(n+1), -\rangle = C(t, \epsilon_{pol})\hbar\Omega(t).
\end{aligned}
\tag{3.23}
$$

Truncating to $n = 2$ reproduces the $5 \times 5$ Hamiltonian of Eq. (3.21).

**Interaction picture.** For numerical simulation it is convenient to transform into the interaction picture with respect to the diagonal part $H_0 = \mathrm{diag}(\frac{p^2}{2m}, \frac{p^2}{2m} + 4\hbar\omega_{rec}, \dots)$. For the five-level truncation, this yields

$$
\bar{H}_{Doppler}^{pol}/\hbar = \begin{pmatrix}
0 & \sqrt{2}\Omega C\, e^{-i4\omega_{rec}t} & 0 & 0 & 0 \\
\sqrt{2}\Omega C\, e^{i4\omega_{rec}t} & 0 & 4\frac{p}{\hbar k_L}\omega_{rec} & \Omega C\, e^{-i12\omega_{rec}t} & 0 \\
0 & 4\frac{p}{\hbar k_L}\omega_{rec} & 0 & 0 & \Omega C\, e^{-i12\omega_{rec}t} \\
0 & \Omega C\, e^{i12\omega_{rec}t} & 0 & 0 & 8\frac{p}{\hbar k_L}\omega_{rec} \\
0 & 0 & \Omega C\, e^{i12\omega_{rec}t} & 8\frac{p}{\hbar k_L}\omega_{rec} & 0
\end{pmatrix},
$$

$$(3.24)$$

where we have introduced the shorthand notation $C \equiv C(t, \epsilon_{pol})$ for compactness, and the ordered basis is given by $\{|p\rangle, |1, +\rangle, |1, -\rangle, |2, +\rangle, |2, -\rangle\}$. This interaction-picture Hamiltonian will be used for the five-level theory presented in the following sections.

## 3.4 Mitigations of AC-Stark Shift and Polarization Errors

In practical implementations of double Bragg diffraction, idealized resonance conditions are modified by light-induced shifts and technical imperfections. Two of the most significant limitations are (i) the time-dependent AC-Stark shift, which alters the effective resonance frequency of the transition, and (ii) polarization errors, which couple undesired momentum states and degrade beam-splitter and mirror efficiencies. Both effects, if uncompensated, lead to reduced interferometric contrast and limit the robustness of DBD-based atom interferometers. To overcome these challenges, one can exploit detuning control strategies, where the frequency difference between the two lattice beams is carefully chosen—or even dynamically varied—to counteract the detrimental shifts. In this section, we investigate these issues in detail. We first analyze how the AC-Stark shift modifies the traditional double-Bragg resonance condition, using both exact numerical simulations and our effective two-level model. We then demonstrate how constant or time-dependent detuning functions can be employed to compensate for AC-Stark shifts and polarization errors, enabling efficient and robust population transfer in realistic experimental regimes.

### 3.4.1 AC-Stark shifted DBD resonance condition

We now turn to the important question of how the resonance condition of double Bragg diffraction is modified by the presence of AC-Stark shifts. In the idealized treatment found in the literature [138, 139], the resonance condition for $j$th-order double Bragg diffraction is given by

$$\Delta\omega = \frac{(2j\hbar k_L)^2}{2jm\hbar} = 4j\,\omega_{rec}. \tag{3.25}$$

This condition assumes vanishing initial momentum and neglects light-induced shifts of the involved states. As we will demonstrate, however, the AC-Stark effect shifts the resonance of both the initial and target states, such that simply setting $\Delta\omega = 4\omega_{rec}$ does not yield perfect population transfer, even in the absence of polarization errors.

To illustrate this effect, let us first consider the simplest case of a square (box) pulse with no polarization error, $\epsilon_{pol} = 0$. The Rabi frequency is modeled as

$$\Omega(t) = \Omega\left[\Theta(t) - \Theta(t - \tau)\right], \tag{3.26}$$

where $\tau$ denotes the pulse duration and $\Theta(t)$ is the Heaviside step function. We quantify the efficiency of a double Bragg beam splitter by the final population of the target state $|1\rangle$ after the pulse, i.e. $P(|1\rangle)$.

Fig. 3.2 shows two-dimensional scans of the efficiency as a function of pulse duration $\tau$ and peak Rabi frequency $\Omega$. The left panel corresponds to exact numerical simulations of the full Hamiltonian (3.1) calculated by a position-space numerical solver of the Schrödinger equation (UATIS [143, 144]) based on the second-order Suzuki-Trotter decomposition [145] with a momentum truncation up to $\pm 10.9\,\hbar k_L$, which we will refer to as the "exact numerical solution", or simply "exact" in the figure legends. The right panel displays the predictions of the effective two-level Hamiltonian (3.16). For the UATIS simulations, we employed a narrow Gaussian initial state with $\sigma_p = 0.01\hbar k_L$ centered at $p_0 = 0$ in order to approximate the plane-wave eigenstate $|0\hbar k_L\rangle$. Both methods reveal regions of high efficiency, highlighted by contour lines above 91%, and show very good qualitative agreement.

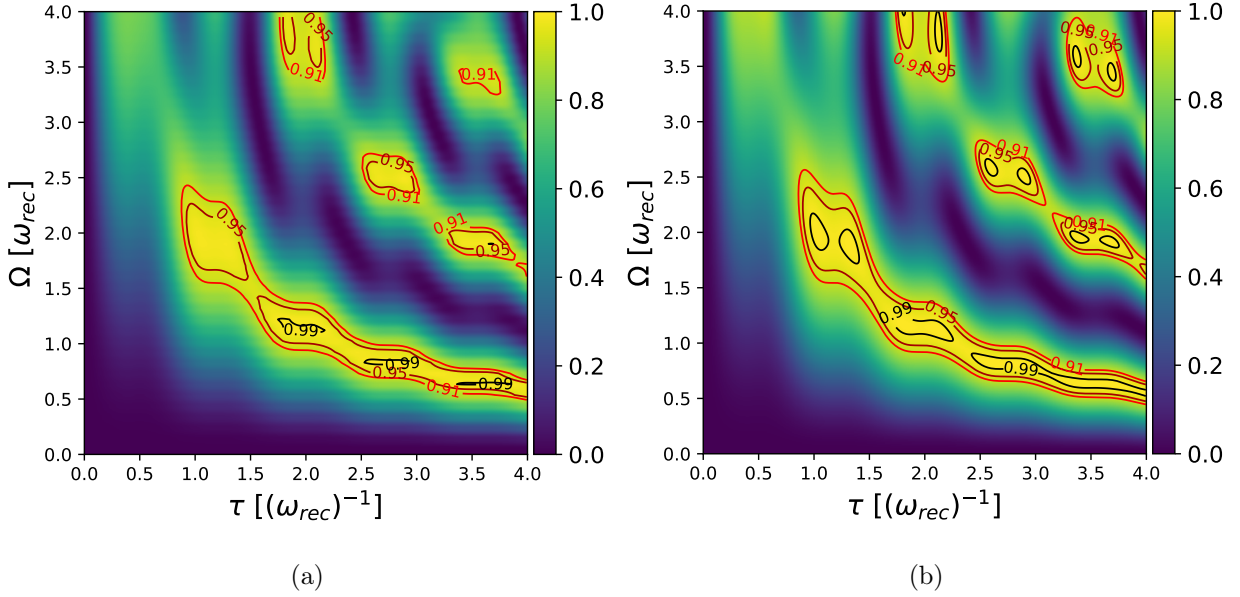

**Figure 3.2** DBD efficiency for a box-pulse as a function of duration $\tau$ and peak Rabi frequency $\Omega$ with $\epsilon_{pol} = 0$ and $\Delta = 0$. (a) Exact numerical solution of the full Hamiltonian (3.1) using UATIS with an initial Gaussian momentum distribution ($\sigma_p = 0.01\,\hbar k_L$, $p_0 = 0$). (b) Dynamics from the effective TLS Hamiltonian (3.16). Contour lines highlight regions with efficiency > 91%.

For a more quantitative comparison, Fig. 3.3 shows the time evolution of the target-state population $P(|1\rangle)$ for a fixed Rabi frequency $\Omega = 2\omega_{rec}$ as a function of $\tau$. We compare three models: the full Hamiltonian (red circles), the effective two-level Hamiltonian (black solid line), and the rotating-wave approximation (RWA) Hamiltonian (blue dashed line). The latter can be derived by transforming Eq. (3.16) into another interaction picture with respect to the diagonal part $H_0 = \text{diag}(0, \Delta)$ and eliminating fast oscillating terms. The

result is

$$\bar{H}_{\mathrm{RWA}} = \hbar \begin{pmatrix} 0 & \frac{\sqrt{2}}{2}\Omega \\ \frac{\sqrt{2}}{2}\Omega & \delta_{diff} \end{pmatrix}, \tag{3.27}$$

with the differential light shift given by

$$\delta_{diff} = -\Delta - \frac{3}{64}\frac{\Omega^2}{\omega_{rec}}. \tag{3.28}$$

The corresponding transition probability follows *Rabi's formula* [146]

$$P_{|0\rangle \to |1\rangle}(t) = \frac{2\Omega^2}{2\Omega^2 + \delta_{diff}^2} \sin^2\left(\frac{\sqrt{2\Omega^2 + \delta_{diff}^2}}{2}t\right). \tag{3.29}$$

From this one recovers the familiar pulse-area condition $\sqrt{2\Omega^2 + \delta_{diff}^2}\, t = \pi$ for a perfect $\pi$-pulse. Evidently, a complete population inversion is only possible when $\delta_{diff} = 0$, where the pulse-area condition reduces to $\sqrt{2}\Omega t = \pi$.

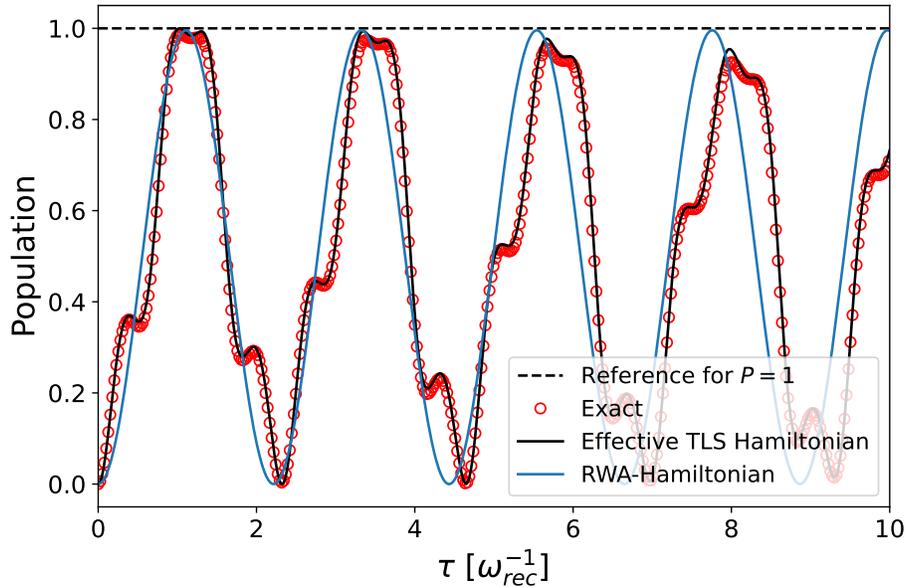

**Figure 3.3** Time evolution of $P(|1\rangle)$ for a box pulse with $\Omega = 2\,\omega_{rec}$. Results are shown for the exact Hamiltonian (3.1) (red circles), the effective TLS Hamiltonian (3.16) (black line), and the RWA Hamiltonian (3.27) with $\delta_{diff} = 0$ (blue line).

As seen in Fig. 3.3, the RWA model fails to reproduce important oscillatory features present in the exact dynamics, while the effective TLS Hamiltonian captures them with excellent accuracy (deviations below 3% over the shown range). This demonstrates that the effective Hamiltonian (3.16) provides a reliable description of DBD in the quasi-Bragg regime, and that AC-Stark shifts must be accounted for when determining the true resonance condition.

## 3.4.2 Polarization errors and their mitigation

As seen in the $(2, 2)$-matrix element of the effective TLS Hamiltonian (3.16), polarization errors do not only couple unwanted momentum states, but also contribute to the AC-Stark shift of the target state. This dual role makes them particularly harmful for high-fidelity beam-splitter operations. In this subsection, we therefore investigate how polarization errors affect double Bragg diffraction and how these detrimental effects can be mitigated by various detuning control strategies.

While we focus on Gaussian pulses in most of the following discussion—since they more closely resemble an ideal two-level system and are less distorted in the first three Rabi cycles under moderate polarization errors (see Fig. 3.4)—we also plot the landscape of the box-pulse efficiency for a qualitative comparison in the upper row of the same figure. The effective Rabi frequency of a Gaussian pulse is parameterized as

$$\Omega(t) = \Omega_R \, e^{-\frac{(t-t_0)^2}{2\tau^2}}, \tag{3.30}$$

with peak amplitude $\Omega_R$, Gaussian pulse width $\tau$, and temporal center $t_0$ (set to zero unless specified otherwise). In the absence of AC-Stark shifts and detuning, the well-known pulse-area condition for DBD [138, 139] can be recovered directly from *Rabi's formula* [146] under the assumption of a slowly varying Rabi frequency $\Omega(t)$. In this case, the transition probability reads

$$P_{|0\rangle \to |1\rangle}(\tau) = \sin^2\left(\frac{\sqrt{2}}{2} \int_{-\infty}^{+\infty} \Omega(t) dt\right) = \sin^2\left(\sqrt{\pi}\, \Omega_R \tau\right), \tag{3.31}$$

from which one can immediately read off the condition for a perfect DBD beam-splitter pulse is $\Omega_R \tau = \sqrt{\pi}/2$.

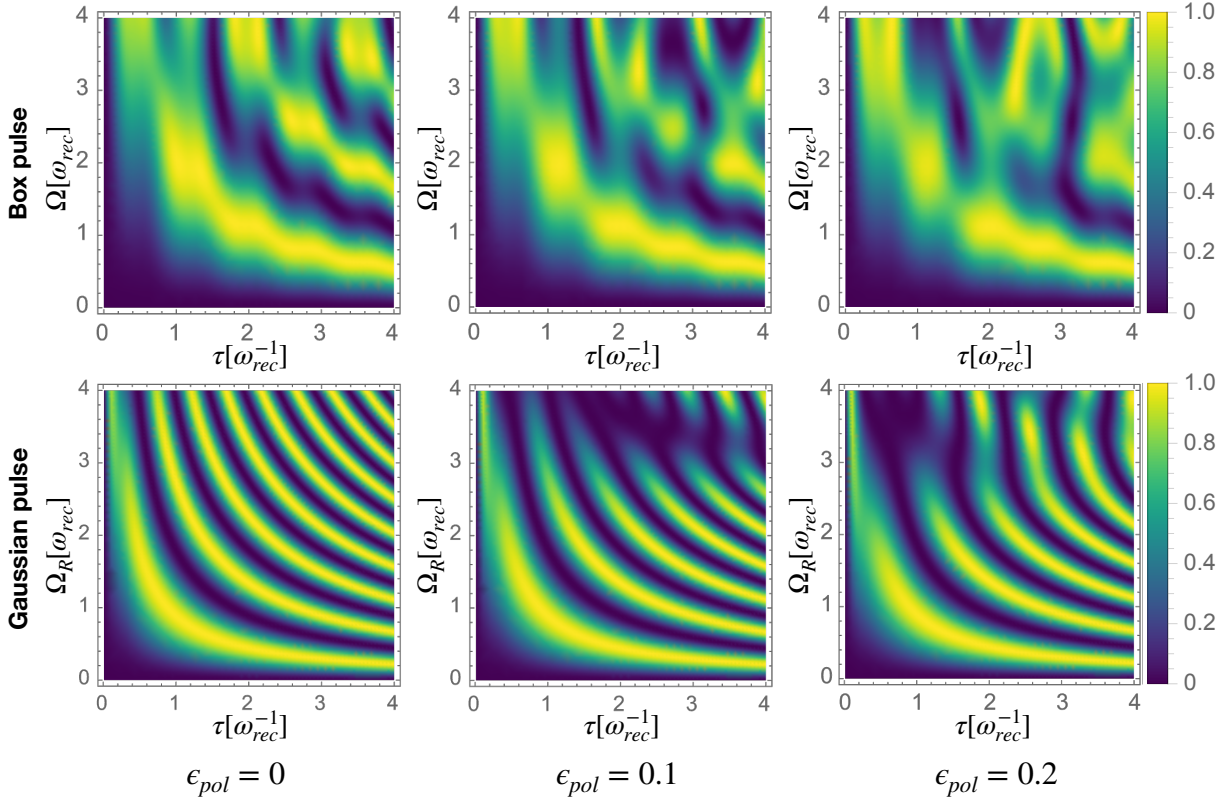

**Figure 3.4** DBD efficiency as a function of pulse width and peak Rabi frequency for box (top row) and Gaussian (bottom row) pulses, under different polarization errors $\epsilon_{pol} = (0, 0.1, 0.2)$. Results are obtained from the effective TLS Hamiltonian (3.16) with $\Delta = 0$.

Fig. 3.4 shows the predicted DBD efficiency for both box and Gaussian pulses under different polarization errors $\epsilon_{pol} = (0, 0.1, 0.2)$, obtained by numerically solving the TLS Hamiltonian (3.16). As the polarization error increases, distortions appear in the Rabi oscillations and the DBD efficiency drops. Gaussian pulses, however, remain closer to the ideal two-level behavior even in the absence of polarization errors ($\epsilon_{pol} = 0$) and are less susceptible to finite $\epsilon_{pol}$ than box pulses, making them the preferred choice for realizing robust atom interferometers. Box-pulse results, shown in the upper row of Fig. 3.4, are nonetheless instructive as they demonstrate how polarization errors quickly degrade the validity of the two-level approximation. To highlight this point, Fig. 3.5(a) compares the time evolution of $P(|1\rangle)$ for a fixed Rabi frequency $\Omega = 2\omega_{rec}$ under different polarization errors, obtained from the TLS Hamiltonian (dashed) and exact numerical simulations (solid).

Clear deviations appear as $\epsilon_{pol}$ increases. Fig. 3.5(b) shows that these deviations correlate with significant population in higher-order momentum states (e.g., $|2\rangle$), which the two-level description does not capture. Thus, while Gaussian pulses remain well described by the TLS Hamiltonian up to moderate polarization errors, box pulses expose the need for a multi-level treatment based on the full Hamiltonian (3.9).

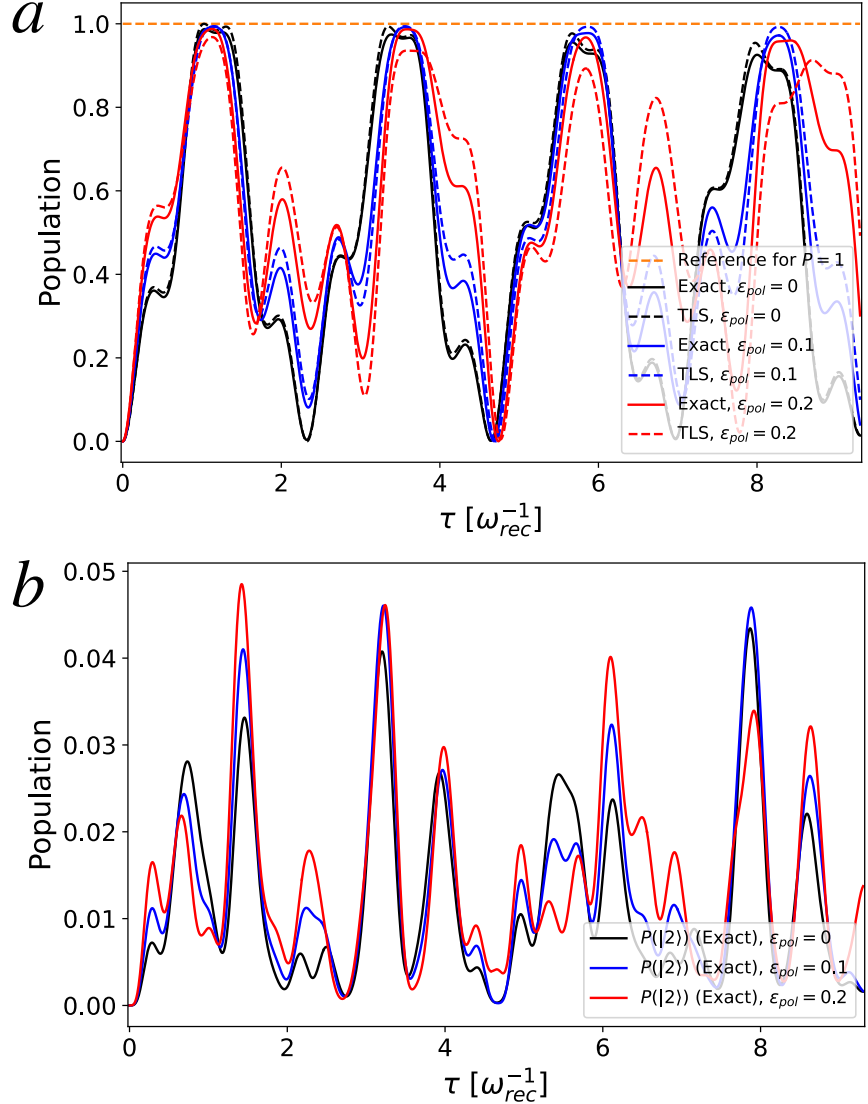

**Figure 3.5** Time evolution of populations for a box-pulse with $\Omega = 2\omega_{rec}$ and $\Delta = 0$. (a) Population in $|1\rangle$ predicted by the effective TLS Hamiltonian (dashed) and exact numerical solution (solid) under polarization errors $\epsilon_{pol} = (0, 0.1, 0.2)$. (b) Population in $|2\rangle$ from the exact solutions, showing higher-order excitations that cause the deviations observed in (a).

Finally, a direct comparison of dynamics predicted by the pulse-area prediction (3.31), the TLS Hamiltonian (3.16), and the exact numerical solution is given in Fig. 3.6 for Gaussian pulses. The pulse-area formula clearly fails to reproduce the exact dynamics even without polarization errors, whereas the TLS Hamiltonian remains accurate up to $\epsilon_{pol} \simeq 0.2$ within the first Rabi cycle. Thus, it provides a reliable framework for developing mitigation strategies. In the remainder of the chapter, we therefore adopt Gaussian pulses for detuning-based mitigation protocols (constant, linear, and OCT), since they retain the essential two-level character needed for analytical insight while minimizing distortions from higher-order couplings. Box pulses, although pedagogically useful, will not be further pursued.

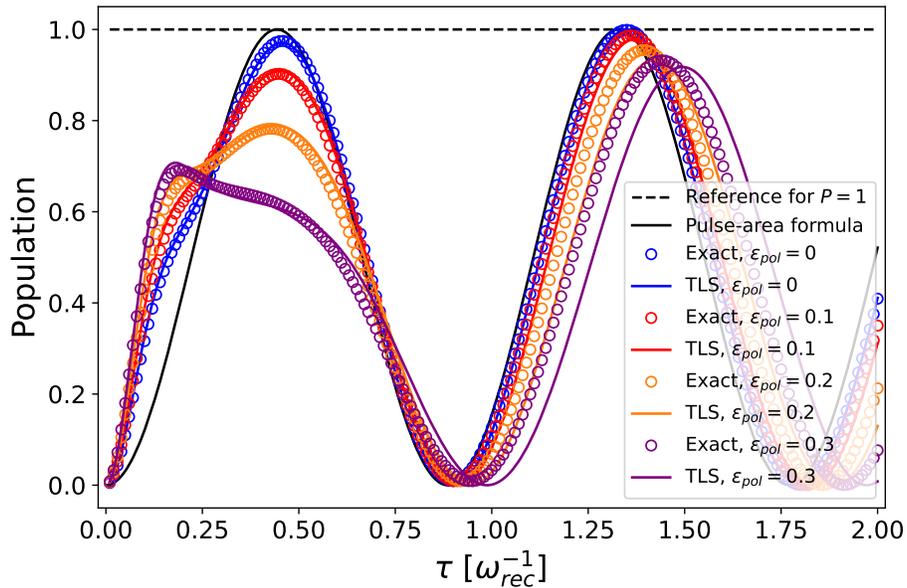

**Figure 3.6** Population in $|1\rangle$ for Gaussian pulses with $\Omega_R = 2\,\omega_{rec}$ as a function of pulse width $\tau$, under different polarization errors. Solid curves: TLS Hamiltonian; circles: exact simulations.

We will now explore three complementary detuning strategies—constant detuning, linear detuning sweeps, and OCT-optimized detuning controls—to assess their relative performance in mitigating AC-Stark shifts and polarization error-induced losses.

**Mitigation with constant detuning**

A straightforward strategy is to offset the resonance condition with a constant detuning $\Delta$. Fig. 3.7 shows for each polarization error $\epsilon_{pol}$ there exists an optimal detuning $\Delta_{opt}$ restoring high efficiency. For example, $\Delta_{opt}/\omega_{rec} = (0.25,\,0.55,\,0.80,\,1.10)$ for $\epsilon_{pol} = (0,\,0.1,\,0.2,\,0.3)$. In typical experimental conditions where $\epsilon_{pol} \lesssim 0.1$, choosing a corresponding $\Delta \lesssim 0.55\,\omega_{rec}$ suffices to recover nearly perfect population transfer. This constant-detuning approach, however, requires $\epsilon_{pol}$ to be well-characterized and stable.

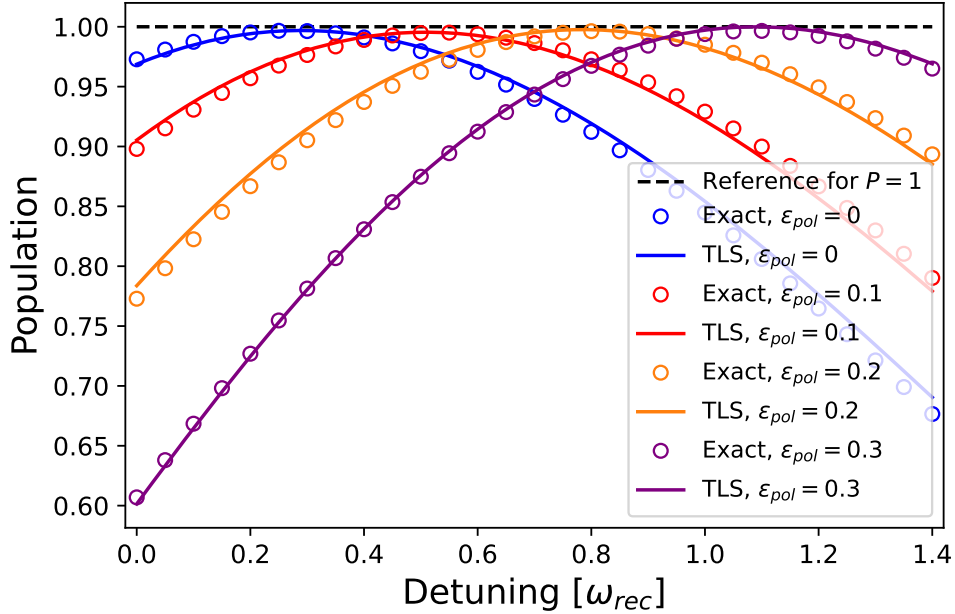

**Figure 3.7** Population in $|1\rangle$ as a function of detuning $\Delta$ for $\tau = 0.47\,\omega_{rec}^{-1}$ and $\Omega_R = 2.0\,\omega_{rec}$, under polarization errors $\epsilon_{pol} = (0,\,0.1,\,0.2,\,0.3)$.

**Mitigation with linear detuning sweeps**

When polarization errors fluctuate from shot to shot or even unknown, a more robust protocol is needed. Inspired by *adiabatic passage* [147–149] in a two-level system, we design a linear detuning sweep

$$\Delta(t)/\omega_{rec} = \frac{1}{2.5\tau}\,(t - t_0 + \tau), \tag{3.32}$$

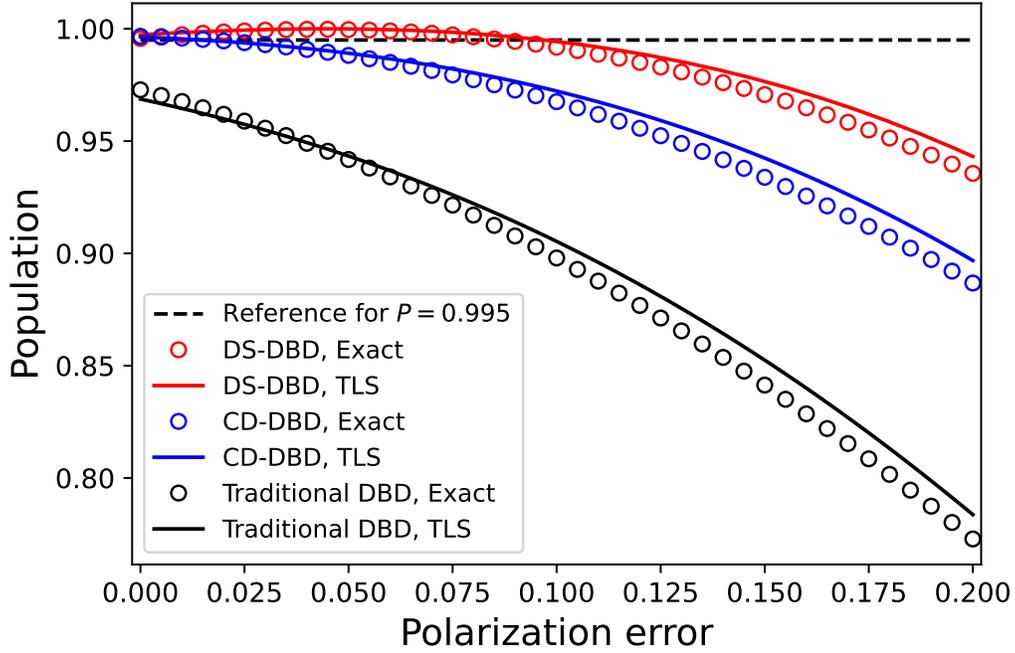

**Figure 3.8** Comparison of Gaussian-pulse DBD efficiency under traditional ($\Delta = 0$), constant detuning (CD-DBD, $\Delta = 0.25\,\omega_{rec}$), and detuning sweep (DS-DBD, Eq. (3.32)). The DS-DBD protocol clearly outperforms the others across $\epsilon_{pol} \in [0, 0.2]$. Dashed line: 99.5% efficiency threshold.

which dynamically traverses the light-shifted resonance during the Gaussian pulse. This detuning-sweep DBD (DS-DBD) protocol achieves efficiencies exceeding 99.5% for polarization errors up to 8.5%, with a peak efficiency of 99.976% at $\epsilon_{pol} = 0.045$. As illustrated in Fig. 3.8, its performance clearly surpasses that of both the traditional protocol with $\Delta = 0$ and the constant-detuning protocol with $\Delta = 0.25\,\omega_{rec}$.

**Mitigation with OCT-based detuning control**

Finally, we explore optimal control theory to design time-dependent detuning profiles beyond linear sweeps. Using Q-CTRL's Boulder Opal package [150], we optimize $\Delta(t)$ together with Gaussian pulse parameters ($\Omega_R, \tau, t_0$) under the constraint $\Delta(t) \leq 4\,\omega_{rec}$. Unlike the TLS model, this requires the five-level Hamiltonian (3.21). The technical details of the OCT optimization and the resulting optimized time-dependent detuning profiles are summarized in Sec. 3.6.

Fig. 3.9 compares the linear detuning sweep given by Eq. (3.32) with an OCT-optimized detuning. While both achieve highly efficient population transfer for small polarization errors, the OCT protocol yields an average target-state population above 99.988% across $\epsilon_{pol} \in [0, 0.1]$, more than an order of magnitude better than the linear sweep. Its near-unity robust performance is maintained up to $\epsilon_{pol} \simeq 0.17$, nearly doubling the tolerance compared to DS-DBD.

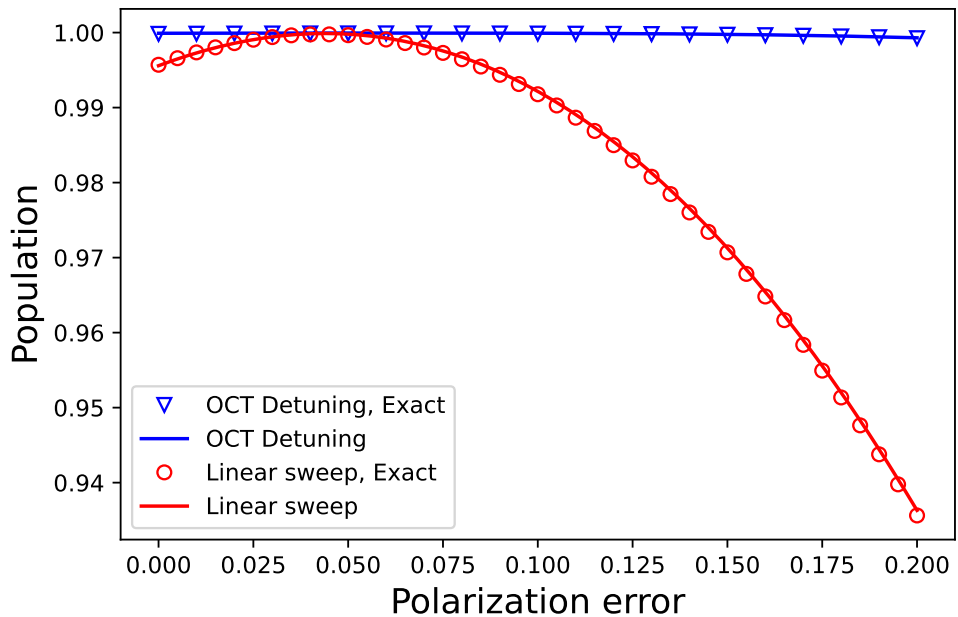

**Figure 3.9** Target state population versus polarization error for Gaussian pulses using linear detuning sweep (red, Eq. (3.32)) and OCT-optimized detuning (blue). Circles/triangles: exact simulations; solid curves: 5-level models. OCT clearly outperforms DS-DBD.

In summary, detuning control offers a hierarchy of strategies: constant detuning suffices for stable and known polarization errors, linear detuning sweeps provide robustness against fluctuations, and OCT-based profiles offer near-perfect compensation across a wide range of polarization errors. This establishes a systematic toolkit for mitigating polarization imperfections in double Bragg diffraction interferometers.

Up until now, we have seen polarization errors as well as AC-Stark shifts, if left uncom-pensated, strongly limit the efficiency of DBD beam splitters, and have shown how constant detuning, linear detuning sweeps, and OCT-based protocols provide a hierarchy of increasingly robust solutions, culminating in near-perfect population transfer even under unknown polarization conditions. Having established these tools, we next turn to another major imperfection in realistic experiments: Doppler detuning arising from finite momentum width and non-zero center-of-mass motion of the atomic ensemble. As we will see in the following section, these Doppler effects require an extended description beyond TLS and call for more sophisticated mitigation strategies complementary to those developed here.

## 3.5 Doppler Effects and Mitigations

### 3.5.1 Losses, momentum selectivity and asymmetry due to Doppler effects

To study Doppler effects on the double Bragg diffraction, we apply the five-level Hamiltonian (3.21) (denoted as 5-LS in the figure legends) to the Gaussian-pulse DBD of an initial momentum eigenstate $|p\rangle$. In Fig. 3.10, we compare the predictions of the five-level Hamiltonian with those of the exact numerical solution. Specifically, we plot the populations in the bare momentum states $\{|p\rangle, |p + 2\hbar k_L\rangle, |p - 2\hbar k_L\rangle\}$ after a Gaussian BS-pulse operated at the traditional resonance condition ($\Delta = 0$). The range of momentum $p$ shown in Fig. 3.10 extends beyond the typical momentum width $\sigma_p = \sqrt{mk_BT}$ of atomic clouds prepared in an AI experiment with delta-kick-cooled (DKC) ultracold atoms, which is usually well below $0.1\,\hbar k_L$ [75, 151–155]. Fig. 3.10 shows that the effective five-level Hamiltonian accurately captures the dynamics of the full Hamiltonian and correctly predicts the losses due to AC-Stark shifts as well as due to *momentum selectivity* which we discuss in detail below.

The *momentum selectivity* in DBD defines a finite acceptance window in momentum space, outside of which atoms remain largely undiffracted. This effect is already evident in Fig. 3.10: as the initial momentum $p$ deviates further from the resonant value $p = 0$, the transition probability into the target states $|p \pm 2\hbar k_L\rangle$ gradually decreases. To better illustrate this effect, we consider a Gaussian BS pulse with twice the width of the earlier example, $\tau = 0.91\,\omega_{rec}^{-1}$, and a reduced peak Rabi frequency of $\Omega_R = 1.0\,\omega_{rec}$. Fig. 3.11a shows the resulting populations in the $\pm 2\hbar k_L$ output ports as a function of the initial momentum $p$. A clear acceptance window is observed, approximately spanning $(-0.1\hbar k_L,\ 0.1\hbar k_L)$, within which efficient diffraction occurs. Outside this window, the transition probability rapidly decreases.

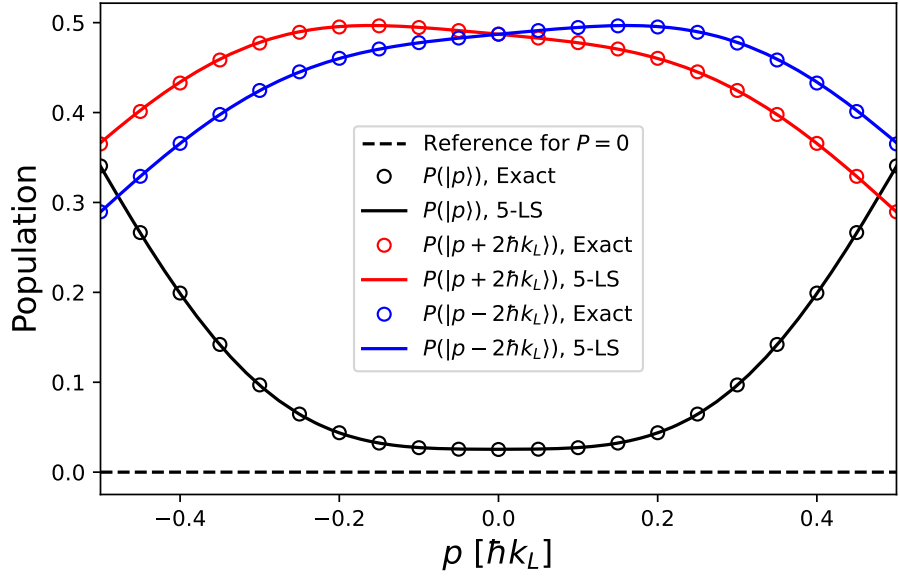

**Figure 3.10** Population in momentum states $\{|p\rangle, |p + 2\hbar k_L\rangle, |p - 2\hbar k_L\rangle\}$ after a Gaussian BS-pulse at classical resonance ($\Delta = 0$) as a function of the initial momentum $p$. The Gaussian-pulse parameters used are $\Omega_R = 2\omega_{rec}$, $\tau = 0.45\omega_{rec}^{-1}$. The initial momentum width used in the exact numerical calculation is $\sigma_p = 0.01\,\hbar k_L$.

The impact of this momentum filtering becomes obvious when an input wave packet has a width comparable to the acceptance window. Figure 3.11b presents the case of an initial Gaussian state with momentum width $\sigma_p = 0.1\,\hbar k_L$, centered at $p_0 = 0$. Here, only the central portion of the wave packet—those momentum modes close to resonance—are efficiently diffracted, while atoms in the high-momentum tails remain in the undiffracted state. Importantly, the results from the five-level model and the exact numerical solution

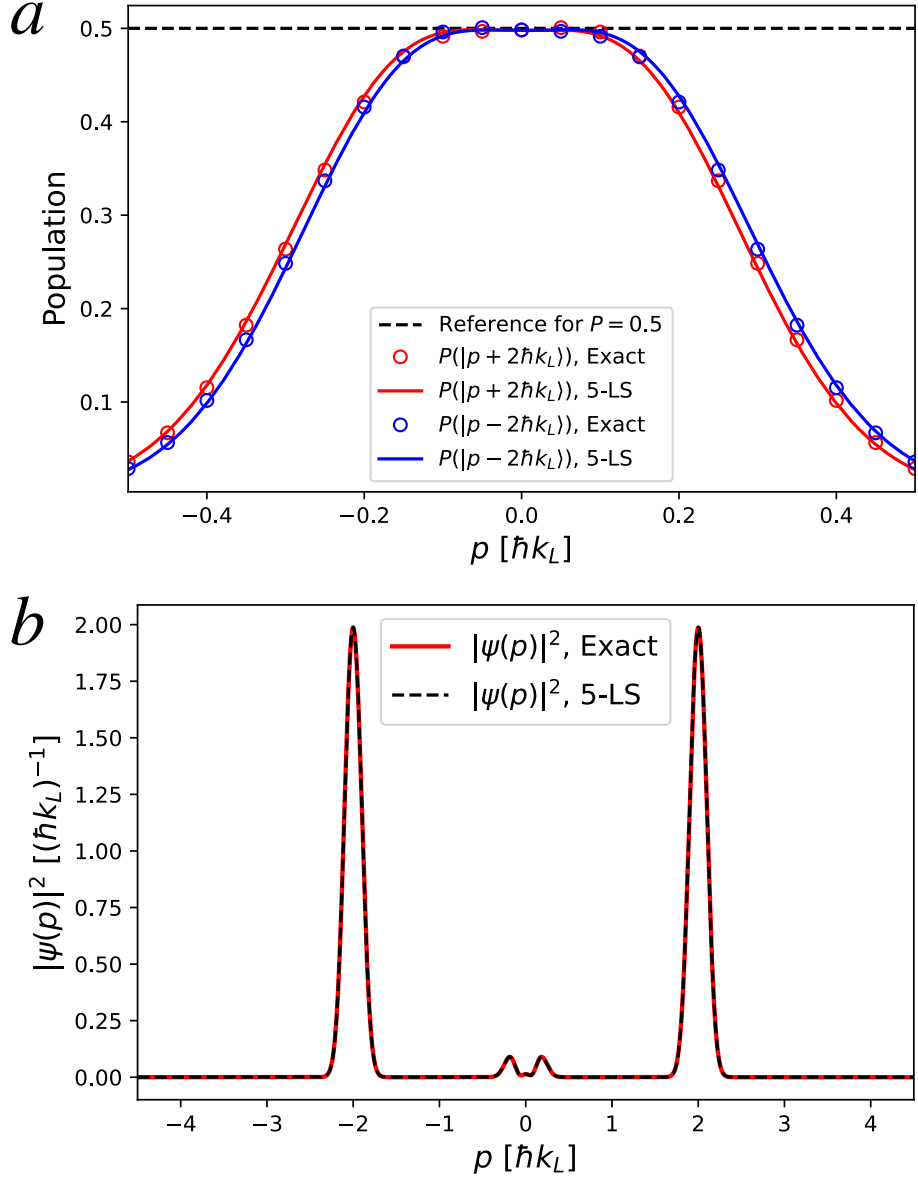

**Figure 3.11** (a) Population in different momentum states after a Gaussian BS pulse as a function of initial momentum $p$ with detuning $\Delta = 0$. The lines are results of the five-level theory and circles are exact numerical results. Red and blue correspond to populations in $|p + 2\hbar k_L\rangle$ and $|p - 2\hbar k_L\rangle$, respectively. (b) Final wave packet in momentum space $|\psi(p)|^2$ after a Gaussian BS pulse showing momentum selectivity due to Doppler effects with an initial momentum width $\sigma_p = 0.1\,\hbar k_L$ and initial COM momentum $p_0 = 0$. The results of the five-level theory and the exact numerical solution are indistinguishable. Pulse parameters: $\Omega_R = 1.0\,\omega_{rec}$, $\tau = 0.91\,\omega_{rec}^{-1}$.

are indistinguishable, confirming that the truncated description provides an accurate and efficient tool for capturing momentum selectivity in DBD.

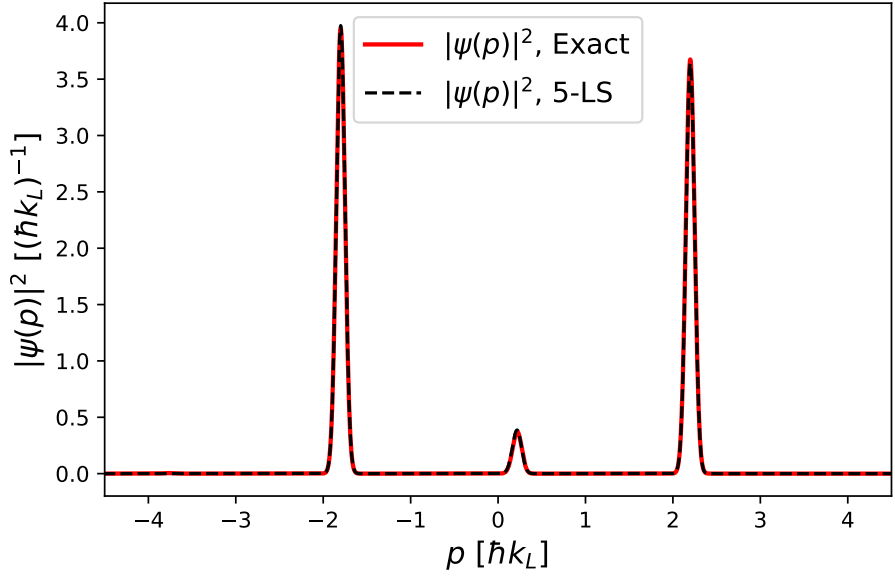

**Figure 3.12** Final wave packet in momentum space $|\psi(p)|^2$ after a BS-pulse showing losses and asymmetry due to Doppler effects with an initial momentum width $\sigma_p = 0.05\,\hbar k_L$ and COM momentum $p_0 = 0.2\,\hbar k_L$. The results of the five-level theory and the exact numerical solution are indistinguishable. Here, the $-2\hbar k_L$ transition is preferred over the $+2\hbar k_L$ transition due to $p_0 > 0$. Pulse parameters: $\Omega_R = 2\omega_{rec}$, $\tau = 0.45\,\omega_{rec}^{-1}$ with $\Delta = 0$.

Another important consequence of the Doppler detuning is the *asymmetry* in the populations between the left and right Bragg diffractions observed for small initial momenta near $p = 0$. This asymmetry is characterized by the slope of $P(|p + 2\hbar k_L\rangle)$ at $p = 0$ after the BS-pulse. Intuitively, this can be understood by noting that the differential light shift $\delta_{diff} = -\Delta - 3\Omega^2/(64\,\omega_{rec})$, derived in Sec. 3.4 for the $p = 0$ case, is negative at the traditional resonance condition ($\Delta = 0$). The quadratic dispersion relation $E(p) = p^2/(2m)$ gives opposite shifts for the two diffraction channels when $p_0 \neq 0$:

$$\Delta E|_L = E(-2\hbar k_L + p_0) - E(p_0) - \left(E(-2\hbar k_L) - E(0)\right) = -\frac{2\hbar k_L}{m}p_0, \qquad (3.33)$$

$$\Delta E|_R = E(2\hbar k_L + p_0) - E(p_0) - \left(E(2\hbar k_L) - E(0)\right) = \frac{2\hbar k_L}{m}p_0, \qquad (3.34)$$

which means the $-2\hbar k_L$ channel moves closer to resonance, while the $+2\hbar k_L$ channel shifts further away. Therefore, for a positive initial momentum $p_0 > 0$, the $-2\hbar k_L$ transition is favored, as shown in Fig. 3.12.

In summary, Doppler detuning leads to two key physical consequences in double Bragg diffraction:

1. **Momentum selectivity:** only atoms with momenta within a finite acceptance window are efficiently diffracted, whereas those in the tails of the momentum distribution largely remain undiffracted.

2. **Asymmetry:** finite center-of-mass momentum causes unequal splitting into the two output ports, with one diffraction direction favored over the other.

### 3.5.2 Doppler effect mitigations via constant or linear detuning control

The Doppler effect introduces both losses and asymmetry in double Bragg beam-splitters, as discussed in the previous subsection. To counteract these detrimental effects, one can employ suitable detuning strategies during the Gaussian pulse. Two simple but instructive options are constant detuning and linear detuning sweeps.

In the first case, applying a constant detuning can effectively eliminate losses around $p = 0$. However, this comes at the cost of enhancing the asymmetry between the $\pm 2\hbar k_L$ output states. Fig. 3.13 illustrates such a scenario for a particular choice of constant detuning $\Delta/\omega_{rec} = 0.345$, showing nearly lossless performance at $p = 0$, but with a clear imbalance between left- and right-diffracted states for finite $p$.

An alternative strategy is to apply a linear detuning sweep across the pulse duration. By slowly sweeping through the shifted resonance, one can suppress the asymmetry while still allowing for some finite losses compared to the constant-detuning case. Fig. 3.14 shows an example of such a linear sweep, $\Delta(t)/\omega_{rec} = (t + 0.9\,\tau)/(5\,\tau)$, which yields more symmetric populations in the two output ports at the expense of reduced transfer efficiency.

These two methods illustrate a trade-off: constant detuning maximizes transfer efficiency but introduces strong asymmetry, whereas linear detuning sweeps suppress asymmetry but at a cost of a finite efficiency loss. In the following subsection, we demonstrate how both limitations can be overcome with optimal-control-based time-dependent detuning.

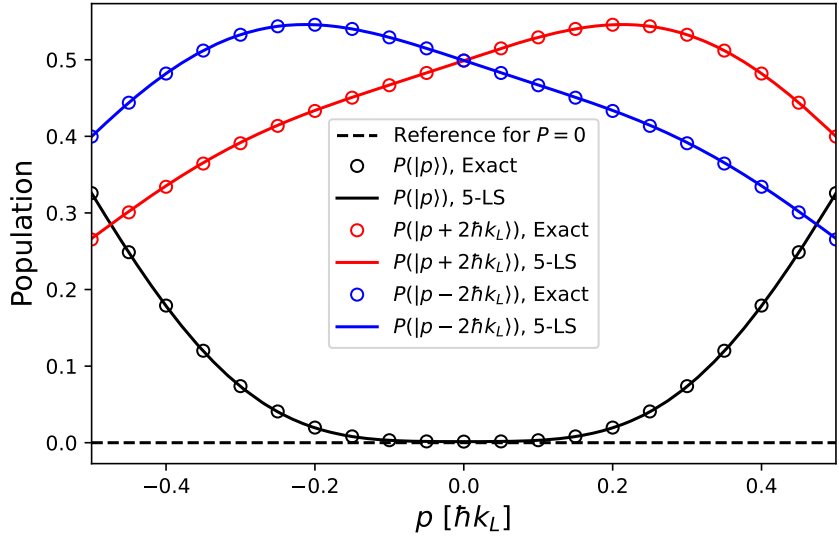

**Figure 3.13** Population in different momentum states after a Gaussian BS pulse as a function of the initial momentum $p$ for a constant detuning $\Delta/\omega_{rec} = 0.345$. Gaussian-pulse parameters: $\Omega_R = 2\,\omega_{rec}$, $\tau = 0.45\,\omega_{rec}^{-1}$. The initial momentum width in the exact numerical calculation is $\sigma_p = 0.01\,\hbar k_L$.

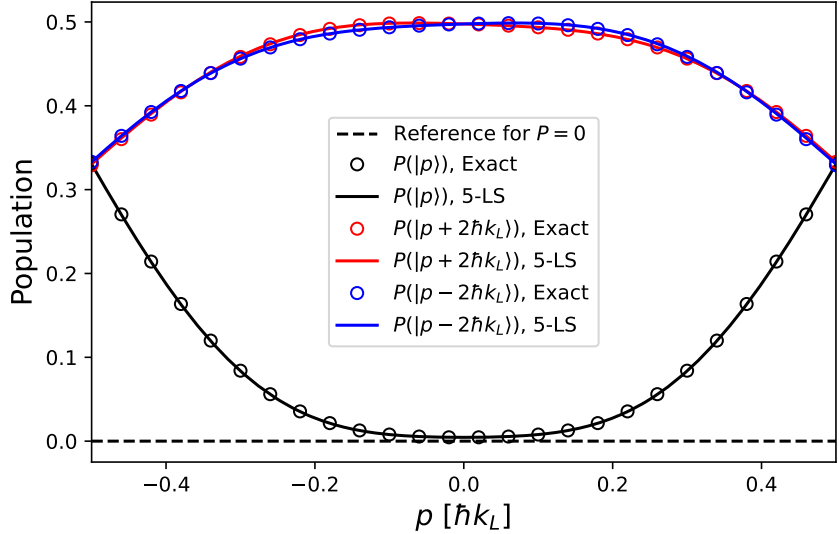

**Figure 3.14** Population in different momentum states after a Gaussian BS pulse as a function of the initial momentum $p$ for a linear detuning sweep $\Delta(t)/\omega_{rec} = (t + 0.9\,\tau)/(5\,\tau)$. Gaussian-pulse parameters: $\Omega_R = 2\,\omega_{rec}$, $\tau = 0.45\,\omega_{rec}^{-1}$. The initial momentum width in the exact numerical calculation is $\sigma_p = 0.01\,\hbar k_L$.

### 3.5.3 Doppler effect mitigation via OCT

To go beyond the limitations of constant or linear detuning strategies, we employ optimal control theory (OCT) to design a generalized time-dependent detuning function that simultaneously minimizes Doppler-induced losses and asymmetry. The optimization is performed by sampling a uniform distribution of initial momenta $p \in [-0.3\hbar k_L,\ 0.3\hbar k_L]$ and minimizing both the deviation from a $50:50$ splitting between $|p+2\hbar k_L\rangle$ and $|p-2\hbar k_L\rangle$, as well as the asymmetry between the two ports. Technical details of the cost function and optimization procedure can be found in Sec. 3.6. The resulting performance of the OCT-optimized detuning is shown in Fig. 3.15, where we plot the final populations in the bare momentum states $|p\rangle$, $|p+2\hbar k_L\rangle$, and $|p-2\hbar k_L\rangle$ as a function of initial momentum $p$. Compared to the results of constant (see Fig. 3.13) and linear detuning (see Fig. 3.14) mitigations, the optimized protocol yields output populations much closer to the ideal $50/50$ distribution over a wide range of momenta. The inset highlights the region $p \in [-0.2\hbar k_L,\ 0.2\hbar k_L]$, showing nearly flat and symmetric populations in the $\pm 2\hbar k_L$ states, with deviations below the 0.1% level.

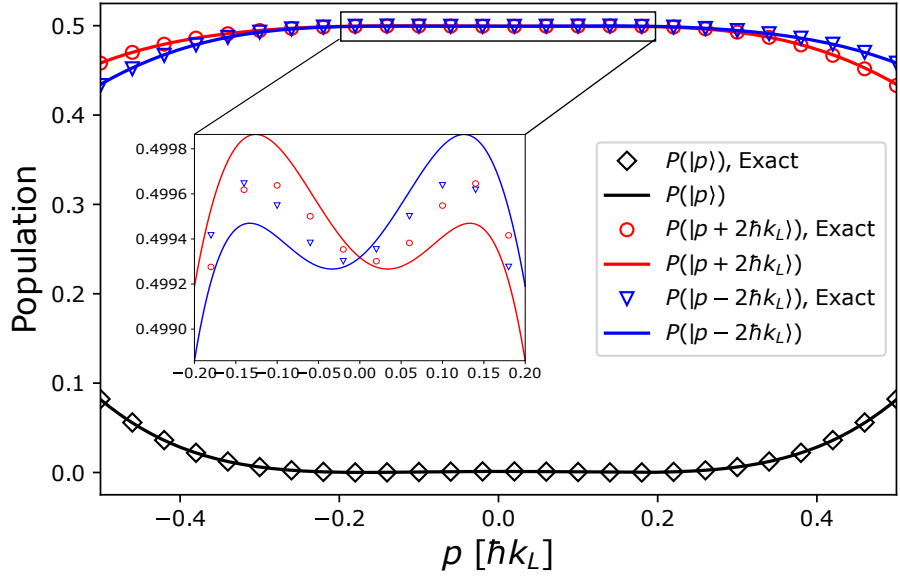

**Figure 3.15** Populations in bare momentum states after a Gaussian BS pulse under the OCT-optimized detuning (exact details are shown in Sec. 3.6). The Gaussian-pulse parameters are $(\Omega_R,\ \tau,\ t_0) = (2.079\,\omega_{rec},\ 0.534\,\omega_{rec}^{-1},\ 2.463\,\omega_{rec}^{-1})$. Inset: zoom-in for $p \in [-0.2\,\hbar k_L,\ 0.2\,\hbar k_L]$ (excluding $|p\rangle$), showing almost perfectly symmetric populations in $|p \pm 2\hbar k_L\rangle$.

At this level of precision, the small differences between exact numerical calculations and the OCT-optimized five-level simulations may arise from two factors: (i) the exact solver assumes a finite initial momentum width ($\sigma_p = 0.01\,\hbar k_L$) while the OCT optimization assumes an infinitely narrow momentum width, and (ii) higher-order momentum states beyond the truncated five-level model may acquire small but non-negligible population during the Gaussian pulse. Nonetheless, the OCT approach offers a clear advantage over constant or linear detuning controls by simultaneously reducing both momentum-selectivity losses and double-diffraction asymmetry.

### 3.5.4 Combined mitigation via OCT

We now turn to the most general case, where both polarization errors and Doppler detuning affect double Bragg diffraction simultaneously. To design a robust mitigation strategy, we employ optimal control theory (OCT) using the full five-level Hamiltonian framework. Specifically, we sample polarization errors uniformly from the interval $\epsilon_{pol} \in [0, 0.1]$, together with initial atomic momenta drawn uniformly from $p \in [-0.3\hbar k_L, 0.3\hbar k_L]$ (as in Sec. 3.5.3). The optimization minimizes both the deviation from a 50/50 population distribution in the output ports $|p \pm 2\hbar k_L\rangle$ and any asymmetry between them, thereby ensuring high-fidelity and balanced beam splitter performance under realistic experimental imperfections.

To quantify the performance, we adopt the *OCT beam-splitter (BS) efficiency*, which extends the DBD efficiency introduced in Sec. 3.4 by including both population transfer and port symmetry. Details of this metric are given in Sec. 3.6. In Fig. 3.16, we compare the OCT BS efficiency for a simple linear detuning sweep [panel (a)] and a fully optimized OCT detuning protocol [panel (b)]. While the linear sweep achieves efficiencies above 0.95 only within a narrow triangular region of the $(p, \epsilon_{pol})$ space, the OCT optimization maintains efficiencies above 0.99 across a much larger square region extending to $|p| \leq 0.18\,\hbar k_L$ and $\epsilon_{pol} \leq 0.12$.

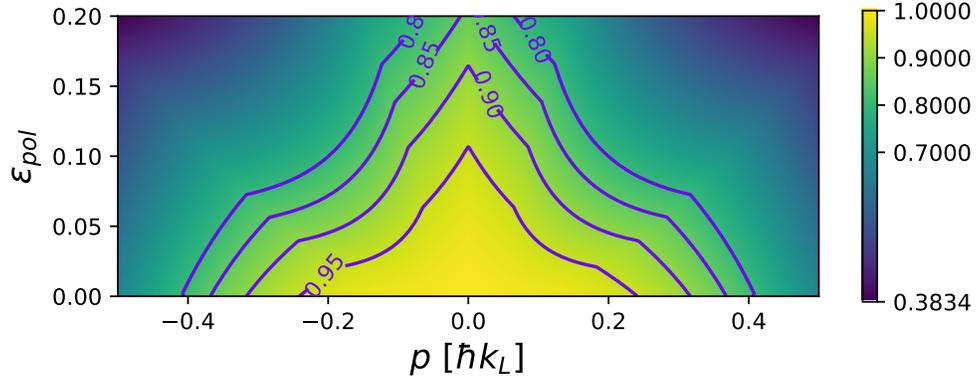

(a) Linear detuning sweep

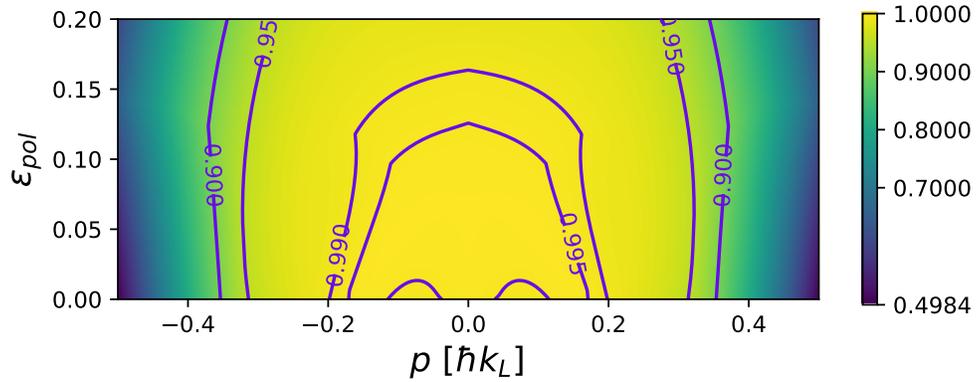

(b) OCT-optimized detuning

**Figure 3.16** OCT beam-splitter efficiency as a function of initial momentum $p$ and polarization error $\epsilon_{pol}$ for two detuning protocols: (a) linear sweep $\Delta(t)/\omega_{rec} = (t + 0.9\tau)/(5\tau)$ with parameters $(\Omega, \tau, t_0) = (2\,\omega_{rec}, 0.45\,\omega_{rec}^{-1}, 0)$; (b) OCT-optimized detuning (corresponding to Fig. 3.21) with parameters $(\Omega, \tau, t_0) = (1.264\,\omega_{rec}, 0.915\,\omega_{rec}^{-1}, 4.065\,\omega_{rec}^{-1})$.

To probe the robustness of the OCT protocol more closely, Fig. 3.17 shows the final momentum populations for a fixed polarization error $\epsilon_{pol} = 0.1$, corresponding to a vertical cut through Fig. 3.16b. For small initial momenta $p$, the output populations remain nearly symmetric and well-balanced. At larger $|p|$, however, the efficiency begins to degrade compared to the purely Doppler-optimized case of Fig. 3.15, reflecting the trade-off introduced by simultaneous optimization over two independent error sources.

Finally, we investigate the experimentally realistic case of a finite-momentum-width input state. For a Gaussian atomic ensemble with $\sigma_p = 0.05\,\hbar k_L$, we adapt the optimization by sampling $p$ from a Gaussian distribution rather than a uniform one. Fig. 3.18 compares the resulting target-port populations for three different OCT protocols: (a) optimization tailored for $\sigma_p = 0.05\,\hbar k_L$ and $\epsilon_{pol} \in [0, 0.1]$ (red line) with an optimized OCT detuning shown in Fig. 3.22; (b) optimization only against polarization errors at $p = 0$ (blue line); and (c) optimization against both Doppler and polarization effects simultaneously (black line). The tailored finite-width optimization (red) yields the highest average performance, with a mean target population of 99.92% across $\epsilon_{pol} \in [0, 0.1]$. By contrast, the $p = 0$ optimization (blue) is more stable for larger polarization errors, while the combined optimization (black) offers balanced robustness across both error sources. Small deviations between OCT predictions and exact numerical solutions arise from population leakage into higher momentum states beyond the five-level model, typically at the $10^{-4}$ level.

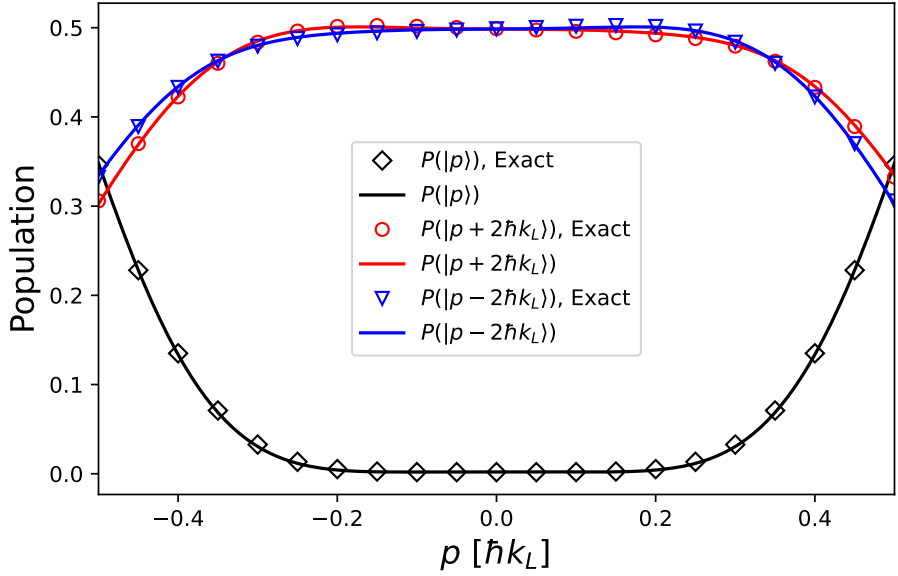

**Figure 3.17** Momentum-state populations after a Gaussian BS pulse under a fixed polarization error $\epsilon_{pol} = 0.1$, shown as a function of initial momentum $p$. The detuning protocol corresponds to the OCT optimization in Fig. 3.21 with parameters $(\Omega, \tau, t_0) = (2.230\,\omega_{rec}, 0.505\,\omega_{rec}^{-1}, 2.970\,\omega_{rec}^{-1})$.

In summary, OCT-based detuning control enables simultaneous mitigation of polarization errors and Doppler detuning, achieving near-unit population transfer and balanced output even for finite-momentum-width atomic ensembles. While trade-offs emerge when optimizing against multiple error sources, the approach remains significantly more robust than either constant or linear detuning strategies.

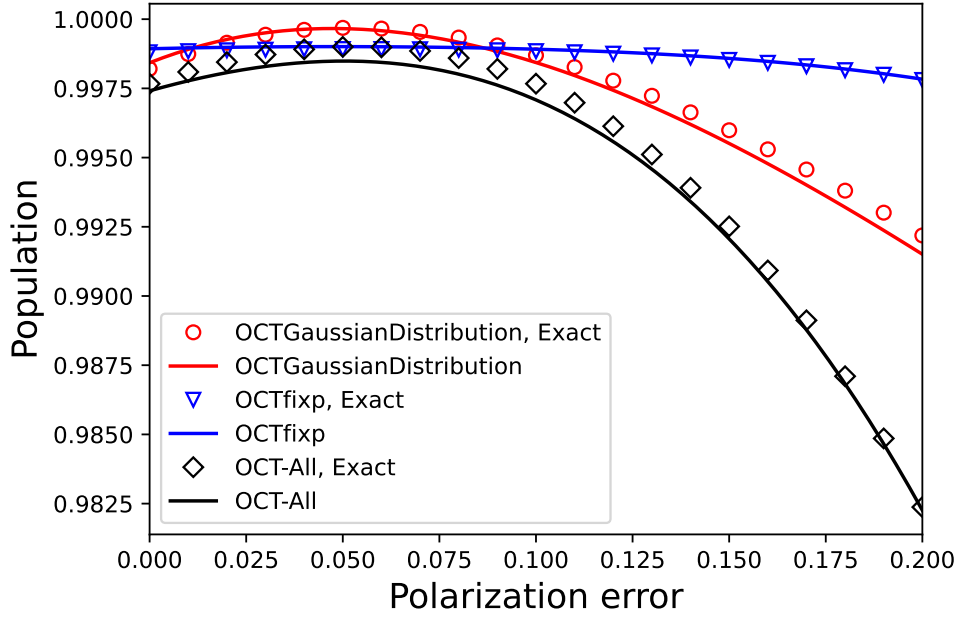

**Figure 3.18** Summed population in the $\pm 2\hbar k_L$ ports after a Gaussian BS pulse for an input Gaussian state with $\sigma_p = 0.05\,\hbar k_L$ and $p_0 = 0$. Three OCT detuning protocols are compared: (a) optimized for finite $\sigma_p$ and polarization errors (red); (b) optimized against polarization errors at $p = 0$ (blue); (c) optimized against both Doppler and polarization errors (black). Lines denote OCT predictions, symbols show exact simulations.

## 3.6 Methods and Technical Details of OCT Optimization

In this section, we summarize the optimization procedure used to obtain the time-dependent detunings presented in the main text. All optimizations were carried out using the full five-level Hamiltonian of double Bragg diffraction [Eq. (3.21)], with the optimization variables chosen as $(\Omega_R,\, \tau,\, t_0,\, \Delta(t))$, where $\Omega_R$ is the peak Rabi frequency, $\tau$ the pulse width, $t_0$ the pulse center, and $\Delta(t)$ the time-dependent detuning. Unless otherwise stated, $\Delta(t)$ was constrained to values $\Delta(t) \leq 4\,\omega_{rec}$.

## Polarization error mitigation with $p = 0$

For atoms prepared with an infinitely narrow momentum distribution around initial momentum $p = 0$, the dynamics decouple into symmetric and antisymmetric subspaces. In this case, the optimization directly targeted the population transfer into the symmetric (or even) state $|1\rangle = (|2\hbar k_L\rangle + |-2\hbar k_L\rangle)/\sqrt{2}$. The resulting optimal Gaussian-pulse parameters are $(\Omega_R, \tau, t_0) = (1.264\omega_{rec}, 0.915\omega_{rec}^{-1}, 4.065\omega_{rec}^{-1})$, and the optimal detuning profile is shown in Fig. 3.19.

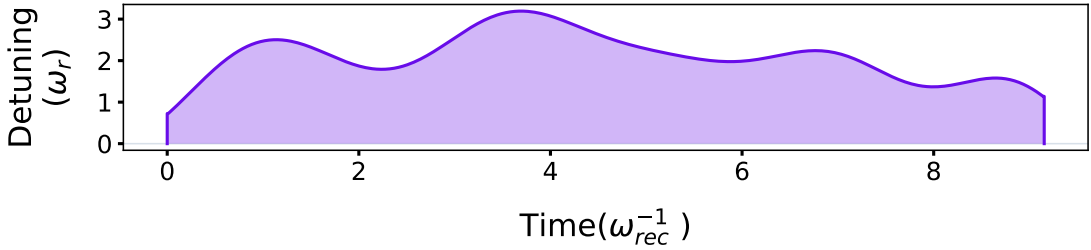

**Figure 3.19** Optimal time-dependent detuning obtained from OCT to mitigate polarization errors at $p = 0$. The optimization was carried out using the 5-level Hamiltonian, targeting maximum population transfer into the target state $|1\rangle$.

## Doppler and combined cases

For finite initial momentum $p \neq 0$, Doppler coupling mixes symmetric and antisymmetric states, so the optimization cost function was defined in terms of the bare momentum populations after the pulse. For a final state $|\psi(t_f)\rangle_{\epsilon_{pol}, p}$ evolved with initial momentum $p$ and polarization error $\epsilon_{pol}$, the cost is given by

$$
\begin{aligned}
Cost(|\psi(t_f)\rangle) = \Big\langle &\Big|0.5 - |\langle p + 2\hbar k_L|\psi(t_f)\rangle|^2\Big| + \Big|0.5 - |\langle p - 2\hbar k_L|\psi(t_f)\rangle|^2\Big| \\
&+ \Big||\langle p + 2\hbar k_L|\psi(t_f)\rangle|^2 - |\langle p - 2\hbar k_L|\psi(t_f)\rangle|^2\Big|\Big\rangle_{\mathcal{E}, P},
\end{aligned}
\tag{3.35}
$$

where $\langle \cdot \rangle_{\mathcal{E}, P}$ denotes averaging over sampled values of the polarization error $\epsilon_{pol} \in \mathcal{E}$ and initial momentum $p \in P$. The OCT beam-splitter efficiency was defined as $1 - Cost$. The first two terms penalize deviations from equal (50%) population transfer, while the third term penalizes asymmetry between the left and right diffraction processes.

For the case of $\epsilon_{pol} = 0$, the cost function is given by Eq. (3.35), but with only one value in $\mathcal{E}$. The results of the optimization are $(\Omega_R, \tau, t_0) = (2.079\omega_{rec}, 0.534\omega_{rec}^{-1}, 2.463\omega_{rec}^{-1})$, and the time-dependent detuning is shown in Fig. 3.20, based on which we obtain the results in Fig. 3.15.

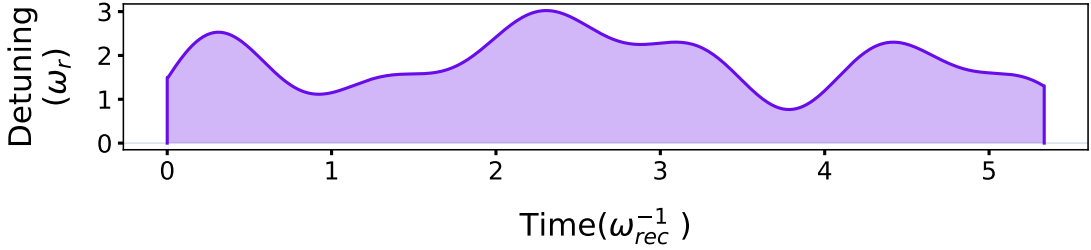

**Figure 3.20** Optimal time-dependent detuning obtained from OCT to mitigate Doppler detuning effects at $\epsilon_{pol} = 0$. The optimization minimizes losses and asymmetry for $p \in [-0.3\hbar k_L, 0.3\hbar k_L]$.

For the case where we have both polarization errors and an initial momentum $p \neq 0$, the results of the optimization are $(\Omega_R, \tau, t_0) = (1.646\omega_{rec}, 0.788\omega_{rec}^{-1}, 4.770\omega_{rec}^{-1})$, and the time-dependent detuning is shown in Fig. 3.21.

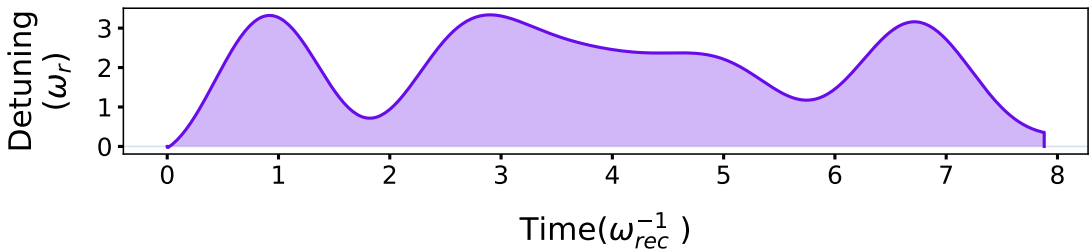

**Figure 3.21** Optimal time-dependent detuning for combined polarization errors and Doppler detuning. The optimization averages over $p \in [-0.3\hbar k_L, 0.3\hbar k_L]$ and $\epsilon_{pol} \in [0, 0.1]$.

For the last case where we have an incoming wave packet with a momentum width of $\sigma_p = 0.05\,\hbar k_L$ and polarization errors, the resulting Gaussian pulse parameters of the optimization are $(\Omega_R, \tau, t_0) = (1.264\,\omega_{rec}, 0.915\,\omega_{rec}^{-1}, 4.065\,\omega_{rec}^{-1})$, and the time-dependent detuning is shown in Fig. 3.22.

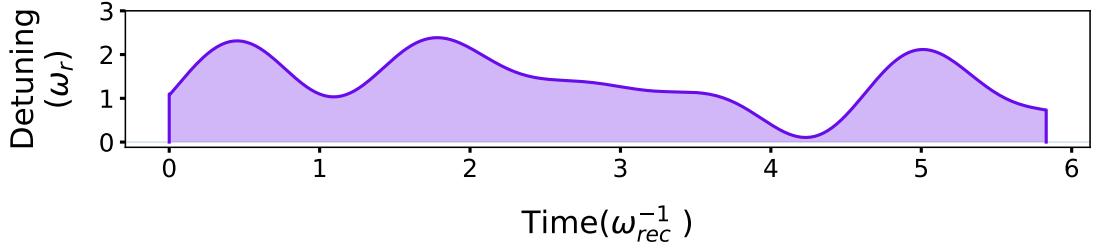

**Figure 3.22** Optimal time-dependent detuning for an incoming atomic ensemble with finite momentum width $\sigma_p = 0.05\,\hbar k_L$ in the presence of polarization errors.

### Notes on implementation

The optimization was performed using the Boulder Opal package by Q-CTRL [150]. Even though we used $(\Omega, \tau, t_0, \Delta(t))$ as our optimization variables, it is worth highlighting that we obtained similar results in the polarization-error-only mitigation if we choose the same pulse parameters as in Sec. 3.4.2, i.e., $(\Omega_R, \tau, t_0) = (2\omega_{rec}, 0.47\omega_{rec}^{-1}, 0)$, and only optimize the time-dependent detuning $\Delta(t)$. In the combined cases of polarization errors and Doppler detuning, allowing all four parameters to vary resulted in significantly improved robustness compared to detuning-only optimizations. For reference, in the case of $^{87}$Rb with wavelength $\lambda = 780.1\,\text{nm}$, the optimized Gaussian pulses correspond to durations on the order of a few hundred microseconds.

## 3.7 Summary and Outlook

In this chapter, we developed a theoretical framework for double Bragg diffraction under realistic experimental conditions, combining effective Hamiltonians with numerical optimization. We showed that for atoms at rest and in the weak-driving regime, a two-level description is sufficient to capture the essential dynamics, including the AC-Stark shift, and agrees well with exact numerical simulations even in the presence of moderate polarization errors.

To incorporate Doppler effects from finite momentum spread and general time-dependent detunings, we extended the analysis to a five-level description. This model accurately reproduces momentum selectivity, population losses, and asymmetry between left and right diffraction ports, and it can be straightforwardly generalized to higher-order or sequential DBD processes [156–160].

Based on these models, we investigated strategies to mitigate AC-Stark shifts, polarization errors, and Doppler detuning. Constant and linear detuning controls were shown to enhance robustness, but optimal control theory yielded the best results, designing time-dependent detuning functions that achieved efficiencies above 99.5% across realistic ranges of polarization errors and momentum spreads. These results highlight the combined power of analytical insight and AI-assisted optimization for implementing high-fidelity large-momentum-transfer pulses in next-generation quantum sensors.

In the next chapter, we extend these robust DBD tools to full Mach–Zehnder interferometer geometries and investigate their impact on fringe contrast in the presence of realistic noise sources.

# Chapter Four

# High-contrast Double Bragg Atom Interferometry

## 4.1 Motivation

Atom-interferometric sensors continue to push the frontiers of precision measurement, yet their full potential remains inhibited by losses and contrast degradation in large-momentum-transfer beam-splitter and mirror operations. In particular, the widely used technique of double Bragg diffraction offers a doubled interferometer scale factor and symmetric momentum-state splitting, but in practice suffers from imperfect pulse efficiency due to Doppler detuning, AC-Stark shifts and polarization imperfections. In Chapter 3, we derived an effective Hamiltonian for DBD in the quasi-Bragg regime and demonstrated that, by controlling the time-dependent detuning of the light pulses, the robustness of beam-splitter operations against polarization error and momentum spread can be dramatically improved. Building on this foundation, in this chapter we turn our attention to the full implementation of detuning-controlled DBD pulses in Mach–Zehnder atom interferometers under acceleration. We will present how tailored detuning profiles—linear sweeps or optimal-control derived shapes—mitigate parasitic transitions and restore high pulse fidelity, thereby enabling DBD to approach and eventually match the contrast and sensitivity of state-of-the-art Raman-based schemes. This enhanced robustness is essential for next-generation inertial sensors, space-borne platforms and tests of fundamental physics that employ long interrogation times, large momentum transfer and relaxed constraints on the atomic source temperature and momentum spread.

## 4.2 Background and Original Contributions

Atom interferometry (AI) has become a cornerstone of precision measurements, enabling the determination of inertial and fundamental physical quantities with unprecedented accuracy. Applications include atomic gravimetry [61–63, 161–164], gravity gradiometry [165–167], rotation and inertial sensing [74, 168–173], precision determinations of fundamental constants [66, 67, 174–176], and searches for new physics beyond the Standard Model [90–94]. A central challenge in pushing the performance of AI is to realize large momentum transfer (LMT) beam splitters and mirrors while maintaining high contrast.

Double Bragg diffraction (DBD) provides a promising route towards high-contrast LMT interferometry. Unlike single Bragg diffraction (SBD) or Raman transitions [55, 99, 177, 178], DBD couples atoms to symmetric momentum states $\pm 2\hbar k_L$ within a single internal ground-state manifold using two counter-propagating optical lattices with orthogonal polarizations [137]. This configuration offers two major advantages: (i) a doubled interferometric scale factor at the same diffraction order, and (ii) immunity to internal-state decoherence channels that limit Raman-based schemes [117, 118]. Moreover, DBD benefits from an intrinsic parity symmetry [179], which suppresses certain noise and systematic shifts, making it attractive for space missions [76, 77, 79, 80] and horizontal interferometers [111, 137]. However, the multi-level nature of Bragg transitions inevitably leads to parasitic diffraction channels and phase shifts [180–182], reducing the efficiency of beam-splitter (BS) and mirror (M) pulses and ultimately limiting interferometric contrast [111, 137, 172].

Interferometer fringe contrast (or visibility) is a key figure of merit. On the one hand, it determines the maximum useful interrogation time $T$ [183] and effective momentum transfer $k_{\text{eff}}$; once these limits are exceeded, contrast losses outweigh statistical gains, eventually washing out the interferometer fringes. On the other hand, the contrast directly impacts the practical acceleration sensitivity of a Mach–Zehnder interferometer, as discussed in Chap. 1.3.3:

$$\delta a \approx \frac{1}{\mathcal{C}\sqrt{N}k_{\text{eff}}T^2}. \tag{4.1}$$

where $N$ denotes the number of uncorrelated atoms detected and $\mathcal{C}$ the fringe contrast [100–103, 105]. Maintaining high contrast in atom interferometers is therefore a prerequisite for achieving high sensitivity and a critical requirement for the development of next-generation quantum sensors capable of reaching hundreds to thousands of $\hbar k_L$ of momentum separation.

In this chapter, we propose and analyze high-contrast Mach–Zehnder atom interferometers based on double Bragg diffraction operating under external acceleration, such as gravity. The key original contributions are:

- **Tri-frequency laser scheme with dynamic detuning control**: We introduce a laser configuration that compensates differential Doppler shifts and suppresses imperfections arising in conventional DBD.

- **Systematic evaluation of detuning-control protocols**: Four detuning-control interferometer strategies—conventional DBD (C-DBD), constant detuning (CD-DBD), linear detuning sweeps (DS-DBD), and a hybrid protocol combining detuning sweeps with optimal control theory (OCT)—are systematically investigated using exact numerical simulations and an effective five-level $S$-matrix model.

- **Demonstration of robustness and high contrast**: The OCT-based protocol achieves the highest robustness, maintaining contrast above 95% under realistic conditions, while the DS-DBD protocol sustains contrast above 90% for well-collimated Bose–Einstein condensates.

- **Practical pathways for quantum sensing applications**: These results establish experimentally viable methods for achieving high-contrast, large-momentum-transfer DBD interferometers with direct relevance to ground-based and space-borne inertial sensing and precision measurements.

In summary, this chapter provides both theoretical insights and practical control strategies for overcoming the main limitations of DBD-based interferometers. The tools developed here lay the foundation for the future robust and high-contrast inertial sensing and precision measurement based on double Bragg diffraction.

## 4.3 Double Bragg Accelerometers and Detuning-Control Strategies

Before turning to the theoretical framework, it is helpful to first visualize the basic configuration of a Mach–Zehnder interferometer based on double Bragg diffraction. Fig. 4.1 shows the laser geometry and the corresponding atomic density evolution for both conventional and optimized DBD schemes. This schematic highlights the emergence of parasitic paths and contrast loss in DBD interferometers—effects that form the central focus of the mitigation strategies developed in this chapter.

### 4.3.1 Tri-frequency double Bragg accelerometers

We now consider a Mach–Zehnder interferometer designed to measure a constant linear acceleration $\vec{g} = g\hat{z}$. In microgravity or weak-acceleration regimes, a standard dual-frequency DBD scheme suffices to drive symmetric momentum transfer in both directions [111, 137]. However, under strong acceleration—such as terrestrial gravity—the situation changes. The atoms experience a differential Doppler shift

$$\nu_g(t) = 2\,k_L g t, \tag{4.2}$$

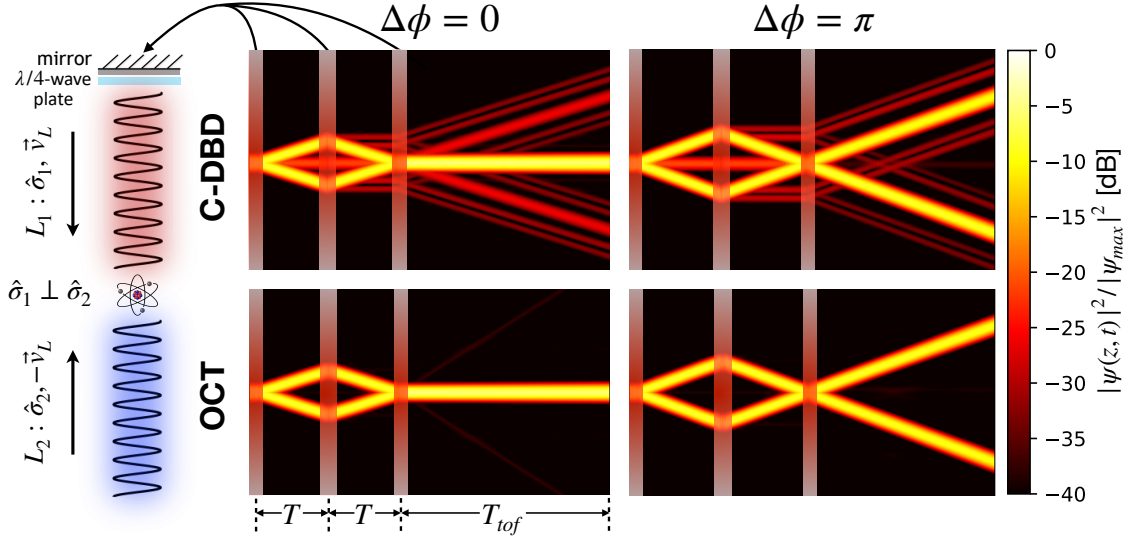

**Figure 4.1** Schematics of a double Bragg atom interferometer under microgravity. Left: Experimental setup of a DBD pulse using counter-propagating optical lattices $L_1$ and $L_2$ with orthogonal polarizations $\hat{\sigma}_1$ and $\hat{\sigma}_2$. Right: Real-space atomic density evolution $|\psi(z,t)|^2$, normalized to its initial maximum $|\psi_{max}|^2 = \max_z |\psi(z,0)|^2$ and shown in decibel units, for conventional (C-DBD) and optimized (OCT) Mach–Zehnder interferometers with phase shifts $\Delta\phi = 0$ (left column) and $\pi$ (right column), adjusted via the interrogation time $T$. Atomic densities are obtained from exact numerical simulations, with red-shaded regions indicating the three DBD pulses.

where $k_L$ is the laser wave number. This time-dependent shift breaks the resonance symmetry of the two counter-propagating Bragg processes, preventing simultaneous excitation of both upward and downward momentum transfers. As a consequence, pulse efficiency and interferometer contrast degrade rapidly.

To overcome this limitation, we propose a tri-frequency retro-reflective laser configuration, shown in Fig. 4.2(a). The idea, first demonstrated in double Raman gravimeters [184], is to replace one of the input frequencies (say $\omega_b$) with a pair $\omega_b \pm \nu_D$, where the detuning

$$\nu_D(t) = 2k_L a_L t \tag{4.3}$$

is dynamically tuned to compensate the Doppler shift $\nu_g(t)$ with a constant $a_L$. Physically, this introduces an effective lattice acceleration $a_L$ that can be tuned to match the atomic acceleration $g$. Due to momentum selectivity, the accelerating atoms couple only to four of

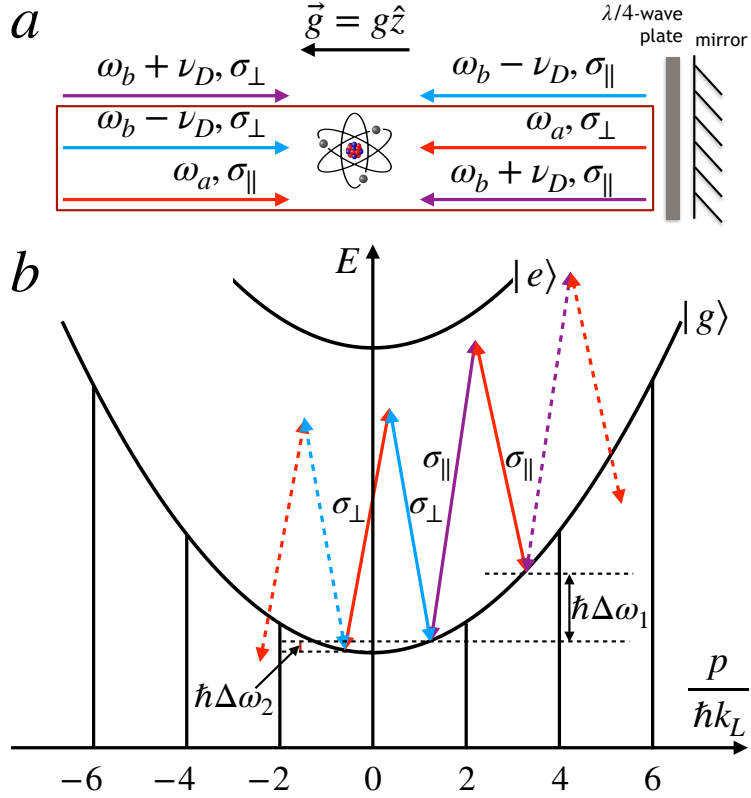

**Figure 4.2** (a) Tri-frequency laser configuration enabling double Bragg diffraction under constant acceleration $\vec{g} = g\hat{z}$. The red box highlights the four resonant beams. (b) Energy-level diagram showing upward and downward Bragg transitions driven by frequency differences $\Delta\omega_{1,2} = \omega_b - \omega_a \pm \nu_D$, with $\nu_D$ dynamically tuned to cancel the Doppler shift $\nu_g = 2k_L g t$.

the six available beams, forming two resonant Bragg lattices highlighted by the red box in Fig. 4.2(a).

**Laboratory-frame Hamiltonian.** The single-particle Hamiltonian describing tri-frequency DBD in the laboratory frame is

$$H_{lab}(t) = \frac{\hat{p}^2}{2m} + 2\hbar\Omega(t)\cos\left(2k_L\hat{z} - \int \nu_D(t)\,dt\right)\left\{\cos\left[\Delta\omega(t)t\right] + \varepsilon_{pol}\right\} - mg\hat{z}, \quad (4.4)$$

where $\Omega(t)$ is the time-dependent effective two-photon Rabi frequency, $\varepsilon_{pol} = |\boldsymbol{\sigma}_\perp^\dagger \boldsymbol{\sigma}_\parallel|$ quantifies polarization imperfections[1], and the last term accounts for the external acceleration. For an MZ sequence driven by three Gaussian pulses (see Fig. 4.1, right panel), $\Omega(t)$ takes

---

[1] For a conservative (worst-case) estimate of polarization-error effects, we neglect the temporal modulation associated with this term.

the form

$$\Omega(t) = \Omega_{BS} e^{-\frac{t^2}{2\tau_{BS}^2}} + \Omega_M e^{-\frac{(t-T)^2}{2\tau_M^2}} + \Omega_{BS} e^{-\frac{(t-2T)^2}{2\tau_{BS}^2}}, \tag{4.5}$$

with beam-splitter (BS) and mirror (M) peak amplitudes $(\Omega_{BS}, \Omega_M)$, widths $(\tau_{BS}, \tau_M)$, and centers at $(0, T, 2T)$.

**Transformation to the twin-lattice COM frame.** The time-dependent Doppler detuning $\nu_D(t) = 2k_L a_L t$ in Eq. (4.4) sets the center-of-mass (COM) frame of the twin Bragg lattices. It is therefore convenient to transform into this twin-lattice COM frame, where these fast oscillations are removed.

We perform this transformation using the time-dependent unitary operator

$$U(t) = \exp\left[i\frac{\hat{p}}{\hbar}\frac{1}{2}a_L t^2 - i\frac{\hat{z}}{\hbar}m a_L t + i\frac{\Phi(t)}{\hbar}\right], \tag{4.6}$$

where $\Phi(t)$ is a phase chosen such that $\dot{\Phi}(t) = m(\frac{1}{2}a_L - g)\frac{1}{2}a_L t^2$, and $a_L$ is the constant acceleration of the twin lattices. The transformed Hamiltonian is defined as

$$H_{COM}(t) = U H_{lab}(t) U^\dagger + i\hbar \dot{U} U^\dagger. \tag{4.7}$$

Since $U(t)$ contains both $\hat{z}$ and $\hat{p}$, we use the Baker–Campbell–Hausdorff (BCH) formula [185–187] and evaluate each term separately.

*Adjoint action on $H_{lab}(t)$.* Applying $U$ to each operator yields

$$U\hat{p}U^\dagger = \hat{p} + m a_L t, \tag{4.8}$$

$$U\hat{z}U^\dagger = \hat{z} + \tfrac{1}{2}a_L t^2. \tag{4.9}$$

Substituting into Eq. (4.4), we find

$$U H_{lab}(t) U^\dagger = \frac{1}{2m}(\hat{p} + m a_L t)^2 + 2\hbar\Omega(t) C(t, \varepsilon_{pol}) \cos(2k_L \hat{z}) - mg\left(\hat{z} + \tfrac{1}{2}a_L t^2\right), \tag{4.10}$$

where $C(t, \varepsilon_{pol}) \equiv \cos[\Delta\omega(t)t] + \varepsilon_{pol}$.

*Derivative term.* For the second contribution, we compute

$$i\hbar \dot{U} U^\dagger$$

$$=i\hbar \frac{d}{dt}\left(e^{i\frac{\hat{p}}{\hbar}\frac{1}{2}a_L t^2 - i\frac{\hat{z}}{\hbar}ma_L t + i\frac{\Phi(t)}{\hbar}}\right)U^\dagger \tag{4.11}$$

$$=i\hbar \frac{d}{dt}\left(e^{i\frac{\hat{p}}{\hbar}\frac{1}{2}a_L t^2} e^{-i\frac{\hat{z}}{\hbar}ma_L t} e^{i\frac{m}{4\hbar}a_L^2 t^3 + i\frac{\Phi(t)}{\hbar}}\right)U^\dagger \tag{4.12}$$

$$=-a_L t\hat{p} + e^{i\frac{\hat{p}}{\hbar}\frac{1}{2}a_L t^2}ma_L \hat{z}e^{-i\frac{\hat{z}}{\hbar}ma_L t}e^{i\frac{m}{4\hbar}a_L^2 t^3 + i\frac{\Phi(t)}{\hbar}}U^\dagger - \frac{3}{4}ma_L^2 t^2 - \dot{\Phi}(t) \tag{4.13}$$

$$=-a_L t\hat{p} + ma_L \hat{z} + ma_L\left[e^{i\frac{\hat{p}}{\hbar}\frac{1}{2}a_L t^2}, \hat{z}\right]e^{-i\frac{\hat{z}}{\hbar}ma_L t}e^{i\frac{m}{4\hbar}a_L^2 t^3 + i\frac{\Phi(t)}{\hbar}}U^\dagger - \frac{3}{4}ma_L^2 t^2 - \dot{\Phi}(t) \tag{4.14}$$

$$=-a_L t\hat{p} + ma_L \hat{z} - \frac{1}{2}ma_L^2 t^2 + \frac{1}{2}mga_L t^2, \tag{4.15}$$

where we have used the operator identity

$$e^{i\frac{\hat{p}}{\hbar}\frac{1}{2}a_L t^2 - i\frac{\hat{z}}{\hbar}ma_L t} = e^{i\frac{\hat{p}}{\hbar}\frac{1}{2}a_L t^2}e^{-i\frac{\hat{z}}{\hbar}ma_L t}e^{i\frac{m}{4\hbar}a_L^2 t^3}, \tag{4.16}$$

from line (4.11) to line (4.12), together with the BCH relation

$$\left[e^{i\frac{\hat{p}}{\hbar}\frac{1}{2}a_L t^2}, \hat{z}\right] = \frac{1}{2}a_L t^2 e^{i\frac{\hat{p}}{2\hbar}a_L t^2}, \tag{4.17}$$

from line (4.14) to line (4.15).

*Final COM-frame Hamiltonian.* Combining Eqs. (4.10) and (4.15), all cross-terms cancel as intended, leaving

$$H_{COM}(t) = \frac{\hat{p}^2}{2m} + 2\hbar\Omega(t)\cos[2k_L\hat{z}]\left\{\cos\left[\Delta\omega(t)t\right] + \varepsilon_{pol}\right\} - m(g - a_L)\hat{z}. \tag{4.18}$$

From Eq. (4.18), we observe that the lattice acceleration $a_L$ effectively reduces the linear potential term, yielding an effective acceleration $g_{\text{eff}} = g - a_L$. By choosing $a_L \simeq g$, the effective acceleration can be made arbitrarily small, such that atoms experience near-microgravity conditions during each light–matter interaction. This is the natural frame in which to apply the effective two-level or five-level double Bragg theory developed in Chap. 3. For the remainder of this chapter, we denote the effective acceleration $g_{eff}$ simply as $g$, and always work in the twin-lattice COM frame unless otherwise specified. If the effective acceleration $g$ is small enough such that

$$|mgz| \ll 2\hbar\Omega_R \quad \text{and} \quad |g| \ll \frac{\hbar k_L}{mt}, \tag{4.19}$$

for all relevant $t$ and $z$ during the interferometer sequence, the last linear potential term in Eq. (4.18) can be neglected during the pulses. This recovers the regime where the effective DBD theory discussed in Chap. 3 can be applied.

During the long free evolution intervals between successive pulses, where the light–atom coupling vanishes ($\Omega(t) \approx 0$), the residual linear acceleration term in Eq. (4.18) cannot be neglected. Its effect is to shift the momentum of the atomic wave packet while simultaneously imprinting a phase. Concretely, the unitary time evolution operator over a free evolution time $T$ acts on a momentum eigenstate $|p\rangle$ as

$$\begin{aligned}
\hat{U}(T)|p\rangle &= \exp\left[-\frac{iT}{\hbar}\left(\frac{\hat{p}^2}{2m} - mg\hat{z}\right)\right]|p\rangle \\
&= \exp\left[-\frac{i}{2m\hbar}\left(Tp^2 + mgT^2 p\right)\right]|p + mgT\rangle \\
&\equiv U(p)\,|p + mgT\rangle.
\end{aligned} \tag{4.20}$$

Here, the momentum state is shifted by $mgT$ due to the constant acceleration, while the prefactor encodes the associated propagation phase. In deriving Eq. (4.20), a global phase proportional to $T^3$ has been omitted, since it is common to all trajectories and thus irrelevant for interferometric phase shift [103].

Importantly, because the acquired phase depends only on the initial momentum $p$, all spatially parallel trajectories accumulate the same phase during the free evolution. Each subsequent DBD pulse then coherently splits and redirects the trajectories into different momentum ports, leading at the final detection port to a superposition of contributions from all intermediate paths. This momentum-dependent phase structure underlies the generic interferometric signal analyzed in the following sections.

### 4.3.2 Detuning-control strategies for high-contrast DBD interferometers

We now present four detuning-control strategies designed to maximize the efficiencies of double Bragg beam-splitter (BS) and mirror (M) pulses for Doppler-broadened atomic wave packets. The performance of each protocol is quantified in terms of integrated pulse efficiencies for realistic input states with finite momentum width. Below, we first introduce the efficiency metrics, and then evaluate the conventional DBD (C-DBD) and constant-detuning DBD (CD-DBD) schemes in detail. The numerical results for all protocols are summarized in Table 4.1 and compared in Fig. 4.11.

**Pulse efficiency definitions**

For a given quasi-momentum $p \in [-\hbar k_L, \hbar k_L]$ in the first Brillouin zone, the efficiency of a BS pulse is defined as

$$\mathcal{F}_{BS}(p) = P_{|p\rangle \to |p+2\hbar k_L\rangle} + P_{|p\rangle \to |p-2\hbar k_L\rangle}, \tag{4.21}$$

i.e., the probability that the atom is transferred into either of the two target ports $|p \pm 2\hbar k_L\rangle$. For the mirror pulse, we distinguish left and right efficiencies as

$$\mathcal{F}_M^+(p) = P_{|p+2\hbar k_L\rangle \to |p-2\hbar k_L\rangle}, \tag{4.22}$$

$$\mathcal{F}_M^-(p) = P_{|p-2\hbar k_L\rangle \to |p+2\hbar k_L\rangle}, \tag{4.23}$$

which are related by parity symmetry of the DBD Hamiltonian in the absence of gravity [179] as $\mathcal{F}_M^-(p) = \mathcal{F}_M^+(-p)$. Without loss of generality, we use $\mathcal{F}_M^+(p)$ as the mirror figure of merit.

For an initial wave packet $|\psi(0)\rangle = \int dp\, \psi(p)|p\rangle$ with compact support in the first Brillouin zone, the integrated BS and M efficiencies are defined as

$$\eta_{BS} = \int_{-\hbar k_L}^{\hbar k_L} dp\, |\psi(p)|^2 \mathcal{F}_{BS}(p), \tag{4.24}$$

$$\eta_M = \int_{-\hbar k_L}^{\hbar k_L} dp\, |\psi(p)|^2 \mathcal{F}_M^+(p). \tag{4.25}$$

These integrated efficiencies directly quantify the pulse performance for experimentally relevant input states with finite momentum width.

**Conventional DBD (C-DBD)**

As a baseline, we first consider the conventional DBD (C-DBD) sequence, in which both the beam splitter (BS) and mirror (M) pulses operate at the standard first-order resonance condition $\Delta\omega = 4\omega_{rec}$ (equivalently $\Delta(t) = 0$) [137–139, 179]. The pulse shapes are Gaussian, with widths individually optimized at fixed peak Rabi frequencies to balance two competing requirements: (i) maximizing the momentum acceptance window, and (ii) suppressing higher-order diffraction losses.

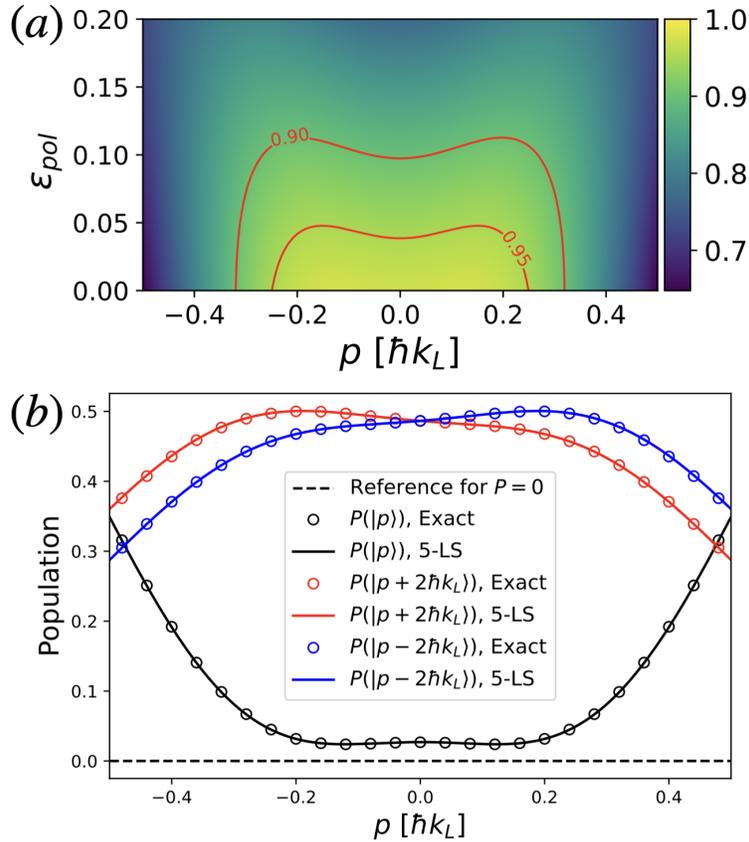

**Figure 4.3** Performance of the conventional DBD (C-DBD) beam-splitter pulse. (a) Efficiency landscape $\mathcal{F}_{BS}(p, \varepsilon_{pol})$ as a function of quasi-momentum and polarization error. (b) Transition probabilities into different momentum states for input $|p\rangle$ at $\varepsilon_{pol} = 0$. Solid lines: five-level theory (5-LS). Markers: exact simulations.

For the beam-splitter, we select $\Omega_{BS} = 2.0\,\omega_{\text{rec}}$ with an optimal width $\tau_{BS} = 0.47\,\omega_{\text{rec}}^{-1}$. For the mirror, the choice $\Omega_M = 2.89\,\omega_{\text{rec}}$ and $\tau_M = 0.64\,\omega_{\text{rec}}^{-1}$ yields the best compromise between selectivity and efficiency. The performance of these pulses is illustrated in Figs. 4.3 and 4.4.

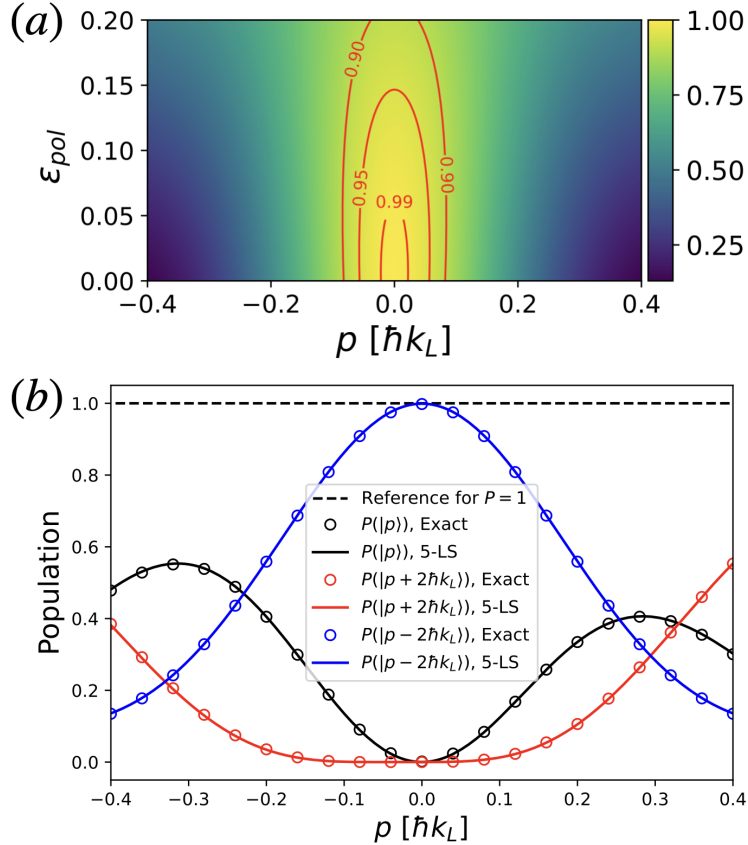

**Figure 4.4** Performance of the conventional DBD (C-DBD) mirror pulse. (a) Mirror efficiency landscape $\mathcal{F}_M(p, \varepsilon_{pol})$. (b) Transition probabilities for input $|p + 2\hbar k_L\rangle$ at $\varepsilon_{pol} = 0$. Comparison between 5-LS and exact simulations.

Figure 4.3(a) shows the 2D landscape of the BS efficiency $\mathcal{F}_{BS}(p, \varepsilon_{pol})$, while Fig. 4.3(b) compares transition probabilities predicted by the five-level model (5-LS) with exact position-space simulations for an input momentum eigenstate. Analogous plots for the mirror pulse are given in Fig. 4.4. For finite-width Gaussian inputs, the integrated efficiencies $\eta_{BS}$ and $\eta_M$ are summarized in Table 4.1. The full C-DBD interferometer consists of a three-pulse BS–M–BS sequence with interrogation time $T$ between pulses. It serves as a reference protocol against which all mitigation strategies are benchmarked in the next section.

**Constant-detuning DBD (CD-DBD)**

A straightforward improvement to C-DBD is to add a fixed detuning $\Delta$ to compensate for AC-Stark shifts and known polarization errors [179]. The BS pulse shares the same Gaussian envelope as in C-DBD but with $\Delta = 0.27\,\omega_{\text{rec}}$ optimized at $p = 0$. The mirror pulse parameters remain unchanged with $\Delta = 0$.

The efficiency landscape for the CD-DBD BS pulse is shown in Fig. 4.5. Compared to C-

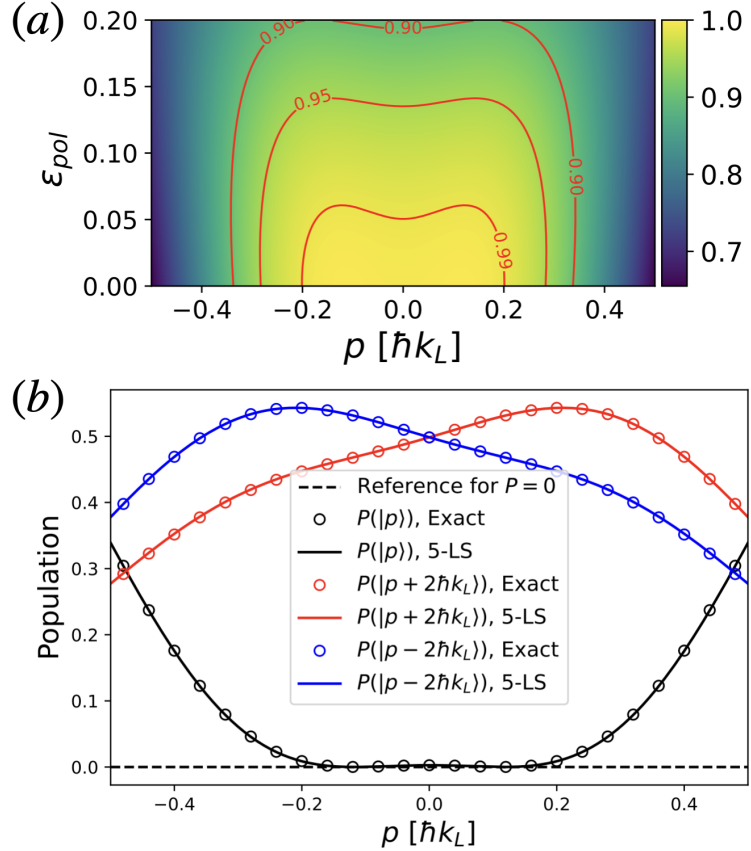

**Figure 4.5** Performance of the constant-detuning DBD (CD-DBD) beam-splitter pulse. (a) Efficiency landscape $\mathcal{F}_{BS}(p, \varepsilon_{pol})$. (b) Transition probabilities into different momentum states for input $|p\rangle$ with $\varepsilon_{pol} = 0$. Comparison between 5-LS and exact simulations.

DBD, the added detuning broadens the momentum acceptance window and increases robustness against parasitic couplings, as evident from the transition probabilities in Fig. 4.5(b). Quantitative efficiencies for Gaussian input states are listed in Table 4.1 and compared in Fig. 4.11.

**Linear-detuning-sweep DBD (DS-DBD)**

The third strategy improves upon C-DBD and CD-DBD by introducing a time-dependent linear detuning sweep, which dynamically compensates resonance shifts during the pulse and thus mitigates both Doppler and AC-Stark effects [179]. The detuning is parameterized as

$$\Delta(t)/\omega_{\text{rec}} = \frac{\alpha}{\tau_{BS|M}}(t - t_0) + \beta, \tag{4.26}$$

where $t_0$ denotes the pulse center and $(\alpha, \beta)$ are sweep parameters. Optimized values are found numerically as $(\alpha_{BS}, \beta_{BS}) = (0.37, 0.315)$ for the beam-splitter and $(\alpha_M, \beta_M) = (0.75, -4)$ for the mirror, while keeping the same Gaussian envelopes as in C-DBD.

This approach is inspired by the principle of *adiabatic passage* in two-level systems [147–149], a robust technique well established in nuclear magnetic resonance [188–190]. Extending this idea to the multi-level setting of DBD through a Magnus expansion in the quasi-Bragg regime [179], the dynamics reduce to an effective two-level system between the bare state

$$|0\rangle = |0\hbar k_L\rangle, \qquad |1\rangle = \frac{|2\hbar k_L\rangle + |-2\hbar k_L\rangle}{\sqrt{2}}, \tag{4.27}$$

as illustrated in Fig. 4.6. A linear sweep then mimics adiabatic transfer between $|0\rangle$ and $|1\rangle$, yielding robust population transfer even in the presence of finite momentum spread and polarization errors.

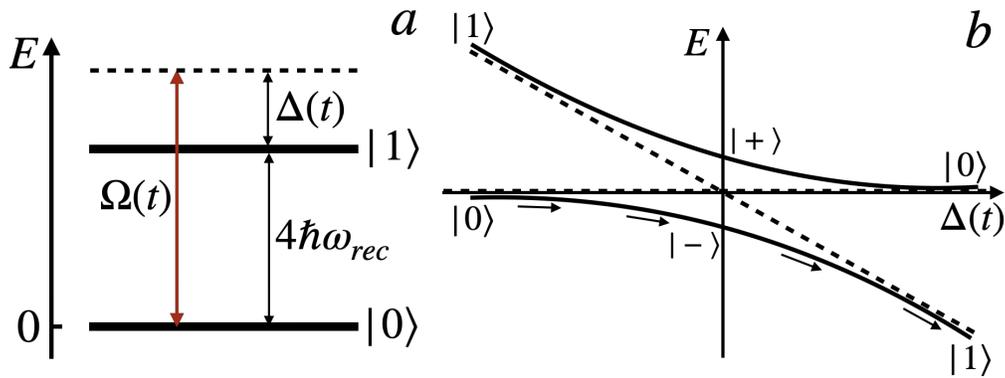

**Figure 4.6** (a) Effective two-level representation of first-order DBD dynamics between $|0\rangle$ and $|1\rangle$. (b) Illustration of a linear detuning sweep enabling robust adiabatic-like population transfer for a beam-splitter pulse.

The performance of the DS-DBD beam-splitter is summarized in Fig. 4.7, which shows its efficiency landscape $\mathcal{F}_{BS}(p, \varepsilon_{pol})$ and representative transition probabilities. For a Gaussian input with $\sigma_p = 0.05\,\hbar k_L$ centered at $p_0 = 0$, the DS-DBD beam-splitter reaches $\eta_{BS} = 99.94\%$, surpassing both C-DBD and CD-DBD and coming within one per mil of unity.

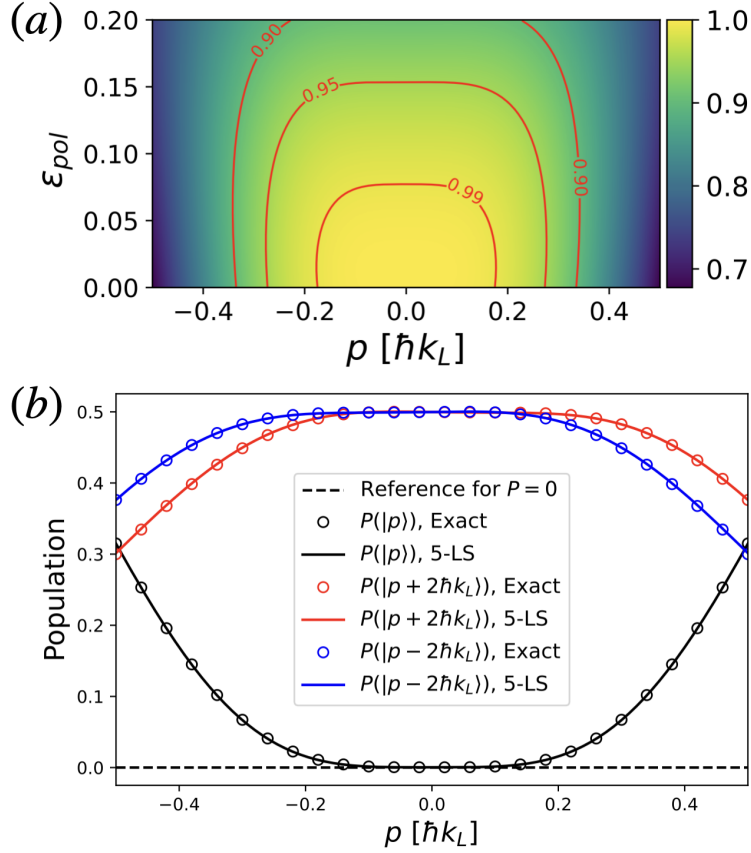

**Figure 4.7** Performance of the DS-DBD beam-splitter pulse. (a) Efficiency landscape $\mathcal{F}_{BS}(p, \varepsilon_{pol})$. (b) Transition probabilities after the DS-DBD BS pulse for an input state $|p\rangle$ with $\varepsilon_{pol} = 0$, compared with exact numerics.

The mirror pulse, in contrast, shows only modest improvement. Figure 4.8(a) presents its efficiency landscape $\mathcal{F}_M(p, \varepsilon_{pol})$, while panel (b) shows the corresponding transition probabilities. For a Gaussian input with $\sigma_p = 0.05\,\hbar k_L$ centered at $p_0 = 2\hbar k_L$, the efficiency is $\eta_M = 97.47\%$, reflecting its intrinsic sensitivity to finite momentum spread. Other monotonic detuning profiles (e.g., sigmoid) were tested but yielded no further gain. Thus, DS-DBD provides a nearly ideal beam-splitter, but its mirror remains the main performance bottleneck. In the next subsection, we address this limitation by applying optimal control theory to design mirror pulses that overcome the shortcomings of DS-DBD.

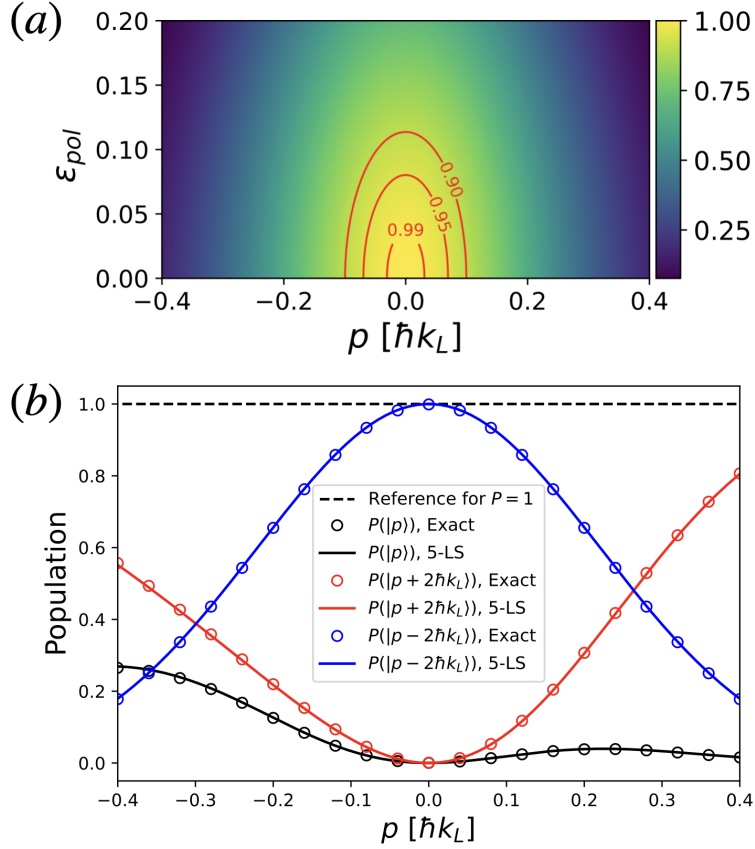

**Figure 4.8** Performance of the DS-DBD mirror pulse. (a) Efficiency landscape $\mathcal{F}_M(p, \varepsilon_{pol})$. (b) Transition probabilities after the DS-DBD mirror pulse for input $|p + 2\hbar k_L\rangle$ at $\varepsilon_{pol} = 0$, compared with exact numerics.

**Optimal control theory-assisted DBD (OCT)**

The final strategy addresses the mirror-pulse bottleneck by applying optimal control theory. Unlike the beam-splitter, the DBD mirror involves a four-photon process,

$$| \pm 2\hbar k_L\rangle \longrightarrow |0\hbar k_L\rangle \longrightarrow |\mp 2\hbar k_L\rangle, \tag{4.28}$$

which makes it particularly sensitive to finite momentum spread and center-of-mass (COM) offsets [172, 191–193].

Using QCTRL's *Boulder Opal* package [194], we simultaneously optimized the Gaussian envelope parameters $(\Omega_M, \tau_M, t_0)$ and a smooth time-dependent detuning $\Delta(t)$. The cost function was defined as an average over initial momenta $p \in [-0.2\hbar k_L, 0.2\hbar k_L]$, ensuring robust transfer for both input states $|\pm 2\hbar k_L\rangle$:

$$\text{Cost}_p = \left\langle \left| 1 - |\langle +2\hbar k_L + p|\psi_{-2}(t_f)\rangle|^2 \right| + \left| 1 - |\langle -2\hbar k_L + p|\psi_{+2}(t_f)\rangle|^2 \right| \right\rangle_P. \tag{4.29}$$

The optimized pulse parameters are

$$(\Omega_M, \tau_M, t_0) = (2.502\,\omega_{\text{rec}}, 1.829\,\omega_{\text{rec}}^{-1}, 3.879\,\omega_{\text{rec}}^{-1}), \tag{4.30}$$

together with the detuning profile shown in Fig. 4.9 (upper panel). The corresponding efficiency landscape $\mathcal{F}_M(p, \varepsilon_{pol})$ is presented in the lower panel, demonstrating robustness against both quasi-momentum spread and polarization errors. As shown in Fig. 4.10, the OCT mirror dynamics agree closely with exact numerical simulations. For a Gaussian input wave packet with $\sigma_p = 0.05\,\hbar k_L$ centered at $p_0 = 2\hbar k_L$, the efficiency reaches $\eta_M = 99.806\%$, a substantial improvement over the DS mirror (see Fig. 4.11). The resulting *hybrid OCT protocol* combines two DS beam-splitters with this OCT-optimized mirror.

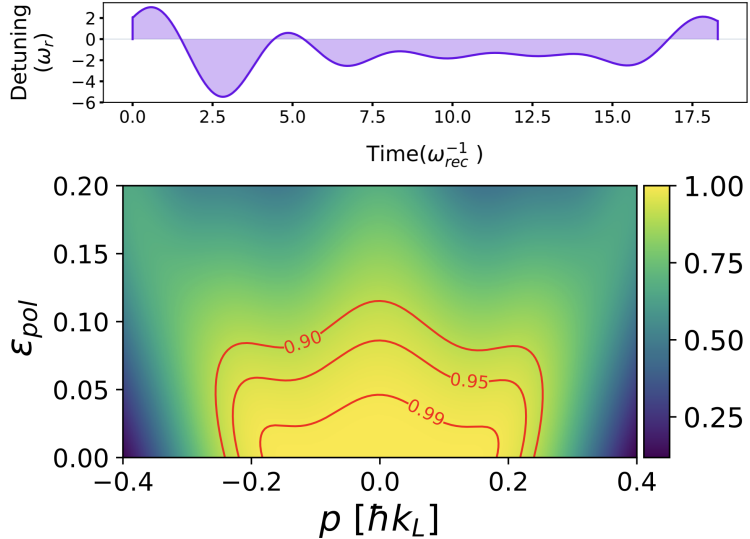

**Figure 4.9** OCT-optimized mirror pulse. Upper: smooth detuning profile $\Delta(t)$. Lower: efficiency landscape $\mathcal{F}_M(p, \varepsilon_{pol})$.

We also explored a fully OCT-optimized Mach–Zehnder sequence, where both BS and M pulses were designed via OCT. Under typical conditions (BEC sources, $\varepsilon_{pol} < 5\%$), no significant improvement was observed compared to the DS+OCT hybrid. A fully OCT-based design becomes advantageous only under harsher conditions, such as large polarization errors or poorly controlled momentum widths.

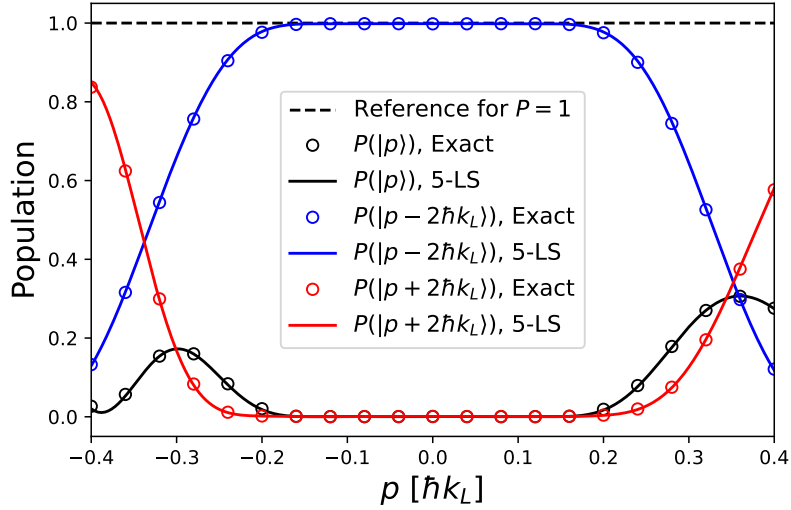

**Figure 4.10** Transition probabilities after the OCT mirror pulse for input $|p+2\hbar k_L\rangle$ with $\varepsilon_{pol} = 0$, compared with exact numerics.

### Summary of BS and M-pulse efficiencies

Table 4.1 and Fig. 4.11 together summarize the performance of the four detuning-control strategies in terms of integrated beam-splitter ($\eta_{BS}$) and mirror ($\eta_M$) efficiencies. All values are obtained for Gaussian input wave packets with momentum width $\sigma_p = 0.05\,\hbar k_L$, centered at $p_0 = 0$ for BS and $p_0 = 2\,\hbar k_L$ for M.

**Table 4.1** Integrated beam-splitter ($\eta_{BS}$) and mirror ($\eta_M$) efficiencies for four detuning-control strategies, evaluated for Gaussian inputs with $\sigma_p = 0.05\,\hbar k_L$.

| Protocol | $\eta_{BS}$ (%) | $\eta_M$ (%) |
|---|---|---|
| C-DBD | 97.35 | 96.43 |
| CD-DBD | 99.76 | 96.43 |
| DS-DBD | 99.94 | 97.47 |
| Hybrid OCT (DS+OCT-M) | 99.94 | 99.81 |

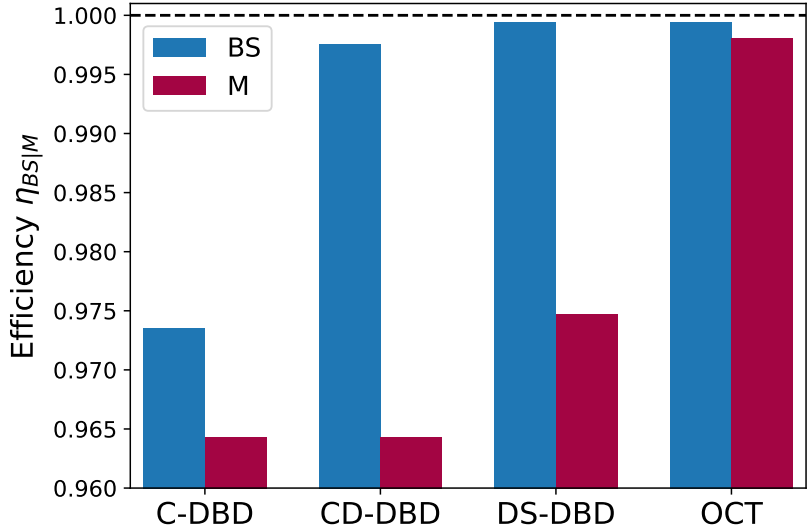

**Figure 4.11** Comparison of beam-splitter (BS) and mirror (M) efficiencies for four protocols: conventional DBD (C-DBD), constant-detuning DBD (CD-DBD), linear detuning-sweep DBD (DS-DBD), and a hybrid strategy combining a detuning-sweep beam splitter with an OCT-optimized mirror (OCT). All efficiencies are calculated for a Gaussian input wave packet with $\sigma_p = 0.05\,\hbar k_L$, centered at $p_0 = 0$ for BS and $p_0 = 2\,\hbar k_L$ for M.

Both the table and the figure reveal a clear progression. Conventional DBD (C-DBD) yields only moderate efficiencies for both BS and M pulses. Introducing constant detuning (CD-DBD) significantly improves the beam-splitter but leaves the mirror unchanged. Linear detuning sweeps (DS-DBD) provide an almost ideal beam-splitter but only modest gains for the mirror. Finally, the hybrid OCT-assisted scheme, which combines a DS beam-splitter with an OCT-optimized mirror, achieves near-unity efficiencies simultaneously for both operations.

This progression highlights the complementary roles of detuning sweeps and OCT: the former delivers near-perfect beam-splitters, while the latter is essential for robust mirrors. Taken together, the hybrid DS+OCT-M protocol emerges as the most effective and experimentally viable strategy for realizing high-contrast DBD interferometers.

In the next section, we evaluate how these pulse-level improvements translate into full Mach–Zehnder interferometer contrast under realistic conditions.

## 4.4   Contrast of DBD Mach–Zehnder Interferometers

In a Mach-Zehnder atom interferometer, one of the key experimental limitations is the fringe contrast. In this section we analyze the contrast achievable with double Bragg diffraction beam-splitters and mirrors, using the detuning-control strategies developed above. The goal is to quantify how experimental imperfections—finite momentum width, center-of-mass (COM) offsets, polarization errors, and lattice-depth fluctuations—affect the interferometer output, and how optimized control can mitigate these effects.

### 4.4.1   Momentum-space description of the DBD Mach–Zehnder interferometer

In spatially unresolved interferometers—such as those with short interrogation times $T$ or long time-of-flight durations $T_{\text{tof}}$ that blur the separation between atomic trajectories—the output ports correspond directly to distinct momentum states. In the far-field limit $T_{\text{tof}} \gg T$, this correspondence becomes exact. The full Mach–Zehnder interferometer can therefore be described entirely in momentum space, with each double Bragg pulse represented by its scattering matrix ($S$-matrix) and the free evolution between pulses encoded by diagonal propagators.

The total scattering matrix of a three-pulse DBD interferometer is given by the time-ordered product

$$S^{\text{tot}} = S^{BS}\, U(2T, T)\, S^{M}\, U(T, 0)\, S^{BS}, \tag{4.31}$$

where $S^{BS}$ and $S^{M}$ represent the beam-splitter and mirror pulse $S$-matrices, respectively, and $U(t_2, t_1)$ is a free-propagation operator encoding the propagation phases of each momentum component accumulated between two adjacent pulses.

Since each DBD pulse couples multiple bare momentum states, we work in a five-level basis

$$\mathcal{B} = \{|p\rangle,\ |p + 2\hbar k_L\rangle,\ |p - 2\hbar k_L\rangle,\ |p + 4\hbar k_L\rangle,\ |p - 4\hbar k_L\rangle\},$$

which is sufficient to capture the relevant dynamics in the quasi-Bragg regime [179]. In this basis, the free-propagation operator is diagonal:

$$U(t_2, t_1) = \mathrm{diag}\left[U(p), U(p + 2\hbar k_L), U(p - 2\hbar k_L), U(p + 4\hbar k_L), U(p - 4\hbar k_L)\right], \quad (4.32)$$

where $U(p)$ is given by Eq. (4.20). The explicit form of a general element of the full interferometer $S$-matrix is then

$$S_{ij}^{\mathrm{tot}}(g, p, T) = \sum_{k,l=1}^{5} B_{il}(p_3)\, U_{ll}(p_2)\, M_{lk}(p_2)\, U_{kk}(p_1)\, B_{kj}(p_1), \quad (4.33)$$

where $p_1 = p$, $p_2 = p + mgT$, and $p_3 = p + 2mgT$. Here $B_{ij}(p)$ and $M_{ij}(p)$ denote matrix elements of the beam-splitter and mirror operations.

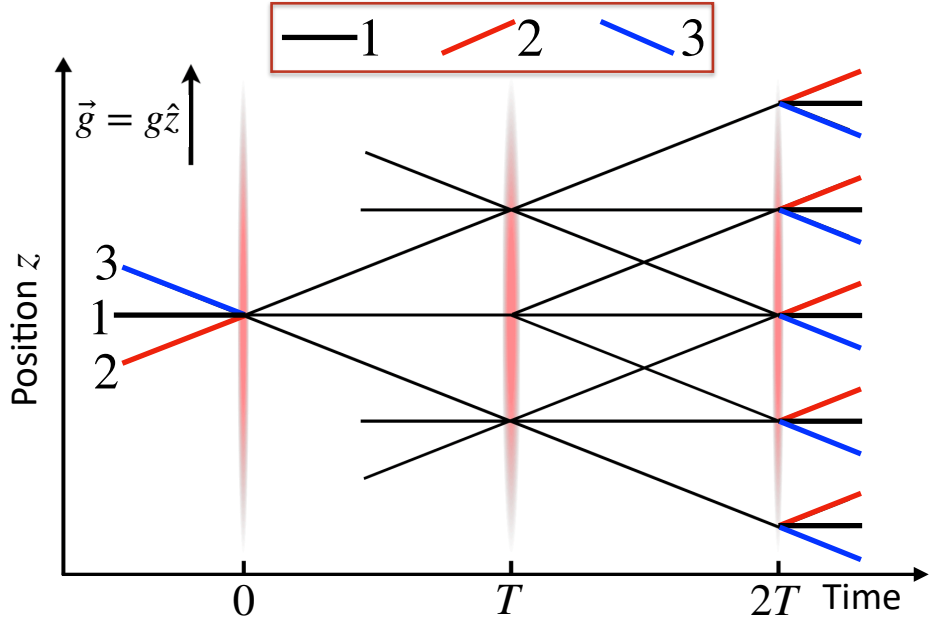

**Figure 4.12** Mach–Zehnder atom interferometer in momentum space implemented with three DBD pulses. Input and output ports are labeled by $i = 1, 2, 3$, corresponding to the momentum states $|p\rangle$, $|p + 2\hbar k_L\rangle$, and $|p - 2\hbar k_L\rangle$, respectively. Higher-order states (e.g. $|\pm 4\hbar k_L\rangle$) are omitted for clarity but are included in the theoretical framework.

Figure 4.12 illustrates the momentum-space topology of the DBD Mach–Zehnder interferometer. The five-level model accurately describes diffraction into higher orders and parasitic paths, while the three lowest momentum states—$|p\rangle$, $|p+2\hbar k_L\rangle$, and $|p-2\hbar k_L\rangle$—correspond to the three primary output ports (black, red, and blue in Fig. 4.12).

After the final pulse, the effective acceleration shifts the output momentum basis to

$$\{|p_3\rangle,\ |p_3 + 2\hbar k_L\rangle,\ |p_3 - 2\hbar k_L\rangle,\ |p_3 + 4\hbar k_L\rangle,\ |p_3 - 4\hbar k_L\rangle\}.$$

For an initial state of the form $|\psi(t = 0)\rangle = \int dp\,\psi(p)|p\rangle$ with $|\psi(p)|^2$ being a normalized momentum distribution with compact support in the first Brillouin zone $[-\hbar k_L,\ \hbar k_L]$, such as a Gaussian $\mathcal{N}(p_0, \sigma_p^2)$ with $p_0,\ \sigma_p \ll \hbar k_L$, the probability amplitudes in the three main output ports are

$$\phi_i(p) = \psi(p)\,S_{i1}^{\text{tot}}(g, p, T), \qquad i = 1, 2, 3, \tag{4.34}$$

and the integrated population in each port is

$$P_i(g, T) = \int_{-\hbar k_L}^{\hbar k_L} \bigl|\psi(p)\,S_{i1}^{\text{tot}}(g, p, T)\bigr|^2 dp. \tag{4.35}$$

Here, $P_1$ corresponds to the central port ($0\hbar k_L$ output port), while $P_2$ and $P_3$ correspond to the $\pm 2\hbar k_L$ output ports. In DBD interferometers, the two symmetric outputs are typically summed to yield a single measurable signal,

$$P_{\pm 2\hbar k}(g, T) = P_2(g, T) + P_3(g, T), \tag{4.36}$$

which is conjugate to the central-port population $P_1(g, T)$. Under ideal beam-splitter and mirror conditions, this summed signal takes the form of a single sinusoidal fringe [156, 171, 195]:

$$P_{\pm 2\hbar k}^{\text{ideal}}(g, T) = \frac{A - \mathcal{C}\cos(4k_L g T^2)}{2}, \tag{4.37}$$

where $A = 1$ and $\mathcal{C} = 1$ represent the ideal offset and contrast, respectively.

## 4.4.2  Definition and extraction of contrast

In realistic conditions, imperfect pulses, parasitic paths, finite momentum width, or polarization errors reduce the visibility of the fringe. The resulting signal $P_{\pm 2\hbar k}(g, T)$ generally contains additional Fourier components [181], but its dominant oscillation remains at frequency $4k_L g$ (when plotted against $T^2$). The contrast, denoted by **C**, is operationally defined as the population difference between the first non-trivial maximum and minimum of this signal during a $T$-scan:

$$\mathbf{C} \equiv P_{\pm 2\hbar k}(g, T_{\max}) - P_{\pm 2\hbar k}(g, T_{\min}), \tag{4.38}$$

where $T_{\max}$ and $T_{\min}$ are the first non-trivial extrema of $P_{\pm 2\hbar k}(g, T)$ at a fixed $g$.

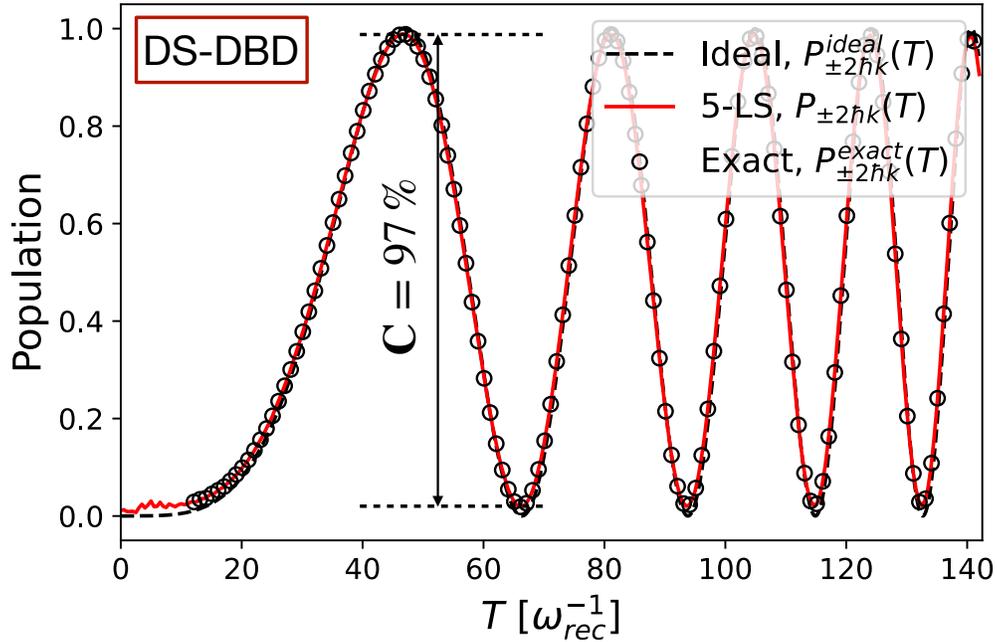

**Figure 4.13** $T$-scan fringe in the $|\pm 2\hbar k_L\rangle$ output ports using the DS-DBD strategy. A contrast of 97% is achieved under $g = 0.000357 k_L^{-1}\omega_{\mathrm{rec}}^2$ for $\sigma_p = 0.05\hbar k_L$ and $p_0 = 0$. The black dashed line represents the ideal sinusoidal fringe from Eq. (4.37).

Figure 4.13 illustrates an example of this procedure for the DS-DBD strategy. The high-contrast sinusoidal modulation of the $\pm 2\hbar k_L$ ports (solid red curve and black circles) reflects the coherent recombination of the two interferometer arms, while small deviations from ideal sinusoidal oscillations (dashed black curve) arise from non-ideal pulse transfer and higher-order momentum coupling.

This operational definition of contrast provides a consistent metric for comparing different detuning-control strategies and assessing their robustness against experimental imperfections in the subsequent sections.

### 4.4.3 Performance of detuning-control strategies

We now compare the contrast robustness of the four detuning-control strategies introduced earlier—C-DBD, CD-DBD, DS-DBD, and the hybrid OCT-assisted protocol—under realistic experimental conditions. The analysis focuses on three principal sources of contrast degradation: the atomic momentum width $\sigma_p$, the initial center-of-mass (COM) momentum $p_0$, and polarization errors $\varepsilon_{\mathrm{pol}}$. Before examining the systematic parameter dependence, it is instructive to visualize how a finite COM offset directly affects the interference signal. Fig. 4.14 shows the $T$-scan fringe for the CD-DBD protocol when a deliberate ini-tial momentum offset of $p_0 = 0.1\,\hbar k_L$ is introduced. Even for a relatively narrow momentum distribution ($\sigma_p = 0.01\,\hbar k_L$), the fringe visibility drops markedly—from above 95% to $\mathbf{C} = 83\%$—demonstrating the sensitivity of conventional DBD sequences to residual Doppler detuning. This example highlights why precise control of COM momentum and improved detuning compensation are essential for maintaining high contrast in realistic interferometers.

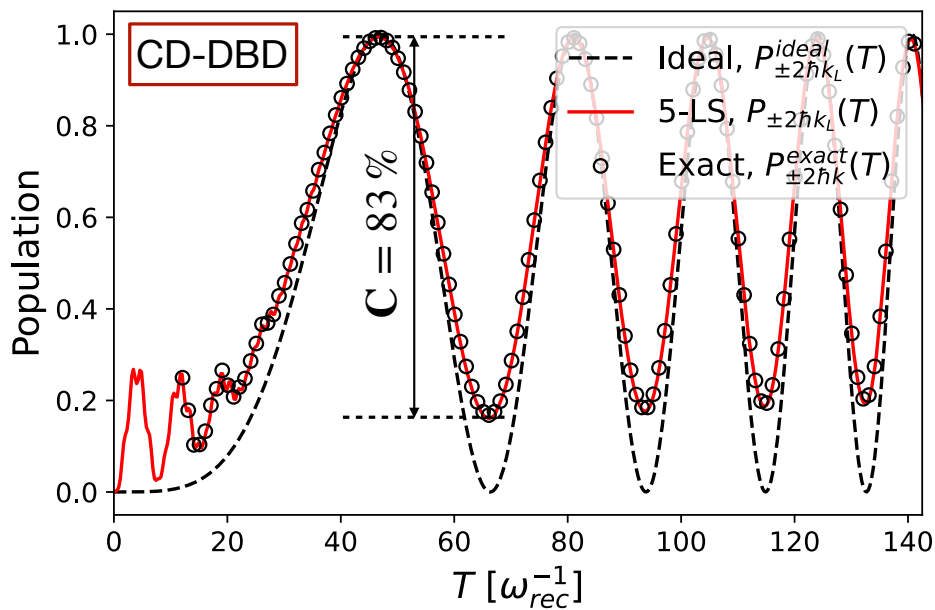

**Figure 4.14** Contrast degradation in the CD-DBD protocol due to an initial COM momentum of $p_0 = 0.1\,\hbar k_L$ and momentum width $\sigma_p = 0.01\,\hbar k_L$. The extracted contrast is $\mathbf{C} = 83\%$.

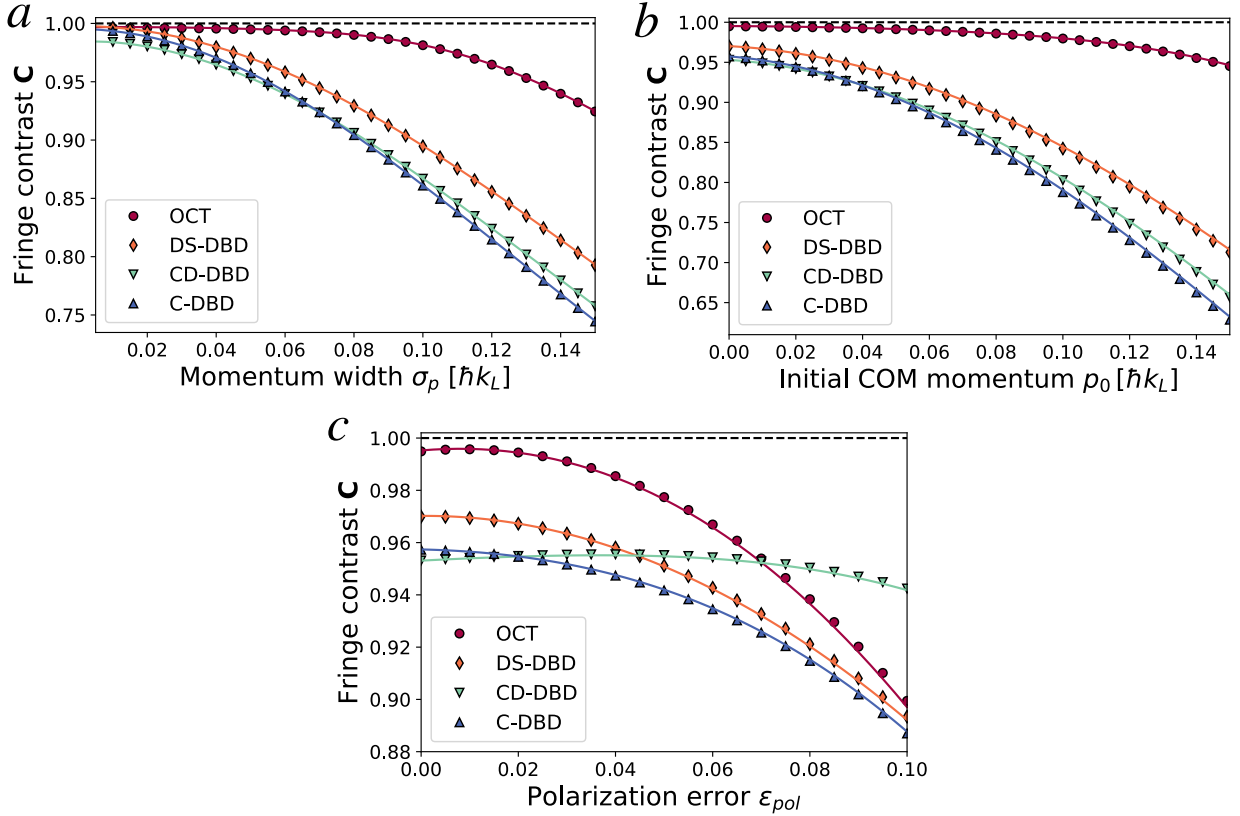

**Figure 4.15** Contrast of double-Bragg Mach–Zehnder interferometers for different detuning-control strategies as a function of (a) atomic momentum width $\sigma_p$, (b) COM momentum $p_0$, and (c) polarization error $\varepsilon_{\text{pol}}$. Symbols: exact numerical simulations; solid lines: five-level $S$-matrix predictions. The DS-DBD and OCT protocols show superior robustness across all parameters.

Building on this qualitative picture, Figs. 4.15(a-c) quantitatively compare the contrast obtained for each detuning-control strategy as a function of (a) momentum width $\sigma_p$, (b) COM momentum $p_0$, and (c) polarization error $\varepsilon_{\text{pol}}$. Exact numerical results (symbols) are shown alongside predictions of the five-level $S$-matrix model (solid lines), exhibiting excellent agreement across all tested regimes.

Under typical experimental conditions with well-controlled polarization errors ($\varepsilon_{\text{pol}} <$ 3%), the relative ranking of performance is

$$\text{OCT} > \text{DS-DBD} > \text{C-DBD} \approx \text{CD-DBD}.$$

The degree of improvement depends on the effective temperature (which sets $\sigma_p$), the COM momentum offset, and polarization balance. For example, at a 1D effective temperature of $2\,\text{nK}$ (corresponding to $\sigma_p \simeq 0.10\,\hbar k_L$ for $^{87}\text{Rb}$), the DS-DBD and OCT schemes achieve contrast improvements of 3.4% and 12.0%, respectively, over the conventional DBD protocol. For colder, better-collimated sources ($\sigma_p \lesssim 0.05\,\hbar k_L$), these relative gains decrease to 1–4%, while for delta-kick-collimated BECs ($\sigma_p \simeq 0.014\,\hbar k_L$), the contrast remains above 95% for OCT and above 90% for DS-DBD. Figures 4.15(a) summarizes these results.

Empirically, the DS-DBD strategy maintains contrast above 90% for momentum widths up to $\sigma_p \leq 0.097\,\hbar k_L$, while the OCT protocol retains above 95% contrast up to $\sigma_p \leq 0.132\,\hbar k_L$. These results establish the hybrid OCT-assisted approach as the most robust and experimentally promising protocol for high-contrast double-Bragg interferometers.

It is important to emphasize that our five-level S-matrix formalism is not limited to spatially unresolved detection. The same framework also encompasses spatially resolved interferometry, where different output trajectories corresponding to the same momentum port are individually distinguishable due to the very large spatial separation. In this case, the full interferometer S-matrix is obtained by restricting the summation indices in Eq. (4.33) to the physically relevant trajectory pairs,

$$S_{ij}^{\text{tot}}(g, p, T) = \sum_{(k,l) \in \mathcal{R}} B_{il}(p_3)\, U_{ll}(p_2)\, M_{lk}(p_2)\, U_{kk}(p_1)\, B_{kj}(p_1), \tag{4.39}$$

where the index set $\mathcal{R} = \{(1,1),\ (2,3),\ (3,2),\ (4,5),\ (5,4)\}$ selectively includes only those momentum-reversal paths that close in space at the final beam-splitter pulse. We have also applied this spatially resolved S-matrix to scenarios relevant to long-baseline interferometers—such as the very-long-baseline atom interferometer (VLBAI) [196]—where the output ports are physically separable in space and can be independently imaged (details can be found in App. B). The resulting contrast trends and robustness properties of the four detuning-control strategies remain qualitatively the same as in the spatially unresolved case, with only minor quantitative differences. This confirms that the conclusions of this chapter are broadly applicable across both detection regimes.

### 4.4.4 Robustness against lattice-depth fluctuations

In addition to atomic and optical imperfections, fluctuations in the peak laser intensity—or equivalently, variations in the peak optical lattice depth—represent a significant practical limitation for achieving high-contrast atom interferometry. Since the effective two-photon Rabi frequency

$$\Omega(t) = \frac{\Omega_a(t)\,\Omega_b(t)}{2\Delta_L} = \frac{\Gamma^2}{2\Delta_L}\,\frac{\sqrt{I_a(t)\,I_b(t)}}{I_{\text{sat}}}, \tag{4.40}$$

where $\Gamma$ is the natural linewidth of the atomic transition and $I_{\text{sat}}$ the saturation intensity, scales with the square root of the two laser intensities $I_a(t)$ and $I_b(t)$, any fluctuation in the laser power $P(t) = I(t)A$, with $A$ denoting the cross-sectional area of the beam, directly alters the instantaneous Rabi frequency. Such variations modify the effective pulse area, introducing systematic errors in the population-transfer efficiency and phase accumulation, which ultimately reduce the interferometric contrast. This effect is particularly pronounced in multi-pulse sequences such as DBD Mach–Zehnder interferometers, where accumulated phase errors compound over successive light–matter interactions.

To quantify this effect, we performed exact numerical simulations of the full Mach–Zehnder interferometer under Gaussian-distributed fluctuations of the peak Rabi frequency for both BS and M pulses,

$$\Omega_{BS|M} \sim \mathcal{N}(\Omega_R,\,\sigma_R^2),$$

where $\Omega_R$ denotes the individually optimized lattice depth and $\sigma_R = \Delta\Omega_R/\Omega_R$ represents the relative standard deviation. For each value of $\sigma_R$, the contrast was averaged over ten independent realizations of the pulse sequence. All simulations were carried out for a Gaussian momentum distribution with $\sigma_p = 0.05\,\hbar k_L$ and zero center-of-mass momentum ($p_0 = 0$), ensuring that contrast variations arise purely from lattice-depth noise.

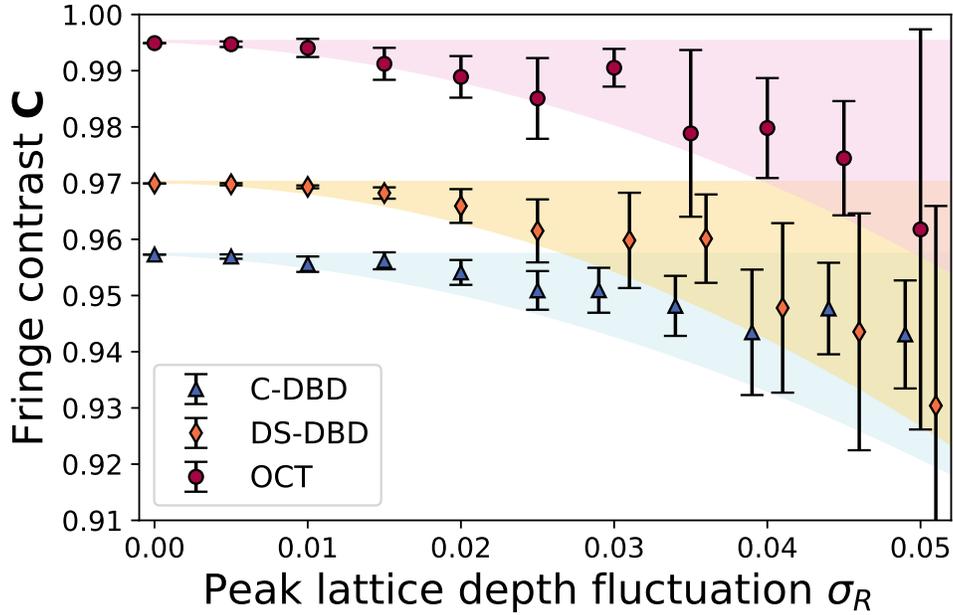

**Figure 4.16** Contrast of DBD Mach–Zehnder interferometers as a function of relative peak lattice-depth fluctuations $\sigma_R$ for different detuning-control strategies. Symbols with error bars denote exact numerical simulations; shaded bands indicate the confidence intervals predicted by the five-level $S$-matrix model.

The results, summarized in Fig. 4.16, show that all protocols are first-order insensitive to small power variations around their respective optimal values. However, beyond a protocol-dependent threshold, the contrast begins to degrade significantly. The five-level $S$-matrix model reproduces the numerical results well, with theoretical confidence intervals (shaded regions) bounded by $\pm 1.05\sigma_R$ around the mean response.

Quantitatively, the DS-DBD strategy starts from a high initial contrast of 97% and maintains above 95% contrast up to relative lattice-depth fluctuations of $\sigma_R \approx 3\%$. In comparison, the conventional C-DBD protocol tolerates fluctuations only up to about $\sigma_R \approx 2.5\%$ for the same contrast threshold. The hybrid OCT-assisted strategy demonstrates the highest resilience, sustaining more than 95% contrast for fluctuations as large as $\sigma_R \approx 4.5\%$. The constant-detuning scheme (CD-DBD) is considerably more sensitive to power noise and is omitted from Fig. 4.16 for clarity.

These findings underscore that optimal control provides not only higher peak contrast but also greater tolerance to laser-power instabilities, an important consideration for long-baseline and space-based interferometers where amplitude noise is difficult to suppress.

## 4.5 Summary and Outlook

In this chapter, we have proposed and analyzed high-contrast Mach–Zehnder atom interferometers based on double Bragg diffraction operating under external acceleration. We introduced a tri-frequency laser configuration that compensates differential Doppler shifts and enables symmetric and high-efficiency momentum transfer even in the presence of strong acceleration, such as gravity. Building on this foundation, we developed and compared four detuning-control strategies—conventional DBD (C-DBD), constant detuning (CD-DBD), linear detuning sweep (DS-DBD), and an optimal-control-assisted protocol (OCT)—designed to mitigate contrast loss arising from experimental imperfections, including momentum spread, center-of-mass motion, polarization errors, and laser-intensity fluctuations.

Using both exact numerical simulations and a five-level $S$-matrix framework, we quantified the robustness of these protocols and identified the dominant mechanisms responsible for contrast degradation. Our results show that appropriate detuning control can elevate the performance of DBD interferometers to a level comparable with, and in some aspects surpassing, that of traditional Raman-based schemes [105, 119, 171, 197, 198]. In particular, the OCT-assisted protocol simultaneously compensates Doppler, AC-Stark, and polarization-induced effects, achieving near-unity beam-splitter and mirror efficiencies and sustaining interferometric contrast above 95% under realistic conditions, including moderate laser power fluctuations.

By overcoming long-standing limitations associated with DBD pulse imperfections, these results establish experimentally viable pathways toward high-contrast, large-momentum-transfer atom interferometers. The detuning-controlled and OCT-optimized techniques developed here bridge the gap between DBD and Raman protocols, opening the door to next-generation quantum sensors for precision inertial measurements and fundamental physics tests—ranging from terrestrial gravimeters and gyroscopes to space-borne interferometers probing gravity, general relativity, and dark-sector physics.

In the next chapter, we extend these non-interacting 1D DBD schemes to realistic three-dimensional interacting Bose–Einstein-condensate interferometers by numerical simulations. There, we investigate how spatial inhomogeneity, mean-field interactions, and trap geometries influence the double Bragg diffraction performance.

# Chapter Five

# Numerical Simulations of Guided Atom Interferometers

## 5.1 Motivation

THE previous chapter established high-contrast double Bragg diffraction (DBD) protocols and detuning-control strategies capable of achieving near-unity beam-splitter and mirror efficiencies. While those results were derived within a quasi-one-dimensional (1D) momentum-space framework, real-world atom interferometers—especially those employing Bose–Einstein condensates (BECs) guided in optical dipole traps—exhibit inherently three-dimensional (3D) dynamics. Finite transverse confinement, mean-field interactions, and optical imperfections such as residual polarization coupling or laser-intensity gradients can all alter interferometric contrast.

To capture these realistic effects, it is necessary to go beyond effective two-level or few-level models and solve the full 3D time-dependent Gross–Pitaevskii equation (GPE) under experimentally relevant conditions. This chapter develops such a simulation framework to model guided BEC interferometers driven by double Bragg diffraction with realistic optical field geometries and atomic interactions. Through this approach, we aim to provide both a predictive and diagnostic tool for analyzing current and future interferometer designs.

## 5.2   Background and Original Contributions

Simulations of BEC interferometers have a long history, ranging from 1D Raman-type models to mean-field studies of Bragg diffraction and guided propagation. However, double Bragg diffraction poses additional challenges: its inherently symmetric twin-lattice coupling makes it more sensitive to spatial mode distortions and polarization imperfections, while 3D propagation in optical guides introduces coupling between longitudinal and transverse degrees of freedom. Previous theoretical work has not fully incorporated these effects in a unified numerical framework.

The original contributions of this chapter are as follows:

- **Modeling of realistic optical potentials.** The simulation includes the full Gaussian intensity profiles of both the guiding and double-Bragg pulse beams, with spatially varying beam waist, optical power, and frequency detuning. This enables a quantitative description of dipole forces, spatial confinement, and residual standing-wave effects.

- **Quantitative study of polarization errors.** We demonstrate how imperfect polarization alignment between counter-propagating lattices introduces a residual standing-wave potential that affects both the guided expansion and DBD efficiency. Simulations reproduce the reduced axial expansion observed in the QUANTUS-1 experiment and quantify the equivalent polarization error.

- **Full 3D simulation of a Mach–Zehnder sequence.** We simulate a complete $\pi/2$–$\pi$–$\pi/2$ Mach–Zehnder interferometer sequence using realistic parameters, including BEC mean-field interactions and optical potentials, achieving computational convergence on modern GPUs within minutes.

These developments bridge the gap between idealized DBD models and realistic guided-atom interferometers, enabling quantitative predictions of interferometric contrast and stability under experimentally relevant imperfections.

## 5.3 The Gross–Pitaevskii Equation with DBD Twin-Lattice Potentials

We consider a BEC serving as the atomic source of a Mach–Zehnder interferometer. The full dynamics during the three light pulses ($\pi/2$–$\pi$–$\pi/2$) and intermediate guided propagation periods are described by the time-dependent Gross–Pitaevskii equation [199]:

$$i\hbar\frac{\partial\Psi(\mathbf{r},t)}{\partial t} = \left(-\frac{\hbar^2}{2m}\nabla^2 + V(\mathbf{r},t) + g|\Psi(\mathbf{r},t)|^2\right)\Psi(\mathbf{r},t), \tag{5.1}$$

where $\Psi(\mathbf{r},t)$ is the condensate wave function, $m$ the atomic mass, and $g = \frac{4\pi\hbar^2 a_s N}{m}$ the nonlinear interaction strength for $N$ atoms with $s$-wave scattering length $a_s$. The total potential consists of the optical guide and the pulsed twin-lattice fields:

$$V(\mathbf{r},t) = V_{\text{guide}}(\mathbf{r},t) + V_{\text{pulse}}(\mathbf{r},t). \tag{5.2}$$

### Optical Potentials

Both components are derived from the corresponding optical intensities, proportional to the dipole interaction strength:

$$V_{\text{guide}}(\mathbf{r},t) = U_{\text{dip}}^{\text{atom}} I_{\text{guide}}(\mathbf{r},t), \tag{5.3}$$

$$V_{\text{pulse}}(\mathbf{r},t) = U_{\text{dip}}^{\text{atom}} I_{\text{pulse}}(\mathbf{r},t), \tag{5.4}$$

where $U_{\text{dip}}^{\text{atom}}$ depends on the atomic polarizability and is given by

$$U_{\text{dip}}^{\text{atom}} = -\frac{\pi c^2}{2}\left[\frac{\Gamma_{D1}}{\omega_{D1}^3}\left(\frac{1}{\omega_{D1}-\omega_L}+\frac{1}{\omega_{D1}+\omega_L}\right)+\frac{\Gamma_{D2}}{\omega_{D2}^3}\left(\frac{2}{\omega_{D2}-\omega_L}+\frac{2}{\omega_{D2}+\omega_L}\right)\right], \tag{5.5}$$

where $\Gamma_{D1,D2}$ and $\omega_{D1,D2}$ are the line widths and frequencies of the D1 and D2 transitions of the atoms, respectively. For $^{87}$Rb atom, the transition line widths are given by $\Gamma_{D1} = 2\pi \times 5.7500 \times 10^6$ Hz and $\Gamma_{D2} = 2\pi \times 6.0666 \times 10^6$ Hz, and the transition frequencies are given by $\omega_{D1} = 2\pi \times 377.107463380 \times 10^{12}$ Hz and $\omega_{D2} = 2\pi \times 384.2304844685 \times 10^{12}$ Hz. The negative sign indicates a red-detuned trap attracting atoms to regions of high intensity.

The guiding intensity is modeled as a Gaussian beam:

$$I_{\text{guide}}(\mathbf{r}, t) = \frac{2P_{\text{guide}}(t)}{\pi w^2(z)} e^{-2(x^2+y^2)/w^2(z)}, \tag{5.6}$$

where $P_{\text{guide}}(t)$ denotes the time-dependent optical power for the guiding beam, and $w(z) = w_0\sqrt{1 + z^2/z_R^2}$ is the beam waist at axial position $z$ with $z_R$ the Rayleigh range.

The twin-lattice potential generating DBD transitions is expressed as:

$$I_{\text{pulse}}(\mathbf{r}, t) = \frac{2P_{\text{pulse}}(t)}{\pi w^2(z)} e^{-2(x^2+y^2)/w^2(z)} \Big[ 1 + \cos(2k_L z)\cos(\Delta\omega t) \Big], \tag{5.7}$$

with $k_L = 2\pi/\lambda_L$ the laser wavenumber and $\Delta\omega = \omega_1 - \omega_2 = 4\omega_{\text{rec}} + \Delta(t)$ the controllable frequency detuning between the two lasers. The interference between the two counter-propagating frequency components produces a time-dependent moving optical lattice that couples momentum states separated by $2\hbar k_L$.

## Numerical Propagation: Split-Step Suzuki–Trotter Scheme

The wave function evolution is computed using a second-order Suzuki–Trotter expansion of the time-evolution operator [200]:

$$\Psi(\mathbf{r}, t + \Delta t) = e^{(\hat{A}+\hat{B})\Delta t}\Psi(\mathbf{r}, t) \approx e^{\hat{A}\Delta t/2} e^{\hat{B}\Delta t} e^{\hat{A}\Delta t/2}\Psi(\mathbf{r}, t), \tag{5.8}$$

where

$$\hat{A} = i\frac{\hbar\nabla^2}{2m}, \qquad \hat{B} = -\frac{i}{\hbar}\left[ V(\mathbf{r}, t) + g|\Psi(\mathbf{r}, t)|^2 \right]. \tag{5.9}$$

This method alternates between kinetic and potential propagation steps in momentum and position space, respectively, ensuring unitarity up to $\mathcal{O}(\Delta t^3)$. The resulting algorithm forms the backbone of the *Universal Atom Interferometer Simulator* (UATIS), enabling efficient 3D simulations of guided BEC interferometers.

## 5.4   Effects of Polarization Errors and Interaction on Guided DBDs

In an ideal DBD configuration, the counter-propagating lattices have orthogonal linear polarizations, ensuring that their interference produces two independent moving lattices rather than a stationary standing wave. However, any residual overlap of the polarization vectors introduces a small standing-wave component proportional to $\epsilon_{\mathrm{pol}} = |\hat{\epsilon}_1^* \cdot \hat{\epsilon}_2| = |\cos\left[2(\pi/4 + \theta_{\mathrm{pol}})\right]|$ when expressed in the angle misalignment of the quarter-wave plate as discussed in Chap. 2.3. This modifies the intensity distribution as

$$I_{\mathrm{pulse}}^{\mathrm{pol}}(\mathbf{r}, t) = \frac{2P_{\mathrm{pulse}}(t)}{\pi w^2(z)} e^{-2(x^2+y^2)/w^2(z)} \cos(2k_L z)[\cos(\Delta\omega t) + \epsilon_{\mathrm{pol}}], \qquad (5.10)$$

introducing a residual standing lattice that perturbs both the guided expansion and the pulse efficiency.

### Impact on Guided Expansion

During guided propagation, even a small residual optical lattice along the guide axis can modify the axial expansion dynamics of the BEC. This residual lattice arises from imperfect polarization control of the counterpropagating guide beams, which introduces a nonzero standing-wave contrast quantified by the polarization error $\epsilon_{\mathrm{pol}}$. When present, this shallow lattice reduces the effective potential curvature experienced by the atoms and thereby slows axial expansion. For sufficiently small errors the effect is modest, yet at higher $\epsilon_{\mathrm{pol}}$ nonlinear distortions can appear, especially when the residual lattice depth becomes comparable to the chemical potential.

To quantify this behavior, Fig. 5.1 compares simulated and experimental axial expansion rates measured in QUANTUS-1. In the experiment, the BEC is first adiabatically loaded into the guide between $t = 0$ and $80$ ms, after which the axial confinement is relaxed and the cloud expands freely until $t = 160$ ms. The expansion is then extracted from the root-mean-square (rms) radius along the guide direction. The comparison shows that the observed reduction in expansion rate matches simulations using a polarization error of $\epsilon_{\mathrm{pol}} = |\cos(\pi/2 + 2 \times 0.01\pi)| \approx 0.0628$, which is consistent with a small misalignment of the quarter-wave plate by approximately $0.01\pi$.

This agreement confirms that residual standing-wave modulation created by polarization errors is sufficient to quantitatively explain the slower guided expansion observed in QUANTUS-1. It also highlights the importance of precise polarization control in experiments where guided propagation is required over long timescales, since even small deviations from ideal circular polarization introduce measurable distortions in the evolution of an interacting BEC.

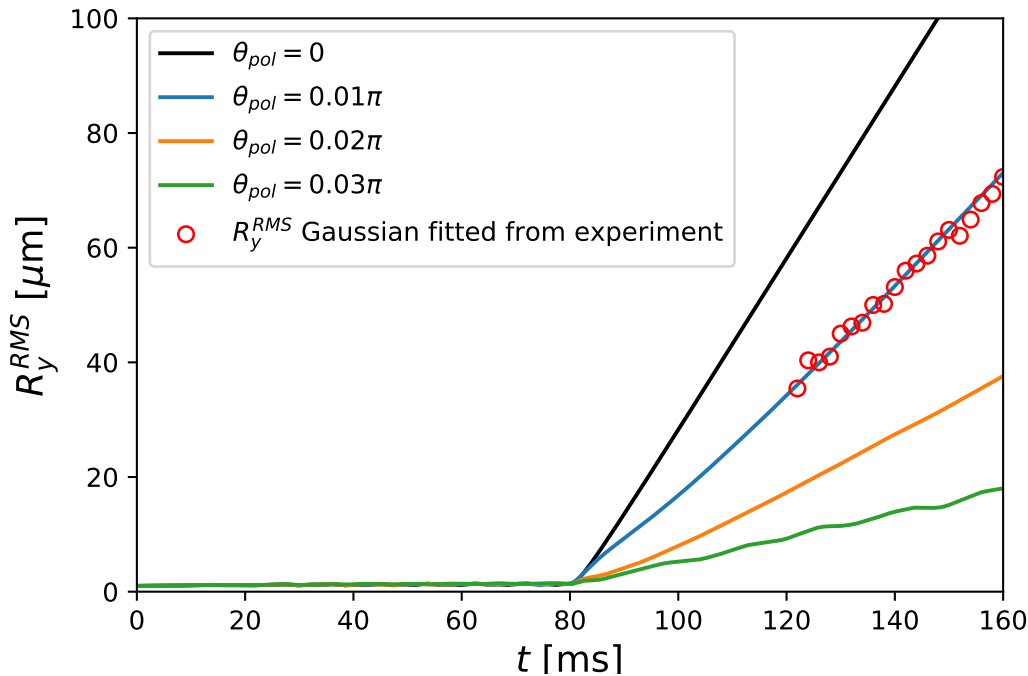

**Figure 5.1 Reduced expansion rate of a BEC along the guide axis due to $\lambda/4$-wave plate angle misalignment $\theta_{\mathrm{pol}}$.** The experimentally measured BEC expansion rate fits best to a simulated polarization error of $\epsilon_{pol} = |\cos(\pi/2 + 2 \times 0.01\pi)| \approx 0.0628$. Experimental data provided by S. Kanthak.

## Impact on Double Bragg Pulses

Polarization errors during DBD pulses play a critical role in determining the diffraction efficiency between the momentum modes $|0\hbar k_L\rangle$ and $|\pm 2\hbar k_L\rangle$. Ideally, the DBD beam splitter couples these three momentum states with well-defined amplitudes and phases. However, a finite $\epsilon_{pol}$ produces an unwanted standing-wave component that modifies the effective coupling Hamiltonian. This leads to reduced transfer efficiency, phase distortions, and population leakage into parasitic higher-order momentum modes that should not be populated in a clean three-port DBD process.

Figs. 5.2(a–b) illustrate the resulting effects by comparing integrated 2D density distributions from full 3D simulations for ideal and imperfect polarization. In the ideal case, the momentum transfer remains confined to the expected diffraction orders and the density distribution maintains a clean, symmetric structure. When polarization errors are introduced, residual standing-wave patterns are visible in the final density, along with clear parasitic trajectories and distortions in the primary ports.

These observations emphasize the need for careful polarization calibration in experiments seeking high-contrast interferometry with DBD pulses. Even small deviations from ideal polarization can induce measurable parasitic structures, degrade beam-splitter fidelity, and ultimately limit interferometer performance.

## Interaction effects on Double Bragg Pulses

Mean-field interactions introduce an additional layer of complexity during DBD diffraction. While the basic three-level model accurately describes low-density condensates in the non-interacting limit, interactions can modify the two-photon resonance condition through density-dependent phase shifts. These shifts become particularly relevant when driving DBD pulses at high Rabi frequencies, where multi-level couplings beyond the three main diffraction orders begin to contribute.

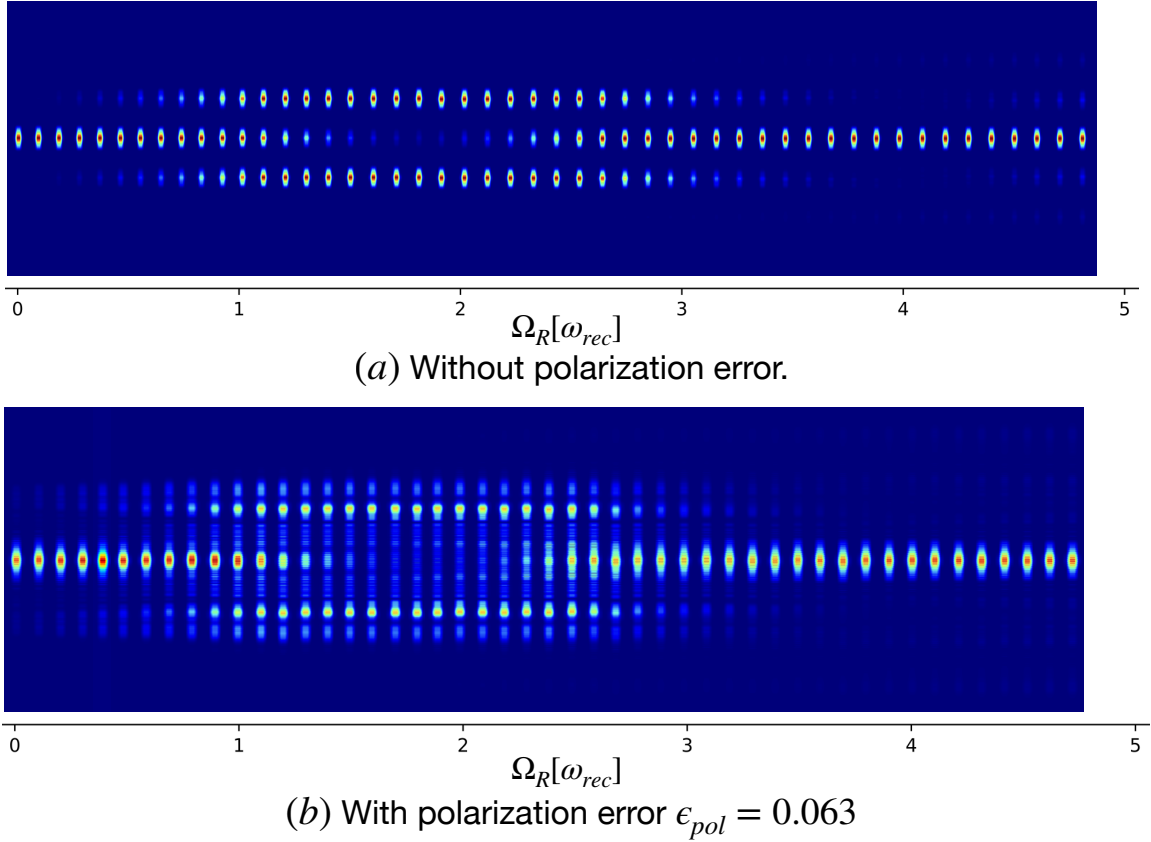

(*a*) Without polarization error.

(*b*) With polarization error $\epsilon_{pol} = 0.063$

**Figure 5.2** Integrated 2D atomic density distributions after a Gaussian DBD pulse for (a) ideal and (b) imperfect polarization. Polarization error introduces additional structures and a reduced double diffraction efficiency.

To investigate this regime, Fig. 5.3 shows an extended Rabi-frequency scan for a condensate with vanishing polarization error but with a finite mean-field interaction corresponding to $N = 50000$ atoms. At large Rabi frequencies, the population distribution exhibits oscillatory behavior characteristic of multi-level coupling. These features can be captured by extending the effective Hamiltonian to include additional diffraction orders up to $\pm 6\hbar k_L$, corresponding to a truncation of Hamiltonian (3.9) under the basis $\{|n\rangle\}$ with $n = 0, 1, 2, 3$.

Although the extended model reproduces the main structure of the population oscillations, small discrepancies remain when comparing with full 3D UATIS simulations. These deviations likely arise because the 1D non-interacting theory cannot capture transverse excitations and mode mixing induced by interactions during the pulse. The interacting BEC can deform transversely, leading to additional phase gradients and mode-dependent detunings that are not present in the simplified model.

Overall, these results highlight that interaction effects can noticeably alter DBD dynamics at high Rabi frequencies. Accurate modeling in this regime may require incorporating both higher diffraction orders and transverse mean-field effects, especially for dense condensates or experiments operating near strong-pulse regimes.

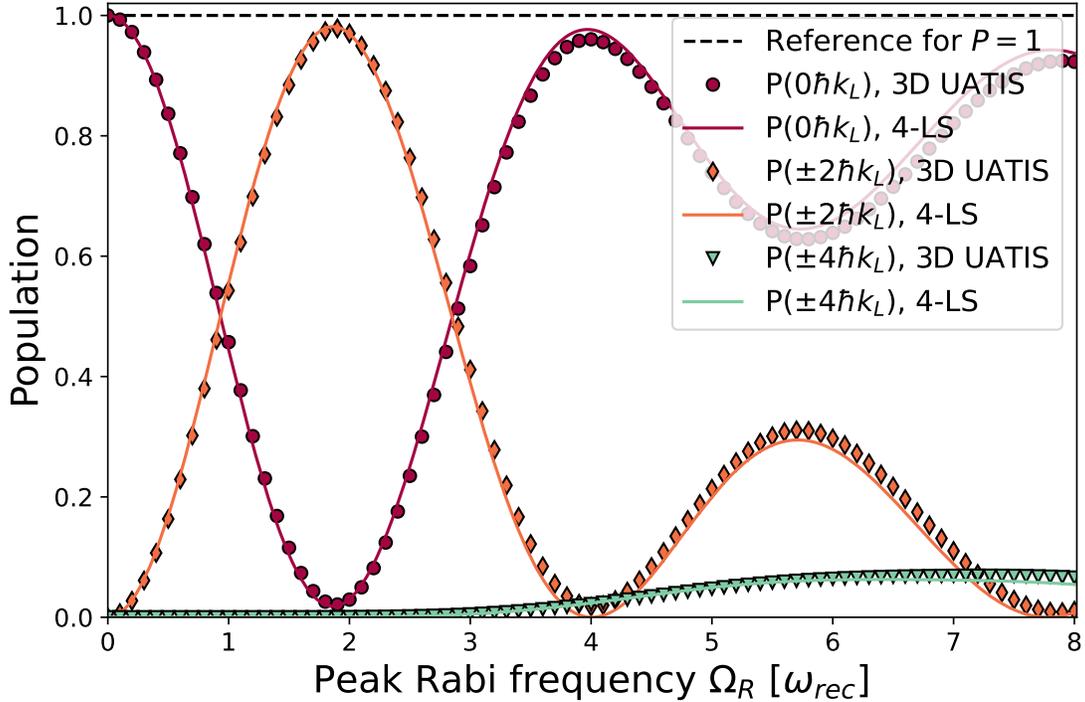

**Figure 5.3 Population distribution among multiple momentum states** after a Gaussian DBD pulse as a function of the peak Rabi frequency $\Omega_R$, at fixed pulse duration $\tau = 0.478, \omega_{\text{rec}}^{-1}$. Symbols represent 3D UATIS simulations with an atom number of $N = 50000$; solid curves show results from a four-level effective Hamiltonian including diffraction orders up to $\pm 6\hbar k_L$. Polarization error in the system is set to zero.

## 5.5 3D Simulation of a Full Mach–Zehnder Interferometer in Free Space

To demonstrate the full capabilities of the Universal Atom Interferometer Simulator (UATIS), we perform a complete three-dimensional simulation of a Mach–Zehnder atom interferometer (MZAI) sequence using a Bose–Einstein condensate (BEC) as the atomic source. The simulation self-consistently incorporates the time-dependent Gross–Pitaevskii equation, including mean-field interactions and the full spatiotemporal laser intensity profiles for all double Bragg diffraction pulses.

The simulated interferometer sequence consists of an initial free expansion of the BEC for $5\,\mathrm{ms}$ after release from the optical trap, followed by a standard $\pi/2$–$\pi$–$\pi/2$ pulse configuration. Each $\pi/2$-pulse has a temporal Gaussian width of $40\,\mu\mathrm{s}$, while the $\pi$-pulse has double the duration $(80\,\mu\mathrm{s})$. The two interrogation intervals between pulses are $T = 40\,\mathrm{ms}$ each, followed by a $60\,\mathrm{ms}$ time of flight before detection, resulting in a total experimental duration of $147\,\mathrm{ms}$. All pulses are modeled as twin-lattice DBD interactions with realistic Gaussian beam profiles and parameters consistent with Table 5.1.

The simulation grid consists of $32 \times 32 \times 16384$ points (about $1.68 \times 10^7$ grid sites) in Cartesian coordinates, evolved using the second-order Suzuki–Trotter split-step algorithm implemented on an NVIDIA RTX 3090 GPU. The total computation time for one full interferometer realization is approximately $20\,\mathrm{minutes}$.

**Density evolution during the interferometric sequence.** Fig. 5.4 shows the integrated 1D atomic density along the interferometer axis as a function of time. The numerical parameters used in the UATIS simulation are summarized in Table 5.1. The three vertical stripes correspond to the three light pulses in the Mach–Zehnder sequence, each driving double Bragg diffraction between the $|0\hbar k_L\rangle$ and the $|\pm 2\hbar k_L\rangle$ momentum states. During the free propagation intervals, the wave packets separate, evolve under mean-field interactions, and subsequently recombine at the output beam splitter, producing an interference signal at the central output port in the absence of gravity.

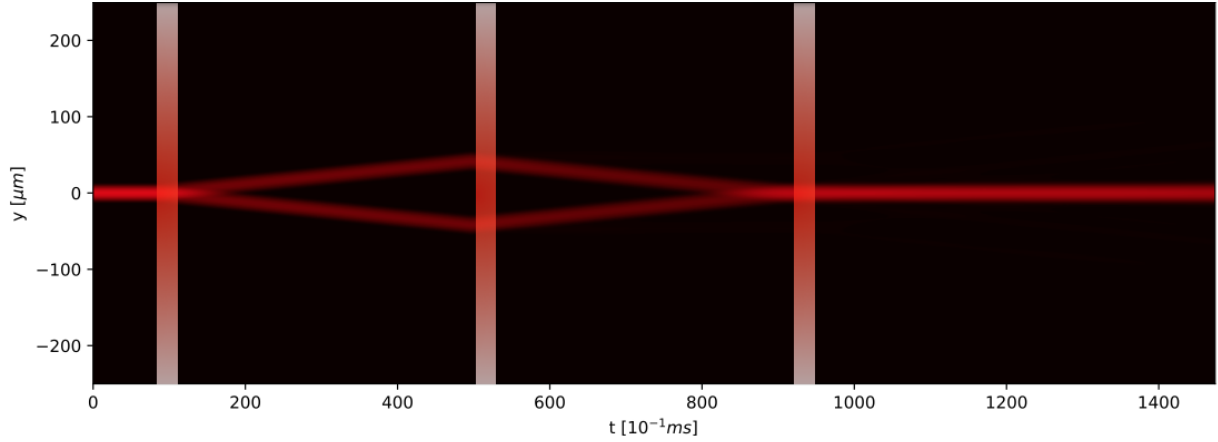

**Figure 5.4** Integrated 1D atomic density along the interferometer axis for the full Mach–Zehnder interferometer in free space. The three vertical stripes indicate the $BS$–$M$–$BS$ DBD pulses, separated by $T = 40\,\mathrm{ms}$. The density modulations reflect coherent splitting, propagation, reflection, and recombination of the atomic wave packets.

**Intermediate density snapshots.** To visualize the dynamics in greater detail, Fig. 5.5 presents integrated 2D atomic density distributions at representative times throughout the interferometric sequence. Panel (a) shows the initial condensate just after release from the trap, while panel (b) depicts the density modulation during the first BS-pulse, where the BEC is coherently split into two momentum components. Panel (c) illustrates the momentum inversion process during the M-pulse, where diffraction couples the counter-propagating wave packets back toward each other, and panel (d) shows the recombination process during the last BS-pulse, where diffraction couples the counter-propagating wave packets back to the initial state and phase shifts are encoded in the population interference. These density maps reveal the clear spatial and temporal symmetry of the DBD-based MZAI and confirm that the interferometer operates in the quasi-Bragg regime, with negligible population in higher-order diffraction states.

**Summary.** The 3D UATIS simulation confirms that realistic DBD pulse parameters can faithfully reproduce a high-contrast Mach–Zehnder interferometer with Bose–Einstein con-densates under free-space conditions. The full spatial dynamics, including mean-field interactions, finite pulse widths, and realistic Gaussian beam geometry, are captured without loss of coherence. The results validate the use of the UATIS framework as a predictive numerical tool for diagnosing and optimizing matter-wave optics in both free-space and guided geometries, as well as for simulating the full sequence of Mach-Zehnder interferometers.

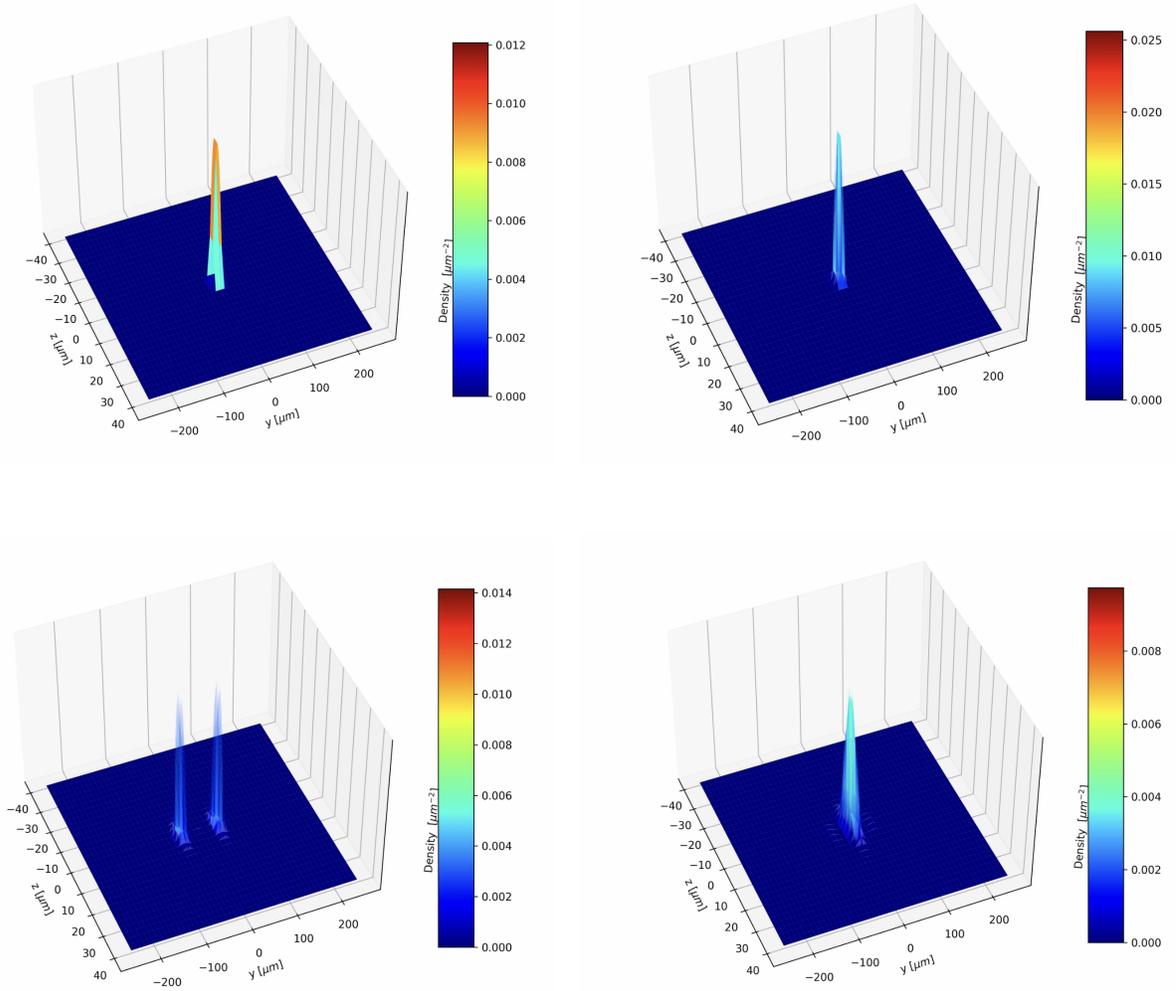

**Figure 5.5** Integrated 2D atomic density at key stages of the Mach–Zehnder interferometer: (a) initial density, (b) during the first *BS*-pulse, (c) during the *M*-pulse, and (d) during the last *BS*-pulse. The symmetric diffraction and recombination dynamics illustrate the coherent evolution of the guided BEC under the DBD pulses.

**Table 5.1 Simulation parameters used in the 3D UATIS model of a Mach–Zehnder interferometer.** All values correspond to typical experimental conditions for $^{87}$Rb at $\lambda_L = 780$ nm. The optical guide and DBD pulse beams share the same Gaussian mode, and mean-field interactions are included through the nonlinear term of the Gross–Pitaevskii equation.

| Parameter | Symbol | Value / Description |
|---|---|---|
| Atomic species | – | $^{87}$Rb ($m = 1.443 \times 10^{-25}$ kg) |
| Wavelength of laser | $\lambda_L$ | 780 nm |
| Recoil frequency | $\omega_{\rm rec}/2\pi$ | 3.77 kHz |
| Beam waist at focus | $w_0$ | 50 $\mu$m |
| Rayleigh range | $z_R$ | 10 mm |
| Peak laser power (guide) | $P_{\rm guide}$ | 1 W |
| Peak laser power (pulse) | $P_{\rm pulse}$ | 40 mW |
| Pulse sequence | – | $BS$–$M$–$BS$ |
| Pulse durations | $\tau_{BS}, \tau_M$ | 40 $\mu$s, 80 $\mu$s |
| Inter-pulse separation | $T$ | 40 ms |
| Time of flight after last pulse | $T_{\rm tof}$ | 60 ms |
| Total evolution time | – | 147 ms |
| Atom number | $N$ | $5 \times 10^4$ |
| $s$-wave scattering length | $a_s$ | $100\, a_0$ ($a_0$ is Bohr radius) |
| Nonlinear interaction strength | $g$ | $\frac{4\pi\hbar^2 a_s N}{m}$ |
| Grid size | – | $32 \times 32 \times 16384$ ($\approx 1.68 \times 10^7$ points) |
| Time step | $\Delta t$ | 0.2 $\mu$s |
| Hardware | – | NVIDIA RTX 3090 GPU |
| Average runtime | – | $\approx 20$ minutes |

## 5.6   Summary and Outlook

In this chapter, we developed a comprehensive 3D simulation framework for guided Bose–Einstein condensate interferometers based on double Bragg diffraction. By solving the full time-dependent Gross–Pitaevskii equation with realistic optical potentials, we have:

- Quantitatively characterized the impact of polarization errors on guided expansion and diffraction efficiency;

- Verified the validity of effective two-level and few-level models under realistic pulse conditions;

- Demonstrated a full Mach–Zehnder sequence in 3D, confirming the feasibility of large-scale, GPU-accelerated simulations.

These results provide a crucial link between analytical pulse optimization and experimental implementation, enabling direct prediction of interferometer performance in realistic geometries. The UATIS platform can be readily extended to include external potentials (e.g., gravity gradients, wavefront aberrations), finite temperature effects, and mean-field-induced phase shifts, paving the way for predictive modeling of guided interferometers in terrestrial and microgravity environments.

# General Conclusion and Outlook

This thesis has developed a general theoretical and computational framework for high-contrast double Bragg diffraction (DBD) atom interferometers, combining analytical models, optimal-control engineering, and large-scale numerical simulations. From the historical foundations of atom interferometry to modern guided Bose–Einstein condensate (BEC) implementations, the work has bridged the gap between idealized light-pulse interferometers and experimentally realizable quantum sensors operating under realistic conditions.

**Chapter 1** traced the evolution of quantum mechanics over the past century and the inevitable birth of atom interferometry. Beginning with Feynman's double-slit *gedanken* experiment on electrons and the original optical Mach-Zehnder interferometer of light, the chapter reviewed the emergence of light-pulse interferometry as a matter-wave counterpart of the optical version and the first experimental realization via stimulated Raman transitions–the so-called *Kasevich–Chu interferometer*–which set the standard for the field of atom interferometry. The fundamental quantum mechanical limits on the sensitivity established there—connecting the acceleration resolution of an interferometer to the effective momentum transfer and interrogation time—provided physical motivation for pursuing large-momentum-transfer techniques with long interrogation times such as double Bragg diffraction in microgravity environment.

**Chapter 2** laid the theoretical foundation for DBD interferometers. Starting from a semi-classical analysis of the de Broglie phase associated with a matter wave and how different phases are encoded during an ideal Mach-Zehnder sequence, it forms the semi-classical picture of a DBD atom interferometer under external acceleration. The general one-dimensional single-particle Hamiltonian of DBD is derived for ideal and imperfect polarizations, forming the basis for the full quantum mechanical description and systematic mitigation strategies developed in later chapters.

**Chapter 3** laid down the new theoretical framework and analyzed a family of *detuning-control strategies* aimed at enhancing the robustness of double Bragg diffraction under AC-Stark shifts, polarization errors and Doppler detunings. Four complementary schemes were investigated: conventional DBD (C-DBD), constant-detuning DBD (CD-DBD), detuning-sweep DBD (DS-DBD), and optimal-control-based DBD (OCT). Using the five-level $S$-matrix formalism and exact numerical simulations, we demonstrated that carefully engineered time-dependent detuning profiles can suppress AC–Stark and Doppler-induced asymmetries while maintaining high DBD efficiency. This chapter thus provided the theoretical bridge between microscopic Hamiltonian control and macroscopic interferometric signal of output-port populations.

**Chapter 4** applied these optimized pulse schemes to full Mach–Zehnder interferometers, quantifying their performance and robustness under realistic conditions. By constructing the total interferometer $S$-matrix and comparing its predictions with exact numerical simulations, we characterized the dependence of fringe contrast on atomic momentum width, center-of-mass (COM) motion, and polarization errors. A systematic study revealed that while conventional DBD loses contrast rapidly for $\sigma_p > 0.05\hbar k_L$, the DS-DBD and OCT strategies sustain contrast above 95% even for $\sigma_p \simeq 0.1\hbar k_L$ and under a few percent of laser-power fluctuations. These results demonstrate that detuning-control strategies can, in principle, elevate DBD interferometers to the same sensitivity class as Raman-based schemes, overcoming a central limitation of earlier implementations.

**Chapter 5** extended the study to fully 3D numerical simulations using the *Universal Atom*

*Interferometer Simulator* (UATIS). By solving the time-dependent Gross–Pitaevskii equation with realistic optical lattice potentials, we investigated the impact of polarization imperfections and mean-field interactions in both guided and free-space configurations. The simulations reproduced experimental observations such as reduced expansion rates and interference visibility degradation due to polarization errors of 5–6%. A complete 3D Mach–Zehnder sequence was then simulated, revealing the coherent splitting, reflection, and recombination dynamics of a BEC under DBD pulses. This validated UATIS as a versatile computational tool for designing future precision sensors.

**Overall conclusions and outlook.** The combined results of this thesis establish that double Bragg diffraction, once regarded as too sensitive to imperfections for high-precision interferometry, can achieve robust, near-ideal performance when supported by advanced detuning control and numerical optimization. The synergy between analytical $S$-matrix modeling and exact numerical simulations has produced a predictive framework applicable across a broad range of experimental platforms—from terrestrial gravimeters to space-borne quantum sensors.

Looking ahead, several directions naturally follow from this work:

- **Guided and hybrid interferometers:** Extending the demonstrated DBD protocols to guided BEC systems will enable vibration-insensitive, long-baseline interferometers for microgravity and transportable applications.

- **Large-momentum-transfer scaling:** Applying OCT and detuning-sweep control to higher-order DBD processes ($8\hbar k_L$, $12\hbar k_L$ and beyond) promises substantial gains in sensitivity.

- **Machine-learning-assisted control:** Data-driven optimization could further enhance robustness against laser-intensity noise, wavefront aberrations, and thermal effects.

- **Applications in precision metrology:** High-contrast DBD interferometers have immediate potential for compact gravimetry, gradiometry, and fundamental tests of general relativity and the equivalence principle.

In summary, the research presented here provides both the conceptual and computational foundation for realizing the next generation of high-fidelity, large-momentum-transfer atom interferometers based on symmetric double Bragg diffraction. By bridging theory, control, and simulation, this work advances DBD interferometry from a theoretical construct to a practical, scalable platform for quantum-enhanced sensing in ground- and space-based environments.

# APPENDICES

# Appendix A

# Magnus Formalism for Deriving Effective Hamiltonians in a Truncated Hilbert Space

In this appendix, we present the general formalism based on the second-order Magnus expansion used to derive effective Hamiltonians in a truncated Hilbert space. This procedure underlies the derivation of the effective two-level system (TLS) Hamiltonian, Eq. (3.16), discussed in Sec. 3.3 of the main text.

## A.1   General framework

Consider a system whose full dynamics are governed by a time-dependent Hamiltonian $H(t)$ defined on an infinite-dimensional Hilbert space $\mathcal{H} = \mathrm{Span}(\{|n\rangle\})$, where $n \in \mathcal{N}$ and $\{|n\rangle\}$ forms a complete orthonormal basis. Suppose that the physically relevant dynamics occur within a restricted subspace $\mathcal{H}_{\mathrm{sub}} = \mathrm{Span}(\{|n\rangle,\, n \in \mathcal{S}\})$, where $\mathcal{S} \subset \mathcal{N}$. We introduce projectors onto the subspace and its complement:

$$\mathbb{P} = \sum_{n \in \mathcal{S}} |n\rangle\langle n|, \qquad \mathbb{Q} = \sum_{n \in \mathcal{N}\backslash\mathcal{S}} |n\rangle\langle n|. \tag{A.1}$$

These satisfy the standard relations

$$\mathbb{P}^2 = \mathbb{P}, \qquad \mathbb{Q}^2 = \mathbb{Q}, \qquad \mathbb{P} + \mathbb{Q} = \hat{\mathbf{1}}. \tag{A.2}$$

The Hamiltonian can be decomposed as

$$H(t) = (\mathbb{P} + \mathbb{Q})\, H(t)\, (\mathbb{P} + \mathbb{Q})$$

$$= \mathbb{P}H(t)\mathbb{P} + \mathbb{P}H(t)\mathbb{Q} + \mathbb{Q}H(t)\mathbb{P} + \mathbb{Q}H(t)\mathbb{Q}$$

$$\equiv H_0(t) + H_1(t), \tag{A.3}$$

where $H_0(t) = \mathbb{P}H(t)\mathbb{P}$ describes dynamics within the subspace, and $H_1(t)$ includes all couplings to the orthogonal complement.

The first-order Magnus expansion yields $H_0(t)$—possibly after applying a rotating-wave approximation—as the lowest-order effective Hamiltonian. The second-order term in Magnus expansion incorporates second-order processes that couple $\mathbb{P}$ and $\mathbb{Q}$ via $H_1(t)$, whose time derivative produces an effective energy shift operator $\Delta_{\mathrm{AC}}(t)$ (AC–Stark shift) acting within the subspace of interests. Therefore, $\Delta_{\mathrm{AC}}(t)$ can be expressed as

$$\frac{\Delta_{\mathrm{AC}}(t)}{\hbar} \equiv i\, \mathbb{P}\, \frac{d}{dt} G_2(t)\, \mathbb{P}$$

$$= -\frac{i}{2\hbar^2} \frac{d}{dt} \int_0^t dt_1 \int_0^{t_1} dt_2 \left[ \mathbb{P}H(t_1)(\mathbb{P}+\mathbb{Q})H(t_2)\mathbb{P} - \mathbb{P}H(t_2)(\mathbb{P}+\mathbb{Q})H(t_1)\mathbb{P} \right] \tag{A.4}$$

$$= -\frac{i}{2\hbar^2} \frac{d}{dt} \int_0^t dt_1 \int_0^{t_1} dt_2 \left[ \mathbb{P}H(t_1)\mathbb{P}H(t_2)\mathbb{P} - \mathbb{P}H(t_2)\mathbb{P}H(t_1)\mathbb{P} \right.$$

$$\left. + \mathbb{P}H(t_1)\mathbb{Q}H(t_2)\mathbb{P} - \mathbb{P}H(t_2)\mathbb{Q}H(t_1)\mathbb{P} \right] \tag{A.5}$$

$$= -\frac{i}{2\hbar^2} \frac{d}{dt} \int_0^t dt_1 \int_0^{t_1} dt_2 \left[ H_0(t_1)H_0(t_2) - H_0(t_2)H_0(t_1) \right.$$

$$\left. + \mathbb{P}H_1(t_1)\mathbb{Q}H_1(t_2)\mathbb{P} - \mathbb{P}H_1(t_2)\mathbb{Q}H_1(t_1)\mathbb{P} \right], \tag{A.6}$$

where terms $\mathbb{P}\dot{G}_2\mathbb{Q} + \mathbb{Q}\dot{G}_2\mathbb{P} + \mathbb{Q}\dot{G}_2\mathbb{Q}$ are neglected due to they drive transitions out of the subspace of interest and are assumed to be far off-resonant. The effective Hamiltonian up to second-order Magnus expansion then reads

$$H_{\mathrm{eff}}(t) = H_0(t) + \Delta_{\mathrm{AC}}(t), \tag{A.7}$$

where rapidly oscillating terms in $\Delta_{\mathrm{AC}}(t)$ can be further discarded via a rotating-wave approximation.

## A.2 Application to the double Bragg Hamiltonian

For the double Bragg problem discussed in the main text, we substitute $\bar{H}(t)$ from Eq. (3.9) as the exact Hamiltonian and choose the subspace $\mathcal{S} = \{0, 1\}$ corresponding to the quasi-Bragg regime. The first-order term becomes

$$
\begin{aligned}
H_0(t) &= (|0\rangle\langle 0| + |1\rangle\langle 1|)\, \bar{H}(t)\, (|0\rangle\langle 0| + |1\rangle\langle 1|) \\
&= \sqrt{2}\, \hbar\Omega(t)\, C(t, \varepsilon_{\text{pol}})\, e^{-i4\omega_{\text{rec}}t}\, |0\rangle\langle 1| + \text{h.c.},
\end{aligned} \tag{A.8}
$$

representing the resonant coupling between the two selected momentum states, where $C(t, \epsilon_{pol}) = \cos\left[(4\omega_{rec} + \Delta)t\right] + \epsilon_{pol}$.

The second-order correction follows from Eq. (A.6):

$$
\begin{aligned}
\frac{\Delta_{\text{AC}}(t)}{\hbar} = -\frac{i}{2\hbar^2}\frac{d}{dt}\int_0^t dt_1 \int_0^{t_1} dt_2 \Big\{ &\left[\bar{H}_{01}(t_1)\bar{H}_{10}(t_2) - \bar{H}_{01}(t_2)\bar{H}_{10}(t_1)\right]|0\rangle\langle 0| \\
+ &\left[\bar{H}_{10}(t_1)\bar{H}_{01}(t_2) - \bar{H}_{10}(t_2)\bar{H}_{01}(t_1) + \bar{H}_{12}(t_1)\bar{H}_{21}(t_2) - \bar{H}_{12}(t_2)\bar{H}_{21}(t_1)\right]|1\rangle\langle 1| \Big\}.
\end{aligned} \tag{A.9}
$$

Using $\bar{H}_{12}(t) = \hbar\Omega(t)C(t, \varepsilon_{\text{pol}})e^{-i12\omega_{\text{rec}}t} = \bar{H}_{21}^\star(t)$ and assuming $\Omega(t)$ varies slowly, we can apply integration by parts and neglect terms proportional to $\dot{\Omega}(t)$. After integration and rotating-wave approximation, we obtain

$$
\left(\frac{\Delta_{\text{AC}}(t)}{\hbar}\right)_{00} = \frac{\Omega^2(t)}{\omega_{\text{rec}}}\left(\frac{1}{4}\epsilon_{pol} - \frac{1}{2}\epsilon_{pol}^2\right), \tag{A.10}
$$

$$
\left(\frac{\Delta_{\text{AC}}(t)}{\hbar}\right)_{11} = \frac{\Omega^2(t)}{\omega_{\text{rec}}}\left[\frac{6\omega_{\text{rec}}^2}{(\Delta - 8\omega_{\text{rec}})(\Delta + 16\omega_{\text{rec}})} - \frac{1}{4}\varepsilon_{\text{pol}} + \frac{5}{12}\varepsilon_{\text{pol}}^2\right]. \tag{A.11}
$$

The two poles at $\Delta = 8\omega_{\text{rec}}$ and $\Delta = -16\omega_{\text{rec}}$ correspond to resonant transitions between $|1\rangle$ and $|2\rangle$ (i.e., between $|\pm 2\hbar k_L\rangle$ and $|\pm 4\hbar k_L\rangle$). To justify truncation to $\mathcal{H}_{\text{sub}} = \text{Span}(|0\rangle, |1\rangle)$, one must operate sufficiently far from these resonances.

In the limit $\Delta \to 0$ (first-order double Bragg resonance), the AC–Stark correction to state $|1\rangle$ simplifies to

$$
\left(\frac{\Delta_{\text{AC}}(t)}{\hbar}\right)_{11} = \frac{\Omega^2(t)}{\omega_{\text{rec}}}\left[-\frac{3}{64} - \frac{1}{4}\varepsilon_{\text{pol}} + \frac{5}{12}\varepsilon_{\text{pol}}^2\right]. \tag{A.12}
$$

Combining Eqs. (A.8), (A.10) and (A.12), we arrive at the effective two-level Hamiltonian presented in Eq. (3.16) of Sec. 3.3.[1] This compact form encapsulates both the resonant coupling and the polarization-dependent AC–Stark shift responsible for the key dynamical features of double Bragg diffraction in the quasi-Bragg regime.

## Summary

The Magnus formalism provides a systematic way to derive effective Hamiltonians governing the dynamics in the physically interesting subspaces while retaining essential higher-order corrections. In the context of double Bragg diffraction, this approach justifies the two-level reduction used in this thesis and reveals the origin of Rabi frequency-dependent shifts (quadratic in $\Omega(t)$) and polarization error-induced shifts. It thus serves as a useful tool for developing physical intuition regarding the detuning-control strategies and optimization techniques introduced in the subsequent chapters. However, the precise regime of validity of the second-order Magnus-type approximation is ultimately confirmed through exact numerical simulations. Consequently, it should not be relied upon indiscriminately for drawing high-precision quantitative conclusions without prior numerical verification.

---

[1] It is worth noting that the above procedure introduces a certain degree of double counting when the resulting effective Hamiltonian is subsequently solved in a time-dependent manner. Alternative derivation schemes have been explored; these affect only the numerical prefactors of the polarization-error-dependent contributions, whereas the polarization-error-free result remains invariant.

# Appendix B

# Mach-Zehnder DBD Interferometers with Spatially Resolved Detections

In this appendix, we present the results obtained for Mach–Zehnder DBD interferometers under a spatially resolved detection scheme. The theoretical description relies on a reduced form of the $S$-matrix from Eq. (4.39). This simplification is justified because the effective mirror efficiency for DBD mirror pulses acting on the high-order momentum states $|p \pm 4\hbar k_L\rangle$ is negligible due to the corresponding momentum transfer requires an eight-photon process. As a result, the matrix elements $M_{45}(p)$ and $M_{54}(p)$ effectively vanish and can be omitted in the resolved-interferometer $S$-matrix. For an input momentum state $|p\rangle$ with $p \in [-\hbar k_L, \hbar k_L]$, only the first column of the total $S$-matrix is relevant for describing the probability amplitudes at the three main output ports, whose explicit form are given by

$$S_{i1}^{\text{tot}}(g, p, T) = B_{i1}(p_3)M_{11}(p_2)B_{11}(p_1)e^{i(\theta_1 + \theta_4)} + B_{i2}(p_3)M_{23}(p_2)B_{31}(p_1)e^{i(\theta_2 + \theta_5)}$$
$$+ B_{i3}(p_3)M_{32}(p_2)B_{21}(p_1)e^{i(\theta_3 + \theta_6)}, \tag{B.1}$$

where $p_1 = p$, $p_2 = p + mgT$, $p_3 = p + 2mgT$ and $\theta_1 = \theta(g, p, T)$, $\theta_2 = \theta(g, p - 2\hbar k_L, T)$, $\theta_3 = \theta(g, p + 2\hbar k_L, T)$, $\theta_4 = \theta(g, p + mgT, T)$, $\theta_5 = \theta(g, p + mgT + 2\hbar k_L, T)$, $\theta_6 = \theta(g, p + mgT - 2\hbar k_L, T)$ with momentum-dependent propagation phase

$$\theta(g, p, T) = -\frac{1}{2m\hbar} \left( Tp^2 + mgT^2 p \right). \tag{B.2}$$

For an input state of the form $|\psi\rangle = \int dp \, \psi(p)|p\rangle$ with $|\psi(p)|^2$ being a normalized momentum distribution with compact support in the first Brillouin zone $[-\hbar k_L, \hbar k_L]$, the final output port populations are given by

$$P_i(g, T) = \int_{-\hbar k_L}^{\hbar k_L} \left| \psi(p) \, S_{i1}^{tot}(g, p, T) \right|^2 \, dp, \tag{B.3}$$

with $i = 1, 2, 3$ corresponding to output port 1, 2 and 3 (see Fig. B.1). Similar to the spatially

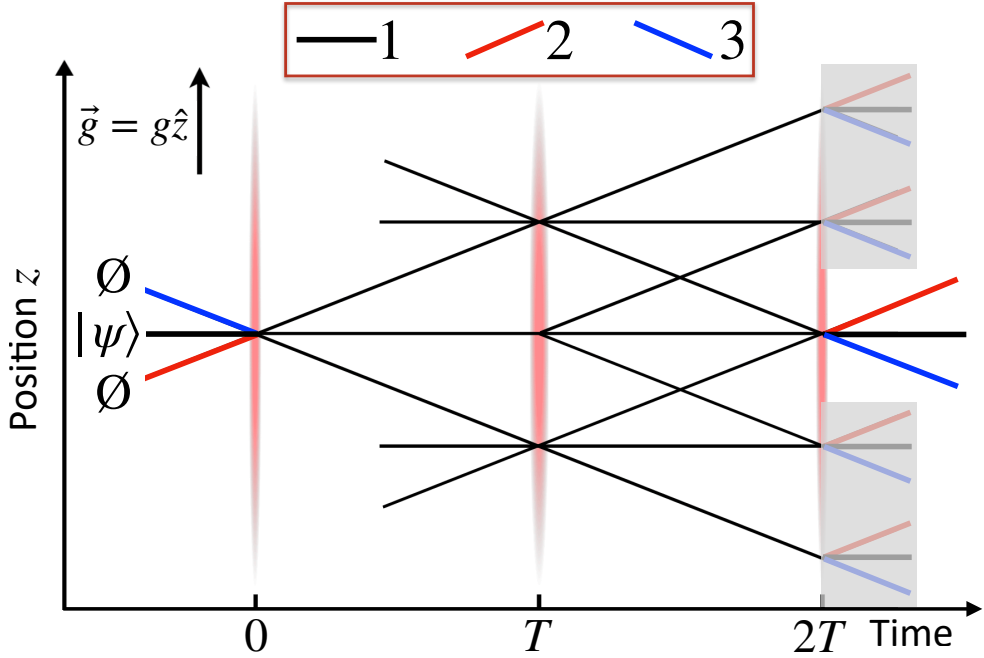

**Figure B.1** Illustration of the spatially resolved detection scheme for a Mach–Zehnder atom interferometer implemented with three DBD pulses. The input and output ports are labeled by $i = 1, 2, 3$, corresponding to the momentum states $|p\rangle$, $|p + 2\hbar k_L\rangle$, and $|p - 2\hbar k_L\rangle$, respectively. In a typical interferometry sequence, only port 1 is populated with the input state $|\psi\rangle = \int dp \, \psi(p)|p\rangle$, while the other two input ports remain empty. The gray rectangles indicate blow-away pulses or coherent absorption used to remove atoms in the undesired output ports.

unresolved case, the conjugated output signals of the full interferometer are

$$P_{0\hbar k_L}(g, T) = P_1(g, T), \tag{B.4}$$

$$P_{\pm 2\hbar k_L}(g, T) = P_2(g, T) + P_3(g, T), \tag{B.5}$$

corresponding to the central output port ($0\hbar k_L$) and the side ports ($\pm 2\hbar k_L$). For spatially resolved detection, it is essential that none of the undesired ports or parasitic trajectories contribute to the final signal. This can be enforced by physically blocking the unwanted trajectories with an aperture, applying a blow-away pulse or by coherently absorbing matter waves that reach unwanted ports [201], as indicated by the gray rectangles forming an absorbing slit at the final beam-splitter pulse in Fig. B.1. In Figs. B.2(a-b), we compare the interferometric signals $P_{0\hbar k_L}(g, T)$ and $P_{\pm 2\hbar k_L}(g, T)$ as functions of the interrogation time $T$, obtained from exact numerical simulations and from the reduced $S$-matrix theory for both the DS-DBD and OCT strategies. The resulting $T$-scan fringe contrasts extracted according to Eq. (4.38) are 97% for the DS-DBD strategy and 99% for the OCT strategy for an initial momentum width of $0.05\hbar k_L$ with $p_0 = 0$ and $\varepsilon_{pol} = 0$. The contrast of the C-DBD or CD-DBD strategy under the same conditions is lower than that of DS-DBD by a few percent and is not shown. In Fig. B.3, we systematically investigate the dependence of the fringe contrast on the initial momentum width up to $\sigma_p = 0.15\,\hbar k_L$. This range spans the temperatures typical for ultracold BECs and even for some thermal atomic samples used in atom interferometry experiments. We observe that across all momentum widths, the OCT and DS-DBD strategies remain significantly more robust against Doppler broadening than the C-DBD or CD-DBD strategies. The contrast advantage of OCT and DS-DBD increases as the initial momentum width (or temperature) grows. Among all four strategies, OCT delivers the highest contrast throughout the entire range of momentum widths considered. The small deviations observed for the C-DBD and CD-DBD strategies between the reduced theory and the exact numerical simulations at large momentum width $\sigma_p > 0.1\hbar k_L$ may arise from the substantial degradation of pulse efficiencies at large momentum spreads, together with imperfect absorption of the unwanted ports in the exact numerical simulations. Above comparison confirms that the relative performance ranking established in Chap. 4.4.3 remains valid for spatially resolved detection,

$$\text{OCT} > \text{DS-DBD} > \text{C-DBD} \approx \text{CD-DBD}.$$

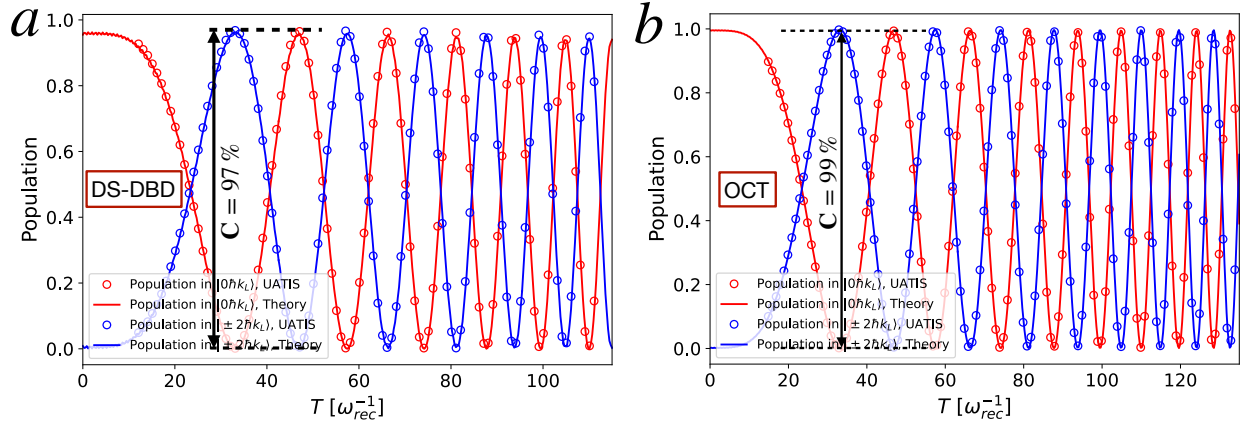

**Figure B.2** DBD Mach–Zehnder interferometer $T$-scan fringes for spatially resolved detection using (a) the DS-DBD strategy and (b) the OCT strategy. The extracted fringe contrasts are $\mathbf{C} = 0.97$ and $\mathbf{C} = 0.99$, respectively, for $g = 0.000714 k_L^{-1} \omega_{\mathrm{rec}}^2$, $\sigma_p = 0.05\hbar k_L$, and $p_0 = 0$. Circles denote results from exact numerical simulations with an absorbing slit placed at the last beam-splitter pulse to remove undesired output ports. Solid curves denote predictions from the reduced $S$-matrix theory of Eq. (B.1). Red and blue colors correspond to the signals $P_{0\hbar k_L}(g, T)$ and $P_{\pm 2\hbar k_L}(g, T)$ in the final output ports, respectively.

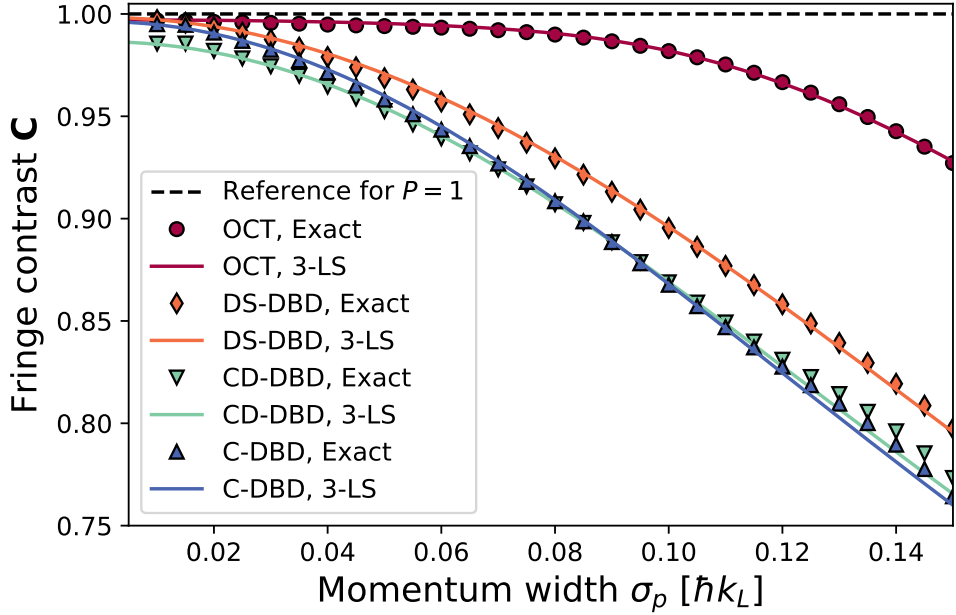

**Figure B.3** Contrast of spatially resolved double-Bragg Mach–Zehnder interferometers for different detuning-control strategies as a function of the initial momentum width for $g = 0.000357k_L^{-1}\omega_{\text{rec}}^2$, $p_0 = 0$, and $\varepsilon_{pol} = 0$. Symbols indicate exact numerical simulations, and solid lines show predictions from the three-level reduced $S$-matrix theory.

Last but not least, we also compare the contrast for different strategies against the polarization error in the system in Fig. B.4. We observe that up to a polarization error of $\varepsilon_{pol} = 0.04$, which is well within the range achievable in experiments with careful alignment of the quarter-wave plate relative to the beam polarizations, the OCT and DS-DBD strategies outperform the C-DBD and CD-DBD strategies. The CD-DBD strategy shows an almost uniform performance of about 95% across the entire range $\varepsilon_{pol} \in [0, 0.1]$. This behavior likely results from the constant detuning $\Delta = 0.27\,\omega_{\mathrm{rec}}$ used for the two beam splitter pulses, which happens to be close to optimal for mitigating polarization errors around $\varepsilon_{pol} = 0.05$. If the DS-DBD and OCT protocols were optimized around a finite target polarization error, they could readily surpass the performance of CD-DBD across the full range of $\varepsilon_{pol}$. Therefore, depending on the typical polarization errors in a given setup, it may be necessary to optimize the protocol around the most probable error values to achieve the best average contrast performance.

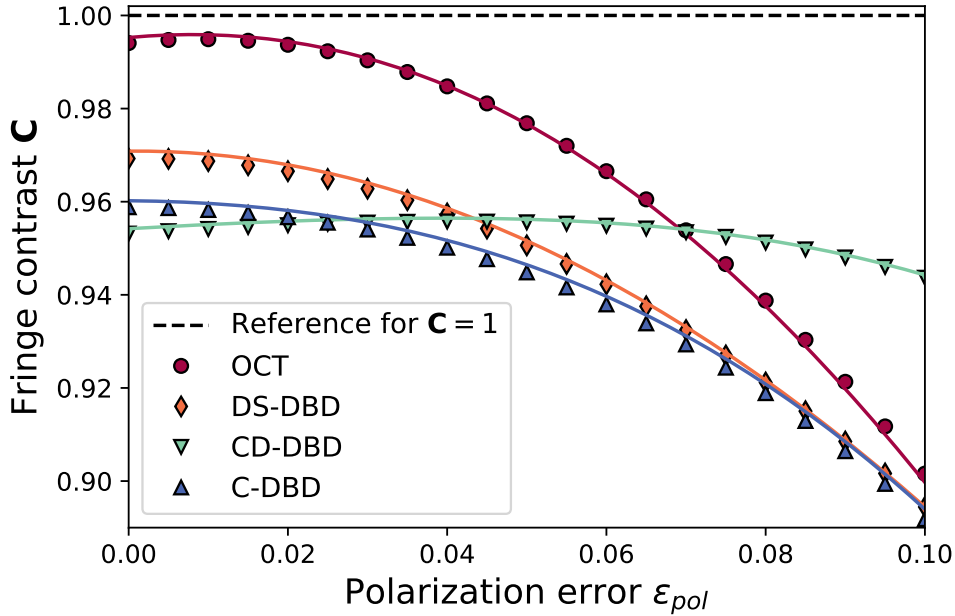

**Figure B.4** Contrast of spatially resolved double-Bragg Mach–Zehnder interferometers for different detuning-control strategies as a function of the polarization error for $g = 0.000357\,k_L^{-1}\omega_{\mathrm{rec}}^2$, $p_0 = 0$, and $\sigma_p = 0.05\,\hbar k_L$. Symbols indicate exact numerical simulations, and solid lines show predictions from the three-level reduced $S$-matrix theory.

# References

1. Heisenberg, W. Über quantentheoretische Umdeutung kinematischer und mechanischer Beziehungen [On the quantum-theoretical reinterpretation of kinematical and mechanical relationships]. *Zeitschrift für Physik* **33,** 879–893. ISSN: 1434-601X. http://dx.doi.org/10.1007/BF01328377 (Dec. 1925).

2. Planck, M. Über eine Verbesserung der Wien'schen Spectralgleichung [On an Improvement of Wien's Equation for the Spectrum]. *Verhandlungen der Deutschen Physikalischen Gesellschaft* **2,** 202–204. https://archive.org/details/verhandlungende01goog/page/n212/mode/2up (Oct. 1900).

3. Planck, M. Zur Theorie des Gesetzes der Energieverteilung im Normalspectrum [On the Theory of the Energy Distribution Law of the Normal Spectrum]. *Verhandlungen der Deutschen Physikalischen Gesellschaft* **2,** 237–245. https://archive.org/details/verhandlungende01goog/page/n246/mode/2up (Dec. 1900).

4. Planck, M. Über das Gesetz der Energieverteilung im Normalspectrum [On the Law of Distribution of Energy in the Normal Spectrum]. *Annalen der Physik* **309,** 553–563. ISSN: 1521-3889. http://dx.doi.org/10.1002/andp.19013090310 (Jan. 1901).

5. Klein, M. J. Max Planck and the beginnings of the quantum theory. *Archive for History of Exact Sciences* **1,** 459–479. ISSN: 1432-0657. http://dx.doi.org/10.1007/BF00327765 (Oct. 1961).

6. Einstein, A. Über einen die Erzeugung und Verwandlung des Lichtes betreffenden heuristischen Gesichtspunkt [On a Heuristic Point of View Concerning the Production and Transformation of Light]. *Annalen der Physik* **322,** 132–148. eprint: https://onlinelibrary.wiley.com/doi/pdf/10.1002/andp.19053220607. https://onlinelibrary.wiley.com/doi/abs/10.1002/andp.19053220607 (1905).

7. Bohr, N. I. On the constitution of atoms and molecules. *The London, Edinburgh, and Dublin Philosophical Magazine and Journal of Science* **26,** 1–25. ISSN: 1941-5990. http://dx.doi.org/10.1080/14786441308634955 (July 1913).

8. Bohr, N. XXXVII. On the constitution of atoms and molecules. *The London, Edinburgh, and Dublin Philosophical Magazine and Journal of Science* **26,** 476–502. ISSN: 1941-5990. http://dx.doi.org/10.1080/14786441308634993 (Sept. 1913).

9. Bohr, N. LXXIII. On the constitution of atoms and molecules. *The London, Edinburgh, and Dublin Philosophical Magazine and Journal of Science* **26,** 857–875. ISSN: 1941-5990. http://dx.doi.org/10.1080/14786441308635031 (Nov. 1913).

10. De Broglie, L. Ondes et quanta [Waves and quanta]. *Comptes Rendus* **177,** 507–510. http://www.academie-sciences.fr/pdf/dossiers/Broglie/Broglie_pdf/CR1923_p507.pdf (1923).

11. De Broglie, L. Recherches sur la théorie des Quanta [On the Theory of Quanta]. *Annales de Physique* **10,** 22–128. ISSN: 1286-4838. http://dx.doi.org/10.1051/anphys/192510030022 (1925).

12. Schrödinger, E. An Undulatory Theory of the Mechanics of Atoms and Molecules. *Phys. Rev.* **28,** 1049–1070. https://link.aps.org/doi/10.1103/PhysRev.28.1049 (6 Dec. 1926).

13. Born, M. Zur Quantenmechanik der Stoßvorgänge [On the quantum mechanics of collision processes]. *Zeitschrift für Physik* **37,** 863–867. ISSN: 1434-601X. http://dx.doi.org/10.1007/BF01397477 (Dec. 1926).

14. Heisenberg, W. über den anschaulichen Inhalt der quantentheoretischen Kinematik und Mechanik [The actual content of quantum theoretical kinematics and mechanics]. *Zeitschrift für Physik* **43,** 172–198. ISSN: 1434-601X. http://dx.doi.org/10.1007/BF01397280 (Mar. 1927).

15. Davisson, C. & Germer, L. H. The Scattering of Electrons by a Single Crystal of Nickel. *Nature* **119,** 558–560. ISSN: 1476-4687. http://dx.doi.org/10.1038/119558a0 (Apr. 1927).

16. Davisson, C. & Germer, L. H. Diffraction of Electrons by a Crystal of Nickel. *Phys. Rev.* **30,** 705–740. https://link.aps.org/doi/10.1103/PhysRev.30.705 (6 Dec. 1927).

17. Davisson, C. J. & Germer, L. H. Reflection of Electrons by a Crystal of Nickel. *Proceedings of the National Academy of Sciences* **14,** 317–322. ISSN: 1091-6490. http://dx.doi.org/10.1073/pnas.14.4.317 (Apr. 1928).

18. Thomson, G. P. & Reid, A. Diffraction of Cathode Rays by a Thin Film. *Nature* **119,** 890–890. ISSN: 1476-4687. http://dx.doi.org/10.1038/119890a0 (June 1927).

19. Dirac, P. A. M. On a Relativistically Invariant Formulation of the Quantum Theory of Wave Fields. *Proceedings of the Royal Society of London. Series A, Containing Papers of a Mathematical and Physical Character* **114,** 243–265. ISSN: 2053-9150. http://dx.doi.org/10.1098/rspa.1927.0039 (Mar. 1927).

20. Tomonaga, S. On a Relativistically Invariant Formulation of the Quantum Theory of Wave Fields. *Progress of Theoretical Physics* **1,** 27–42. ISSN: 0033-068X. http://dx.doi.org/10.1143/PTP.1.27 (Aug. 1946).

21. Koba, Z., Tati, T. & Tomonaga, S.-i. On a Relativistically Invariant Formulation of the Quantum Theory of Wave Fields. II: Case of Interacting Electromagnetic and Electron Fields. *Progress of Theoretical Physics* **2,** 101–116. ISSN: 1347-4081. http://dx.doi.org/10.1143/ptp/2.3.101 (Oct. 1947).

22. Koba, Z., Tati, T. & Tomonaga, S.-i. On a Relativistically Invariant Formulation of the Quantum Theory of Wave Fields. III: Case of Interacting Electromagnetic and Electron Fields. *Progress of Theoretical Physics* **2,** 198–208. ISSN: 1347-4081. http://dx.doi.org/10.1143/ptp/2.4.198 (Dec. 1947).

23. Tomonaga, S.-I. & Oppenheimer, J. R. On Infinite Field Reactions in Quantum Field Theory. *Phys. Rev.* **74,** 224–225. https://link.aps.org/doi/10.1103/PhysRev.74.224 (2 July 1948).

24. Schwinger, J. On Quantum-Electrodynamics and the Magnetic Moment of the Electron. *Phys. Rev.* **73,** 416–417. https://link.aps.org/doi/10.1103/PhysRev.73.416 (4 Feb. 1948).

25. Schwinger, J. Quantum Electrodynamics. I. A Covariant Formulation. *Phys. Rev.* **74,** 1439–1461. https://link.aps.org/doi/10.1103/PhysRev.74.1439 (10 Nov. 1948).

26. Schwinger, J. Quantum Electrodynamics. II. Vacuum Polarization and Self-Energy. *Phys. Rev.* **75,** 651–679. https://link.aps.org/doi/10.1103/PhysRev.75.651 (4 Feb. 1949).

27. Schwinger, J. Quantum Electrodynamics. III. The Electromagnetic Properties of the Electron–Radiative Corrections to Scattering. *Phys. Rev.* **76,** 790–817. https://link.aps.org/doi/10.1103/PhysRev.76.790 (6 Sept. 1949).

28. Feynman, R. P. Space-Time Approach to Non-Relativistic Quantum Mechanics. *Rev. Mod. Phys.* **20,** 367–387. https://link.aps.org/doi/10.1103/RevModPhys.20.367 (2 Apr. 1948).

29. Feynman, R. P. Space-Time Approach to Quantum Electrodynamics. *Phys. Rev.* **76,** 769–789. https://link.aps.org/doi/10.1103/PhysRev.76.769 (6 Sept. 1949).

30. Feynman, R. P. The Theory of Positrons. *Phys. Rev.* **76,** 749–759. https://link.aps.org/doi/10.1103/PhysRev.76.749 (6 Sept. 1949).

31. Feynman, R. P. Mathematical Formulation of the Quantum Theory of Electromagnetic Interaction. *Phys. Rev.* **80,** 440–457. https://link.aps.org/doi/10.1103/PhysRev.80.440 (3 Nov. 1950).

32. Dyson, F. J. The Radiation Theories of Tomonaga, Schwinger, and Feynman. *Phys. Rev.* **75,** 486–502. https://link.aps.org/doi/10.1103/PhysRev.75.486 (3 Feb. 1949).

33. Dyson, F. J. The *S* Matrix in Quantum Electrodynamics. *Phys. Rev.* **75,** 1736–1755. https://link.aps.org/doi/10.1103/PhysRev.75.1736 (11 June 1949).

34. Yang, C. N. & Mills, R. L. Conservation of Isotopic Spin and Isotopic Gauge Invariance. *Phys. Rev.* **96,** 191–195. https://link.aps.org/doi/10.1103/PhysRev.96.191 (1 Oct. 1954).

35. Glashow, S. L. Partial-symmetries of weak interactions. *Nuclear Physics* **22,** 579–588. ISSN: 0029-5582. http://dx.doi.org/10.1016/0029-5582(61)90469-2 (Feb. 1961).

36. Salam, A. & Ward, J. C. On a gauge theory of elementary interactions. *Il Nuovo Cimento* **19,** 165–170. ISSN: 1827-6121. http://dx.doi.org/10.1007/BF02812723 (Jan. 1961).

37. Goldstone, J., Salam, A. & Weinberg, S. Broken Symmetries. *Phys. Rev.* **127,** 965–970. https://link.aps.org/doi/10.1103/PhysRev.127.965 (3 Aug. 1962).

38. Higgs, P. Broken symmetries, massless particles and gauge fields. *Physics Letters* **12,** 132–133. ISSN: 0031-9163. http://dx.doi.org/10.1016/0031-9163(64)91136-9 (Sept. 1964).

39. Higgs, P. W. Broken Symmetries and the Masses of Gauge Bosons. *Phys. Rev. Lett.* **13,** 508–509. https://link.aps.org/doi/10.1103/PhysRevLett.13.508 (16 Oct. 1964).

40. Weinberg, S. A Model of Leptons. *Phys. Rev. Lett.* **19,** 1264–1266. https://link.aps.org/doi/10.1103/PhysRevLett.19.1264 (21 Nov. 1967).

41. Rauch, H., Treimer, W. & Bonse, U. Test of a single crystal neutron interferometer. *Physics Letters A* **47,** 369–371. ISSN: 0375-9601. http://dx.doi.org/10.1016/0375-9601(74)90132-7 (Apr. 1974).

42. Werner, S. A., Colella, R., Overhauser, A. W. & Eagen, C. F. Observation of the Phase Shift of a Neutron Due to Precession in a Magnetic Field. *Phys. Rev. Lett.* **35,** 1053–1055. https://link.aps.org/doi/10.1103/PhysRevLett.35.1053 (16 Oct. 1975).

43. Colella, R., Overhauser, A. W. & Werner, S. A. Observation of Gravitationally Induced Quantum Interference. *Phys. Rev. Lett.* **34,** 1472–1474. https://link.aps.org/doi/10.1103/PhysRevLett.34.1472 (23 June 1975).

44. Rauch, H. & Werner, S. A. *Neutron Interferometry : Lessons in Experimental Quantum Mechanics, Second Edition* ISBN: 9780198712510. http://dx.doi.org/10.1093/acprof:oso/9780198712510.001.0001 (Oxford University Press, Jan. 2015).

45. Greenberger, D. M. The neutron interferometer as a device for illustrating the strange behavior of quantum systems. *Rev. Mod. Phys.* **55,** 875–905. https://link.aps.org/doi/10.1103/RevModPhys.55.875 (4 Oct. 1983).

46. Zeilinger, A., Gähler, R., Shull, C. G., Treimer, W. & Mampe, W. Single- and double-slit diffraction of neutrons. *Rev. Mod. Phys.* **60,** 1067–1073. https://link.aps.org/doi/10.1103/RevModPhys.60.1067 (4 Oct. 1988).

47. Chu, S., Hollberg, L., Bjorkholm, J. E., Cable, A. & Ashkin, A. Three-dimensional viscous confinement and cooling of atoms by resonance radiation pressure. *Phys. Rev. Lett.* **55,** 48–51. https://link.aps.org/doi/10.1103/PhysRevLett.55.48 (1 July 1985).

48. Lett, P. D. *et al.* Observation of Atoms Laser Cooled below the Doppler Limit. *Phys. Rev. Lett.* **61,** 169–172. https://link.aps.org/doi/10.1103/PhysRevLett.61.169 (2 July 1988).

49. Cohen-Tannoudji, C. N. & Phillips, W. D. New Mechanisms for Laser Cooling. *Physics Today* **43,** 33–40. ISSN: 1945-0699. http://dx.doi.org/10.1063/1.881239 (Oct. 1990).

50. Chu, S. Nobel Lecture: The manipulation of neutral particles. *Rev. Mod. Phys.* **70,** 685–706. https://link.aps.org/doi/10.1103/RevModPhys.70.685 (3 July 1998).

51. Cohen-Tannoudji, C. N. Nobel Lecture: Manipulating atoms with photons. *Rev. Mod. Phys.* **70,** 707–719. https://link.aps.org/doi/10.1103/RevModPhys.70.707 (3 July 1998).

52. Phillips, W. D. Nobel Lecture: Laser cooling and trapping of neutral atoms. *Rev. Mod. Phys.* **70,** 721–741. https://link.aps.org/doi/10.1103/RevModPhys.70.721 (3 July 1998).

53. Cornell, E. A. & Wieman, C. E. Nobel Lecture: Bose-Einstein condensation in a dilute gas, the first 70 years and some recent experiments. *Rev. Mod. Phys.* **74,** 875–893. https://link.aps.org/doi/10.1103/RevModPhys.74.875 (3 Aug. 2002).

54. Ketterle, W. Nobel lecture: When atoms behave as waves: Bose-Einstein condensation and the atom laser. *Rev. Mod. Phys.* **74,** 1131–1151. https://link.aps.org/doi/10.1103/RevModPhys.74.1131 (4 Nov. 2002).

55. Kasevich, M. & Chu, S. Atomic interferometry using stimulated Raman transitions. *Phys. Rev. Lett.* **67,** 181–184. https://link.aps.org/doi/10.1103/PhysRevLett.67.181 (2 July 1991).

56. Kozuma, M. *et al.* Coherent Splitting of Bose-Einstein Condensed Atoms with Optically Induced Bragg Diffraction. *Phys. Rev. Lett.* **82,** 871–875. https://link.aps.org/doi/10.1103/PhysRevLett.82.871 (5 Feb. 1999).

57. Oberthaler, M. K. *et al.* Dynamical diffraction of atomic matter waves by crystals of light. *Phys. Rev. A* **60,** 456–472. https://link.aps.org/doi/10.1103/PhysRevA.60.456 (1 July 1999).

58. Ben Dahan, M., Peik, E., Reichel, J., Castin, Y. & Salomon, C. Bloch Oscillations of Atoms in an Optical Potential. *Phys. Rev. Lett.* **76,** 4508–4511. https://link.aps.org/doi/10.1103/PhysRevLett.76.4508 (24 June 1996).

59. Peik, E., Ben Dahan, M., Bouchoule, I., Castin, Y. & Salomon, C. Bloch oscillations of atoms, adiabatic rapid passage, and monokinetic atomic beams. *Phys. Rev. A* **55,** 2989–3001. https://link.aps.org/doi/10.1103/PhysRevA.55.2989 (4 Apr. 1997).

60. Peters, A., Chung, K. Y. & Chu, S. Measurement of gravitational acceleration by dropping atoms. *Nature* **400,** 849–852. ISSN: 1476-4687. http://dx.doi.org/10.1038/23655 (Aug. 1999).

61. Peters, A., Chung, K. Y. & Chu, S. High-precision gravity measurements using atom interferometry. *Metrologia* **38,** 25–61. ISSN: 0026-1394. http://dx.doi.org/10.1088/0026-1394/38/1/4 (Feb. 2001).

62. Hu, Z.-K. *et al.* Demonstration of an ultrahigh-sensitivity atom-interferometry absolute gravimeter. *Phys. Rev. A* **88,** 043610. https://link.aps.org/doi/10.1103/PhysRevA.88.043610 (4 Oct. 2013).

63. Altin, P. A. *et al.* Precision atomic gravimeter based on Bragg diffraction. *New Journal of Physics* **15,** 023009. ISSN: 1367-2630. http://dx.doi.org/10.1088/1367-2630/15/2/023009 (Feb. 2013).

64. Hardman, K. S. *et al.* Simultaneous Precision Gravimetry and Magnetic Gradiometry with a Bose-Einstein Condensate: A High Precision, Quantum Sensor. *Phys. Rev. Lett.* **117,** 138501. https://link.aps.org/doi/10.1103/PhysRevLett.117.138501 (13 Sept. 2016).

65. Bouchendira, R., Cladé, P., Guellati-Khélifa, S., Nez, F. c. & Biraben, F. c. New Determination of the Fine Structure Constant and Test of the Quantum Electrodynamics. *Phys. Rev. Lett.* **106,** 080801. https://link.aps.org/doi/10.1103/PhysRevLett.106.080801 (8 Feb. 2011).

66. Parker, R. H., Yu, C., Zhong, W., Estey, B. & Müller, H. Measurement of the fine-structure constant as a test of the Standard Model. *Science* **360,** 191–195. eprint: https://www.science.org/doi/pdf/10.1126/science.aap7706. https://www.science.org/doi/abs/10.1126/science.aap7706 (2018).

67. Morel, L., Yao, Z., Cladé, P. & Guellati-Khélifa, S. Determination of the fine-structure constant with an accuracy of 81 parts per trillion. *Nature* **588,** 61–65. ISSN: 1476-4687. http://dx.doi.org/10.1038/s41586-020-2964-7 (Dec. 2020).

68. Fray, S., Diez, C. A., Hänsch, T. W. & Weitz, M. Atomic Interferometer with Amplitude Gratings of Light and Its Applications to Atom Based Tests of the Equivalence Principle. *Phys. Rev. Lett.* **93,** 240404. https://link.aps.org/doi/10.1103/PhysRevLett.93.240404 (24 Dec. 2004).

69. Schlippert, D. *et al.* Quantum Test of the Universality of Free Fall. *Phys. Rev. Lett.* **112,** 203002. https://link.aps.org/doi/10.1103/PhysRevLett.112.203002 (20 May 2014).

70. Tarallo, M. G. *et al.* Test of Einstein Equivalence Principle for 0-Spin and Half-Integer-Spin Atoms: Search for Spin-Gravity Coupling Effects. *Phys. Rev. Lett.* **113,** 023005. https://link.aps.org/doi/10.1103/PhysRevLett.113.023005 (2 July 2014).

71. Zhou, L. *et al.* Test of Equivalence Principle at $10^{-8}$ Level by a Dual-Species Double-Diffraction Raman Atom Interferometer. *Phys. Rev. Lett.* **115,** 013004. https://link.aps.org/doi/10.1103/PhysRevLett.115.013004 (1 July 2015).

72. Duan, X.-C. *et al.* Test of the Universality of Free Fall with Atoms in Different Spin Orientations. *Phys. Rev. Lett.* **117,** 023001. https://link.aps.org/doi/10.1103/PhysRevLett.117.023001 (2 July 2016).

73. Asenbaum, P., Overstreet, C., Kim, M., Curti, J. & Kasevich, M. A. Atom-Interferometric Test of the Equivalence Principle at the $10^{-12}$ Level. *Phys. Rev. Lett.* **125,** 191101. https://link.aps.org/doi/10.1103/PhysRevLett.125.191101 (19 Nov. 2020).

74. Geiger, R. *et al.* Detecting inertial effects with airborne matter-wave interferometry. *Nature Communications* **2.** ISSN: 2041-1723. http://dx.doi.org/10.1038/ncomms1479 (Sept. 2011).

75. Müntinga, H. *et al.* Interferometry with Bose-Einstein Condensates in Microgravity. *Phys. Rev. Lett.* **110,** 093602. https://link.aps.org/doi/10.1103/PhysRevLett.110.093602 (9 Feb. 2013).

76. Barrett, B. *et al.* Dual matter-wave inertial sensors in weightlessness. *Nature Communications* **7.** ISSN: 2041-1723. http://dx.doi.org/10.1038/ncomms13786 (Dec. 2016).

77. Becker, D. *et al.* Space-borne Bose–Einstein condensation for precision interferometry. *Nature* **562,** 391–395. ISSN: 1476-4687. http://dx.doi.org/10.1038/s41586-018-0605-1 (Oct. 2018).

78. Lachmann, M. D. *et al.* Ultracold atom interferometry in space. *Nature Communications* **12.** ISSN: 2041-1723. http://dx.doi.org/10.1038/s41467-021-21628-z (Feb. 2021).

79. Gaaloul, N. *et al.* A space-based quantum gas laboratory at picokelvin energy scales. *Nature Communications* **13.** ISSN: 2041-1723. http://dx.doi.org/10.1038/s41467-022-35274-6 (Dec. 2022).

80. Elliott, E. R. *et al.* Quantum gas mixtures and dual-species atom interferometry in space. *Nature* **623,** 502–508. ISSN: 1476-4687. http://dx.doi.org/10.1038/s41586-023-06645-w (Nov. 2023).

81. He, M. *et al.* The space cold atom interferometer for testing the equivalence principle in the China Space Station. *npj Microgravity* **9.** ISSN: 2373-8065. http://dx.doi.org/10.1038/s41526-023-00306-y (July 2023).

82. Li, J. *et al.* Realization of a cold atom gyroscope in space. *National Science Review* **12.** ISSN: 2053-714X. http://dx.doi.org/10.1093/nsr/nwaf012 (Jan. 2025).

83. Dimopoulos, S., Graham, P. W., Hogan, J. M., Kasevich, M. A. & Rajendran, S. Gravitational wave detection with atom interferometry. *Physics Letters B* **678,** 37–40. ISSN: 0370-2693. http://dx.doi.org/10.1016/j.physletb.2009.06.011 (July 2009).

84. Graham, P. W., Hogan, J. M., Kasevich, M. A. & Rajendran, S. New Method for Gravitational Wave Detection with Atomic Sensors. *Phys. Rev. Lett.* **110,** 171102. https://link.aps.org/doi/10.1103/PhysRevLett.110.171102 (17 Apr. 2013).

85. Graham, P. W., Hogan, J. M., Kasevich, M. A. & Rajendran, S. Resonant mode for gravitational wave detectors based on atom interferometry. *Phys. Rev. D* **94,** 104022. https://link.aps.org/doi/10.1103/PhysRevD.94.104022 (10 Nov. 2016).

86. Canuel, B. *et al.* Exploring gravity with the MIGA large scale atom interferometer. *Scientific Reports* **8.** ISSN: 2045-2322. http://dx.doi.org/10.1038/s41598-018-32165-z (Sept. 2018).

87. Zhan, M.-S. *et al.* ZAIGA: Zhaoshan long-baseline atom interferometer gravitation antenna. *International Journal of Modern Physics D* **29,** 1940005. ISSN: 1793-6594. http://dx.doi.org/10.1142/S0218271819400054 (July 2019).

88. Canuel, B. *et al.* ELGAR—a European Laboratory for Gravitation and Atom-interferometric Research. *Classical and Quantum Gravity* **37,** 225017. https://doi.org/10.1088/1361-6382/aba80e (Oct. 2020).

89. Abe, M. *et al.* Matter-wave Atomic Gradiometer Interferometric Sensor (MAGIS-100). *Quantum Science and Technology* **6,** 044003. https://doi.org/10.1088/2058-9565/abf719 (July 2021).

90. Geraci, A. A. & Derevianko, A. Sensitivity of Atom Interferometry to Ultralight Scalar Field Dark Matter. *Phys. Rev. Lett.* **117,** 261301. https://link.aps.org/doi/10.1103/PhysRevLett.117.261301 (26 Dec. 2016).

91. Stadnik, Y. V. & Flambaum, V. V. Enhanced effects of variation of the fundamental constants in laser interferometers and application to dark-matter detection. *Phys. Rev. A* **93,** 063630. https://link.aps.org/doi/10.1103/PhysRevA.93.063630 (6 June 2016).

92. Graham, P. W., Kaplan, D. E., Mardon, J., Rajendran, S. & Terrano, W. A. Dark matter direct detection with accelerometers. *Phys. Rev. D* **93,** 075029. https://link.aps.org/doi/10.1103/PhysRevD.93.075029 (7 Apr. 2016).

93. Arvanitaki, A., Graham, P. W., Hogan, J. M., Rajendran, S. & Van Tilburg, K. Search for light scalar dark matter with atomic gravitational wave detectors. *Phys. Rev. D* **97,** 075020. https://link.aps.org/doi/10.1103/PhysRevD.97.075020 (7 Apr. 2018).

94. Du, Y., Murgui, C., Pardo, K., Wang, Y. & Zurek, K. M. Atom interferometer tests of dark matter. *Phys. Rev. D* **106,** 095041. https://link.aps.org/doi/10.1103/PhysRevD.106.095041 (9 Nov. 2022).

95. Richard P. Feynman, R. B. L. & Sands, M. *The Feynman Lectures on Physics, Vol. III: Quantum Mechanics* (Addison-Wesley (1965 edition)).

96. Bohr, N. Das Quantenpostulat und die neuere Entwicklung der Atomistik [The Quantum Postulate and the Recent Development of Atomistics]. *Die Naturwissenschaften* **16,** 245–257. ISSN: 1432-1904. http://dx.doi.org/10.1007/BF01504968 (Apr. 1928).

97. Zehnder, L. Ein neuer Interferenzrefraktor [A new interferometer]. *Zeitschrift für Instrumentenkunde* **11,** 275–285. https://archive.org/details/zeitschriftfrin11gergoog (Aug. 1891).

98. Mach, L. Über einen Interferenzrefraktor [About an interferometer]. *Zeitschrift für Instrumentenkunde* **12,** 89–93. https://archive.org/details/zeitschriftfrin14gergoog (Mar. 1892).

99. Kasevich, M. & Chu, S. Measurement of the gravitational acceleration of an atom with a light-pulse atom interferometer. *Applied Physics B Photophysics and Laser Chemistry* **54.** ISSN: 07217269 (5 1992).

100. Storey, P. & Cohen-Tannoudji, C. The Feynman path integral approach to atomic interferometry: A tutorial. *J. Phys. II* **4,** 1999–2027 (1994).

101. Schleich, W. P., Greenberger, D. M. & Rasel, E. M. Redshift Controversy in Atom Interferometry: Representation Dependence of the Origin of Phase Shift. *Phys. Rev. Lett.* **110,** 010401. https://link.aps.org/doi/10.1103/PhysRevLett.110.010401 (1 Jan. 2013).

102. Bongs, K., Launay, R. & Kasevich, M. High-order inertial phase shifts for time-domain atom interferometers. *Applied Physics B* **84,** 599–602. ISSN: 1432-0649. http://dx.doi.org/10.1007/s00340-006-2397-5 (Aug. 2006).

103. Kritsotakis, M., Szigeti, S. S., Dunningham, J. A. & Haine, S. A. Optimal matter-wave gravimetry. *Phys. Rev. A* **98,** 023629. https://link.aps.org/doi/10.1103/PhysRevA.98.023629 (2 Aug. 2018).

104. Itano, W. M. *et al.* Quantum projection noise: Population fluctuations in two-level systems. *Phys. Rev. A* **47,** 3554–3570. https://link.aps.org/doi/10.1103/PhysRevA.47.3554 (5 May 1993).

105. Le Gouët, J. *et al.* Limits to the sensitivity of a low noise compact atomic gravimeter. *Applied Physics B* **92,** 133–144. ISSN: 1432-0649. http://dx.doi.org/10.1007/s00340-008-3088-1 (June 2008).

106. Anders, F. *et al.* Momentum Entanglement for Atom Interferometry. *Phys. Rev. Lett.* **127,** 140402. https://link.aps.org/doi/10.1103/PhysRevLett.127.140402 (14 Sept. 2021).

107. Malia, B. K., Wu, Y., Martínez-Rincón, J. & Kasevich, M. A. Distributed quantum sensing with mode-entangled spin-squeezed atomic states. *Nature* **612,** 661–665. ISSN: 1476-4687. http://dx.doi.org/10.1038/s41586-022-05363-z (Nov. 2022).

108. Greve, G. P., Luo, C., Wu, B. & Thompson, J. K. Entanglement-enhanced matter-wave interferometry in a high-finesse cavity. *Nature* **610,** 472–477. ISSN: 1476-4687. http://dx.doi.org/10.1038/s41586-022-05197-9 (Oct. 2022).

109. Fuderer, L. A., Hope, J. J. & Haine, S. A. Hybrid method of generating spin-squeezed states for quantum-enhanced atom interferometry. *Phys. Rev. A* **108,** 043722. https://link.aps.org/doi/10.1103/PhysRevA.108.043722 (4 Oct. 2023).

110. Pezzè, L., Smerzi, A., Oberthaler, M. K., Schmied, R. & Treutlein, P. Quantum metrology with nonclassical states of atomic ensembles. *Rev. Mod. Phys.* **90,** 035005. https://link.aps.org/doi/10.1103/RevModPhys.90.035005 (3 Sept. 2018).

111. Gebbe, M. *et al.* Twin-lattice atom interferometry. *Nature Communications* **12.** ISSN: 2041-1723. http://dx.doi.org/10.1038/s41467-021-22823-8 (May 2021).

112. Gauguet, A. *et al.* Off-resonant Raman transition impact in an atom interferometer. *Phys. Rev. A* **78,** 043615. https://link.aps.org/doi/10.1103/PhysRevA.78.043615 (4 Oct. 2008).

113. Carraz, O. *et al.* Phase shift in an atom interferometer induced by the additional laser lines of a Raman laser generated by modulation. *Phys. Rev. A* **86,** 033605. https://link.aps.org/doi/10.1103/PhysRevA.86.033605 (3 Sept. 2012).

114. Zhou, M.-K. *et al.* Effect of the Gaussian distribution of both atomic cloud and laser intensity in an atom gravimeter. *Phys. Rev. A* **93,** 053615. https://link.aps.org/doi/10.1103/PhysRevA.93.053615 (5 May 2016).

115. Zhou, H. *et al.* Impact of additional sidebands generated by a tapered amplifier on an atom interferometer. *Optics Letters* **47,** 4945. ISSN: 1539-4794. http://dx.doi.org/10.1364/OL.469783 (Sept. 2022).

116. Hu, Q.-Q. *et al.* Mapping the absolute magnetic field and evaluating the quadratic Zeeman-effect-induced systematic error in an atom interferometer gravimeter. *Phys. Rev. A* **96,** 033414. https://link.aps.org/doi/10.1103/PhysRevA.96.033414 (3 Sept. 2017).

117. Ozeri, R. *et al.* Hyperfine Coherence in the Presence of Spontaneous Photon Scattering. *Phys. Rev. Lett.* **95,** 030403. https://link.aps.org/doi/10.1103/PhysRevLett.95.030403 (3 July 2005).

118. Uys, H. *et al.* Decoherence due to Elastic Rayleigh Scattering. *Phys. Rev. Lett.* **105,** 200401. https://link.aps.org/doi/10.1103/PhysRevLett.105.200401 (20 Nov. 2010).

119. Lévèque, T., Gauguet, A., Michaud, F., Pereira Dos Santos, F. & Landragin, A. Enhancing the Area of a Raman Atom Interferometer Using a Versatile Double-Diffraction Technique. *Phys. Rev. Lett.* **103,** 080405. https://link.aps.org/doi/10.1103/PhysRevLett.103.080405 (8 Aug. 2009).

120. Schleich, W. P., Greenberger, D. M. & Rasel, E. M. A representation-free description of the Kasevich–Chu interferometer: a resolution of the redshift controversy. *New Journal of Physics* **15,** 013007. https://dx.doi.org/10.1088/1367-2630/15/1/013007 (Jan. 2013).

121. Müller, H., Peters, A. & Chu, S. A precision measurement of the gravitational redshift by the interference of matter waves. *Nature* **463,** 926–929. ISSN: 1476-4687. http://dx.doi.org/10.1038/nature08776 (Feb. 2010).

122. Wolf, P. *et al.* Atom gravimeters and gravitational redshift. *Nature* **467,** E1–E1. ISSN: 1476-4687. http://dx.doi.org/10.1038/nature09340 (Sept. 2010).

123. Müller, H., Peters, A. & Chu, S. Müller, Peters amp; Chu reply. *Nature* **467,** E2–E2. ISSN: 1476-4687. http://dx.doi.org/10.1038/nature09341 (Sept. 2010).

124. Sinha, S. & Samuel, J. Atom interferometry and the gravitational redshift. *Classical and Quantum Gravity* **28,** 145018. https://doi.org/10.1088/0264-9381/28/14/145018 (June 2011).

125. Hohensee, M. A., Chu, S., Peters, A. & Müller, H. Equivalence Principle and Gravitational Redshift. *Physical Review Letters* **106.** ISSN: 1079-7114. http://dx.doi.org/10.1103/PhysRevLett.106.151102 (Apr. 2011).

126. Wolf, P. *et al.* Does an atom interferometer test the gravitational redshift at the Compton frequency? *Classical and Quantum Gravity* **28,** 145017. ISSN: 1361-6382. http://dx.doi.org/10.1088/0264-9381/28/14/145017 (June 2011).

127. Hohensee, M. A., Chu, S., Peters, A. & Müller, H. Comment on: 'Does an atom interferometer test the gravitational redshift at the Compton frequency?' *Classical and Quantum Gravity* **29,** 048001. ISSN: 1361-6382. http://dx.doi.org/10.1088/0264-9381/29/4/048001 (Jan. 2012).

128. Wolf, P. *et al.* Reply to comment on: 'Does an atom interferometer test the gravitational redshift at the Compton frequency?' *Classical and Quantum Gravity* **29,** 048002. https://doi.org/10.1088/0264-9381/29/4/048002 (Jan. 2012).

129. Schleich, W. P., Greenberger, D. M. & Rasel, E. M. Redshift Controversy in Atom Interferometry: Representation Dependence of the Origin of Phase Shift. *Phys. Rev. Lett.* **110,** 010401. https://link.aps.org/doi/10.1103/PhysRevLett.110.010401 (1 Jan. 2013).

130. Di Pumpo, F., Friedrich, A., Ufrecht, C. & Giese, E. Universality-of-clock-rates test using atom interferometry with $T^3$ scaling. *Phys. Rev. D* **107,** 064007. https://link.aps.org/doi/10.1103/PhysRevD.107.064007 (6 Mar. 2023).

131. Overstreet, C., Asenbaum, P. & Kasevich, M. A. Physically significant phase shifts in matter-wave interferometry. *American Journal of Physics* **89,** 324–332. ISSN: 1943-2909. http://dx.doi.org/10.1119/10.0002638 (Mar. 2021).

132. Zych, M., Costa, F., Pikovski, I. & Brukner, Č. Quantum interferometric visibility as a witness of general relativistic proper time. *Nature Communications* **2.** ISSN: 2041-1723. http://dx.doi.org/10.1038/ncomms1498 (Oct. 2011).

133. Rosi, G. *et al.* Quantum test of the equivalence principle for atoms in coherent superposition of internal energy states. *Nature Communications* **8.** ISSN: 2041-1723. http://dx.doi.org/10.1038/ncomms15529 (June 2017).

134. Herrmann, S., Dittus, H. & Lämmerzahl, C. Testing the equivalence principle with atomic interferometry. *Classical and Quantum Gravity* **29,** 184003. ISSN: 1361-6382. http://dx.doi.org/10.1088/0264-9381/29/18/184003 (Aug. 2012).

135. Asenbaum, P., Overstreet, C., Kim, M., Curti, J. & Kasevich, M. A. Atom-Interferometric Test of the Equivalence Principle at the $10^{-12}$ Level. *Physical Review Letters* **125.** ISSN: 1079-7114. http://dx.doi.org/10.1103/PhysRevLett.125.191101 (Nov. 2020).

136. Ufrecht, C. *et al.* Atom-interferometric test of the universality of gravitational redshift and free fall. *Phys. Rev. Res.* **2,** 043240. https://link.aps.org/doi/10.1103/PhysRevResearch.2.043240 (4 Nov. 2020).

137. Ahlers, H. *et al.* Double Bragg Interferometry. *Phys. Rev. Lett.* **116,** 173601. https://link.aps.org/doi/10.1103/PhysRevLett.116.173601 (17 Apr. 2016).

138. Giese, E., Roura, A., Tackmann, G., Rasel, E. M. & Schleich, W. P. Double Bragg diffraction: A tool for atom optics. *Phys. Rev. A* **88,** 053608. https://link.aps.org/doi/10.1103/PhysRevA.88.053608 (5 Nov. 2013).

139. Giese, E. Mechanisms of matter-wave diffraction and their application to interferometers: Mechanisms of matter-wave diffraction and their application to interferometers. *Fortschritte der Physik* **63,** 337–410. ISSN: 0015-8208. http://dx.doi.org/10.1002/prop.201500020 (June 2015).

140. Magnus, W. On the exponential solution of differential equations for a linear operator. *Communications on Pure and Applied Mathematics* **7,** 649–673. ISSN: 1097-0312. http://dx.doi.org/10.1002/cpa.3160070404 (Nov. 1954).

141. Blanes, S., Casas, F., Oteo, J. & Ros, J. The Magnus expansion and some of its applications. *Physics Reports* **470,** 151–238. ISSN: 0370-1573. http://dx.doi.org/10.1016/j.physrep.2008.11.001 (Jan. 2009).

142. Abend, S. *et al.* Atom-Chip Fountain Gravimeter. *Phys. Rev. Lett.* **117,** 203003. https://link.aps.org/doi/10.1103/PhysRevLett.117.203003 (20 Nov. 2016).

143. Fitzek, F. *et al.* Universal atom interferometer simulation of elastic scattering processes. *Scientific Reports* **10.** ISSN: 2045-2322. http://dx.doi.org/10.1038/s41598-020-78859-1 (Dec. 2020).

144. Seckmeyer, S. J., Struckmann, C., Müller, G., Kirsten-Siemß, J.-N. & Gaaloul, N. *FFTArray: A Python Library for the Implementation of Discretized Multi-Dimensional Fourier Transforms* 2025. arXiv: 0902.0885 [physics.comp-ph]. https://arxiv.org/abs/2508.03697.

145. Suzuki, M. Fractal decomposition of exponential operators with applications to many-body theories and Monte Carlo simulations. *Physics Letters A* **146,** 319–323. ISSN: 0375-9601. http://dx.doi.org/10.1016/0375-9601(90)90962-N (June 1990).

146. Rabi, I. I. Space Quantization in a Gyrating Magnetic Field. *Phys. Rev.* **51,** 652–654. https://link.aps.org/doi/10.1103/PhysRev.51.652 (8 Apr. 1937).

147. Steck, D. A. *Quantum and Atom Optics* (available online at http://steck.us/teaching (unpublished), revision 0.16.4, 7 May 2025).

148. Landau, L. D. On the theory of energy transfer. II. *Phys. J. Sov. Union* **2,** 46–51 (1932).

149. Zener, C. & Fowler, R. H. Non-adiabatic crossing of energy levels. *Proceedings of the Royal Society of London. Series A, Containing Papers of a Mathematical and Physical Character* **137,** 696–702. eprint: https://royalsocietypublishing.org/doi/pdf/10.1098/rspa.1932.0165. https://royalsocietypublishing.org/doi/abs/10.1098/rspa.1932.0165 (1932).

150. Ball, H. *et al.* Software tools for quantum control: improving quantum computer performance through noise and error suppression. *Quantum Science and Technology* **6,** 044011. ISSN: 2058-9565. http://dx.doi.org/10.1088/2058-9565/abdca6 (Sept. 2021).

151. Chu, S., Bjorkholm, J. E., Ashkin, A., Gordon, J. P. & Hollberg, L. W. Proposal for optically cooling atoms to temperatures of the order of $10^{6K}$. *Optics Letters* **11,** 73. ISSN: 1539-4794. http://dx.doi.org/10.1364/OL.11.000073 (Feb. 1986).

152. Ammann, H. & Christensen, N. Delta Kick Cooling: A New Method for Cooling Atoms. *Phys. Rev. Lett.* **78,** 2088–2091. https://link.aps.org/doi/10.1103/PhysRevLett.78.2088 (11 Mar. 1997).

153. Morinaga, M., Bouchoule, I., Karam, J.-C. & Salomon, C. Manipulation of Motional Quantum States of Neutral Atoms. *Phys. Rev. Lett.* **83,** 4037–4040. https://link.aps.org/doi/10.1103/PhysRevLett.83.4037 (20 Nov. 1999).

154. Kovachy, T. *et al.* Matter Wave Lensing to Picokelvin Temperatures. *Phys. Rev. Lett.* **114,** 143004. https://link.aps.org/doi/10.1103/PhysRevLett.114.143004 (14 Apr. 2015).

155. Deppner, C. *et al.* Collective-Mode Enhanced Matter-Wave Optics. *Phys. Rev. Lett.* **127,** 100401. https://link.aps.org/doi/10.1103/PhysRevLett.127.100401 (10 Aug. 2021).

156. Abend, S. *Atom-chip gravimeter with Bose-Einstein condensates* PhD thesis (Gottfried Wilhelm Leibniz Universität Hannover, 2017). https://www.repo.uni-hannover.de/handle/123456789/8974.

157. Müller, H., Chiow, S.-w., Long, Q., Herrmann, S. & Chu, S. Atom Interferometry with up to 24-Photon-Momentum-Transfer Beam Splitters. *Phys. Rev. Lett.* **100,** 180405. https://link.aps.org/doi/10.1103/PhysRevLett.100.180405 (18 May 2008).

158. Chiow, S.-w., Kovachy, T., Chien, H.-C. & Kasevich, M. A. 102$\hbar k$ Large Area Atom Interferometers. *Phys. Rev. Lett.* **107,** 130403. https://link.aps.org/doi/10.1103/PhysRevLett.107.130403 (13 Sept. 2011).

159. Kovachy, T., Chiow, S.-w. & Kasevich, M. A. Adiabatic-rapid-passage multiphoton Bragg atom optics. *Phys. Rev. A* **86,** 011606. https://link.aps.org/doi/10.1103/PhysRevA.86.011606 (1 July 2012).

160. Kovachy, T. *et al.* Quantum superposition at the half-metre scale. *Nature* **528,** 530–533. ISSN: 1476-4687. http://dx.doi.org/10.1038/nature16155 (Dec. 2015).

161. Peters, A., Chung, K. Y. & Chu, S. Measurement of gravitational acceleration by dropping atoms. *Nature* **400,** 849–852. ISSN: 1476-4687. http://dx.doi.org/10.1038/23655 (Aug. 1999).

162. Poli, N. *et al.* Precision Measurement of Gravity with Cold Atoms in an Optical Lattice and Comparison with a Classical Gravimeter. *Phys. Rev. Lett.* **106,** 038501. https://link.aps.org/doi/10.1103/PhysRevLett.106.038501 (3 Jan. 2011).

163. Karcher, R., Imanaliev, A., Merlet, S. & Santos, F. P. D. Improving the accuracy of atom interferometers with ultracold sources. *New Journal of Physics* **20,** 113041. https://dx.doi.org/10.1088/1367-2630/aaf07d (Nov. 2018).

164. Szigeti, S. S., Nolan, S. P., Close, J. D. & Haine, S. A. High-Precision Quantum-Enhanced Gravimetry with a Bose-Einstein Condensate. *Phys. Rev. Lett.* **125,** 100402. https://link.aps.org/doi/10.1103/PhysRevLett.125.100402 (10 Sept. 2020).

165. Snadden, M., McGuirk, J., Bouyer, P., Haritos, K. & Kasevich, M. Measurement of the Earth's Gravity Gradient with an Atom Interferometer-Based Gravity Gradiometer. *Physical Review Letters* **81,** 971–974. ISSN: 1079-7114. http://dx.doi.org/10.1103/PhysRevLett.81.971 (Aug. 1998).

166. McGuirk, J. M., Foster, G. T., Fixler, J. B., Snadden, M. J. & Kasevich, M. A. Sensitive absolute-gravity gradiometry using atom interferometry. *Phys. Rev. A* **65**, 033608. https://link.aps.org/doi/10.1103/PhysRevA.65.033608 (3 Feb. 2002).

167. Rosi, G. *et al.* Measurement of the Gravity-Field Curvature by Atom Interferometry. *Phys. Rev. Lett.* **114**, 013001. https://link.aps.org/doi/10.1103/PhysRevLett.114.013001 (1 Jan. 2015).

168. Gustavson, T. L., Bouyer, P. & Kasevich, M. A. Precision Rotation Measurements with an Atom Interferometer Gyroscope. *Phys. Rev. Lett.* **78**, 2046–2049. https://link.aps.org/doi/10.1103/PhysRevLett.78.2046 (11 Mar. 1997).

169. Stockton, J. K., Takase, K. & Kasevich, M. A. Absolute Geodetic Rotation Measurement Using Atom Interferometry. *Phys. Rev. Lett.* **107**, 133001. https://link.aps.org/doi/10.1103/PhysRevLett.107.133001 (13 Sept. 2011).

170. Gautier, R. *et al.* Accurate measurement of the Sagnac effect for matter waves. *Science Advances* **8**, eabn8009. eprint: https://www.science.org/doi/pdf/10.1126/sciadv.abn8009. https://www.science.org/doi/abs/10.1126/sciadv.abn8009 (2022).

171. D'Armagnac de Castanet, Q. *et al.* Atom interferometry at arbitrary orientations and rotation rates. *Nature Communications* **15**. ISSN: 2041-1723. http://dx.doi.org/10.1038/s41467-024-50804-0 (July 2024).

172. Stolzenberg, K. *et al.* Multi-Axis Inertial Sensing with 2D Matter-Wave Arrays. *Phys. Rev. Lett.* **134**, 143601. https://link.aps.org/doi/10.1103/PhysRevLett.134.143601 (14 Apr. 2025).

173. Pelluet, C. *et al.* Atom interferometry in an Einstein Elevator. *Nature Communications* **16**. ISSN: 2041-1723. http://dx.doi.org/10.1038/s41467-025-60042-7 (May 2025).

174. Fixler, J. B., Foster, G. T., McGuirk, J. M. & Kasevich, M. A. Atom Interferometer Measurement of the Newtonian Constant of Gravity. *Science* **315**, 74–77. eprint: https://www.science.org/doi/pdf/10.1126/science.1135459. https://www.science.org/doi/abs/10.1126/science.1135459 (2007).

175. Lamporesi, G., Bertoldi, A., Cacciapuoti, L., Prevedelli, M. & Tino, G. M. Determination of the Newtonian Gravitational Constant Using Atom Interferometry. *Physical Review Letters* **100**. ISSN: 1079-7114. http://dx.doi.org/10.1103/PhysRevLett.100.050801 (Feb. 2008).

176. Rosi, G., Sorrentino, F., Cacciapuoti, L., Prevedelli, M. & Tino, G. M. Precision measurement of the Newtonian gravitational constant using cold atoms. *Nature* **510**, 518–521. ISSN: 1476-4687. http://dx.doi.org/10.1038/nature13433 (June 2014).

177. Torii, Y. *et al.* Mach-Zehnder Bragg interferometer for a Bose-Einstein condensate. *Phys. Rev. A* **61**, 041602. https://link.aps.org/doi/10.1103/PhysRevA.61.041602 (4 Feb. 2000).

178. Bordé, C. Atomic interferometry with internal state labelling. *Physics Letters A* **140,** 10–12. ISSN: 0375-9601. https://www.sciencedirect.com/science/article/pii/0375960189905379 (1989).

179. Li, R., Martínez-Lahuerta, V. J., Seckmeyer, S., Hammerer, K. & Gaaloul, N. Robust double Bragg diffraction via detuning control. *Phys. Rev. Res.* **6,** 043236. https://link.aps.org/doi/10.1103/PhysRevResearch.6.043236 (4 Dec. 2024).

180. Hartmann, S. *et al.* Regimes of atomic diffraction: Raman versus Bragg diffraction in retroreflective geometries. *Phys. Rev. A* **101,** 053610. https://link.aps.org/doi/10.1103/PhysRevA.101.053610 (5 May 2020).

181. Kirsten-Siemß, J.-N. *et al.* Large-Momentum-Transfer Atom Interferometers with $\mu$rad-Accuracy Using Bragg Diffraction. *Phys. Rev. Lett.* **131,** 033602. https://link.aps.org/doi/10.1103/PhysRevLett.131.033602 (3 July 2023).

182. Lahuerta, V. J. M., Kirsten-Siemß, J.-N., Hammerer, K. & Gaaloul, N. *Diffraction phase-free Bragg atom interferometry* 2025. arXiv: 2505.23921 [physics.atom-ph]. https://arxiv.org/abs/2505.23921.

183. Estey, B., Yu, C., Müller, H., Kuan, P.-C. & Lan, S.-Y. High-Resolution Atom Interferometers with Suppressed Diffraction Phases. *Phys. Rev. Lett.* **115,** 083002. https://link.aps.org/doi/10.1103/PhysRevLett.115.083002 (8 Aug. 2015).

184. Malossi, N. *et al.* Double diffraction in an atomic gravimeter. *Phys. Rev. A* **81,** 013617. https://link.aps.org/doi/10.1103/PhysRevA.81.013617 (1 Jan. 2010).

185. Campbell, J. E. On a Law of Combination of Operators bearing on the Theory of Continuous Transformation Groups. *Proceedings of the London Mathematical Society* **s1-28,** 381–390. ISSN: 0024-6115. http://dx.doi.org/10.1112/plms/s1-28.1.381 (Nov. 1896).

186. Campbell, J. E. On a Law of Combination of Operators (Second Paper) *. *Proceedings of the London Mathematical Society* **s1-29,** 14–32. ISSN: 0024-6115. http://dx.doi.org/10.1112/PLMS/S1-29.1.14 (Nov. 1897).

187. Rossmann, W. *Lie Groups: An Introduction Through Linear Groups* ISBN: 9781383030990. http://dx.doi.org/10.1093/oso/9780198596837.001.0001 (Oxford University PressOxford, Jan. 2002).

188. Bloch, F. & Siegert, A. Magnetic Resonance for Nonrotating Fields. *Phys. Rev.* **57,** 522–527. https://link.aps.org/doi/10.1103/PhysRev.57.522 (6 Mar. 1940).

189. Bloch, F. Nuclear Induction. *Physical Review* **70,** 460–474. ISSN: 0031-899X. http://dx.doi.org/10.1103/PhysRev.70.460 (Oct. 1946).

190. Abragam, A. *The principles of nuclear magnetism* **32** (Oxford university press, 1961).

191. Szigeti, S. S., Debs, J. E., Hope, J. J., Robins, N. P. & Close, J. D. Why momentum width matters for atom interferometry with Bragg pulses. *New Journal of Physics* **14,** 023009. https://dx.doi.org/10.1088/1367-2630/14/2/023009 (Feb. 2012).

192. Müller, H., Chiow, S.-w., Long, Q., Herrmann, S. & Chu, S. Atom Interferometry with up to 24-Photon-Momentum-Transfer Beam Splitters. *Phys. Rev. Lett.* **100,** 180405. https://link.aps.org/doi/10.1103/PhysRevLett.100.180405 (18 May 2008).

193. Chiow, S.-w., Kovachy, T., Chien, H.-C. & Kasevich, M. A. 102ℏk Large Area Atom Interferometers. *Phys. Rev. Lett.* **107,** 130403. https://link.aps.org/doi/10.1103/PhysRevLett.107.130403 (13 Sept. 2011).

194. Ball, H. *et al.* Software tools for quantum control: improving quantum computer performance through noise and error suppression. *Quantum Science and Technology* **6,** 044011. https://doi.org/10.1088/2058-9565/abdca6 (2021).

195. Roura, A., Zeller, W. & Schleich, W. P. Overcoming loss of contrast in atom interferometry due to gravity gradients. *New Journal of Physics* **16,** 123012. ISSN: 1367-2630. http://dx.doi.org/10.1088/1367-2630/16/12/123012 (Dec. 2014).

196. Abend, S. *et al.* Terrestrial very-long-baseline atom interferometry: Workshop summary. *AVS Quantum Science* **6.** ISSN: 2639-0213. http://dx.doi.org/10.1116/5.0185291 (May 2024).

197. Delhuille, R. *et al.* Fringe Contrast in Mach–Zehnder Atom Interferometers. *Acta Physica Polonica B* **33,** 2157 (Aug. 2002).

198. López-Monjaraz, C., Peña Vega, H., Jiménez-García, K., López Romero, J. M. & Corzo, N. V. Impact of the -pulse shape on the contrast of thermal Atom Interferometers. *Physica Scripta* **99,** 125414. ISSN: 1402-4896. http://dx.doi.org/10.1088/1402-4896/ad92b2 (Nov. 2024).

199. Pethick, C. J. & Smith, H. *Bose-Einstein Condensation in Dilute Gases (2nd ed.)* ISBN: 9780511802850 (Cambridge: Cambridge University Press, Jan. 2011).

200. Suzuki, M. Fractal decomposition of exponential operators with applications to many-body theories and Monte Carlo simulations. *Physics Letters A* **146,** 319–323. ISSN: 0375-9601. http://dx.doi.org/10.1016/0375-9601(90)90962-N (June 1990).

201. Müllers, A. *et al.* Coherent perfect absorption of nonlinear matter waves. *Science Advances* **4,** eaat6539. eprint: https://www.science.org/doi/pdf/10.1126/sciadv.aat6539. https://www.science.org/doi/abs/10.1126/sciadv.aat6539 (2018).